\documentclass[fleqn,usenatbib]{mnras}

\usepackage[T1]{fontenc}
\usepackage{ae,aecompl}

\usepackage{graphicx}	% Including figure files
\usepackage{amsmath}	% Advanced maths commands
\usepackage{amssymb}	% Extra maths symbols
\usepackage{lipsum} 
\usepackage{deluxetable}
\usepackage{rotating}
\usepackage{longtable}
\usepackage{xcolor}
\usepackage{mathrsfs}
\usepackage{caption}
\usepackage{subcaption}
\usepackage{lastpage}

\newcommand{\nhone}{N({\rm H\,{\scriptsize I}})}

\newcommand{\sbunit}{\rm erg \, s^{-1}\, cm^{-2}\,arcsec^{-2}}
\newcommand{\ang}{\text{\AA}}

\newcommand{\lya}{${\rm Ly\alpha}$}

\newcommand{\halpha}{H$\alpha$}
\newcommand{\hbeta}{H$\beta$}

\providecommand{\kms}{\,\ensuremath{\rm{km\,s}^{-1}}}

\newcommand{\mstar}{M_{\rm star}}
\newcommand{\msun}{{\rm M}_\odot}

\title[Resolved galactic superwinds at $z>3$]{Resolved galactic superwinds reconstructed around their host galaxies at z>3}

\author[M. C. Chen et al.]{Mandy C. Chen$^{1}$\thanks{E-mail: mandychen@astro.uchicago.edu}, Hsiao-Wen Chen$^{1}$, Max Gronke$^{2}$\thanks{Hubble Fellow}, Michael Rauch$^{3}$, and \newauthor  Tom Broadhurst$^{4,5,6}$
\\
$^{1}$Department of Astronomy and Astrophysics, The University of Chicago, Chicago, IL 60637, USA\\
$^{2}$Department of Physics \& Astronomy, Johns Hopkins University, Baltimore, MD 21218, USA\\
$^{3}$Carnegie Observatories, 813 Santa Barbara Street, Pasadena, CA 91101, USA\\
$^{4}$Department of Theoretical Physics, University of the Basque Country UPV/EHU, Bilbao, Spain\\
$^{5}$Donostia International Physics Center (DIPC), 20018 Donostia,Spain\\
$^{6}$IKERBASQUE, Basque Foundation for Science, Bilbao, Spain
}

\date{Accepted XXX. Received YYY; in original form ZZZ}

\pubyear{2021}

\begin{document}
\label{firstpage}
\pagerange{\pageref{firstpage}--\pageref{lastpage}}
\maketitle

\begin{abstract}
This paper presents a detailed analysis of two giant Lyman-alpha (Ly$\alpha$) arcs detected near galaxies at $z=3.038$ and $z=3.754$ lensed by the massive cluster MACS\,1206$-$0847 ($z=0.44$).  The \lya\ nebulae revealed in deep MUSE observations exhibit a double-peak profile with a dominant red peak, indicating expansion/outflowing motions.  One of the arcs stretches over $1'$ around the cluster Einstein radius, resolving the velocity field of the line-emitting gas on kpc scales around three star-forming galaxies of $0.3$--$1.6\,L_*$ at $z=3.038$.  The second arc spans $15''$ in size, roughly centered around two low-mass \lya\ emitters of $\approx 0.03\,L_*$ at $z=3.754$. All three galaxies in the $z=3.038$ group exhibit prominent damped \lya\ absorption (DLA) and several metal absorption lines, in addition to nebular emission lines such as \ion{He}{II}$\lambda\,1640$ and C\,III]$\lambda\lambda$1906, 1908.  Extended \lya\ emission appears to emerge from star-forming regions with suppressed surface brightness at the center of each galaxy.  Significant spatial variations in the \lya\ line profile are observed which, when unaccounted for in the integrated line, leads to biased constraints for the underlying gas kinematics.  The observed spatial variations indicate the presence of a steep velocity gradient in a continuous flow of high column density gas from star-forming regions into a low-density halo environment. A detailed inspection of available galaxy spectra shows no evidence of AGN activity in the galaxies, and the observed \lya\ signals are primarily explained by resonant scattering. The study presented in this paper shows that spatially-resolved imaging spectroscopy provides the most detailed insights yet into the kinematics of galactic superwinds associated with star-forming galaxies.
%thought to be responsible for the chemical enrichment in the intergalactic medium. 
%These observations provide the most detailed insights yet into the kinematics of galactic superwinds associated with star-forming galaxies thought to be responsible for the chemical enrichment in the intergalactic medium.

\end{abstract}


%%%%%%%%%%%%%%%%%%%%%%%%%%%%%%%%%%%%%%%%%%%%%%%%%%

%%%%%%%%%%%%%%%%% BODY OF PAPER %%%%%%%%%%%%%%%%%%

\section{Introduction}
The formation and evolution of galaxies are intimately connected to the properties of the circumgalactic medium (CGM).  Characterizing the interactions between galaxies and their surrounding gas, such as gas infall and outflows, is a critical step toward improving our still patchy understanding of the life cycles of baryons and galaxy evolution over cosmic time.  But because of the low-density nature of the CGM, studying such tenuous gas has historically relied on absorption spectroscopy along individual QSO sightlines.  Over the past few decades, absorption-line studies have yielded sensitive, mostly one-dimensional constraints on the gas density, temperature, metallicity and kinematics in the circumgalactic space \citep[see the review by][and references therein]{Chen2017, Tumlinson2017, Rudie2019}. 
However, uncertainties remain in connecting gas to galaxies in the absence of a spatially-resolved two-dimensional map of the gas.  To access the spatial information of gas distribution in the CGM, direct detections of the tenuous gas in emission provide a promising avenue. 
The hydrogen \lya\ line, being the strongest emission line expected of photo-ionized gas at a temperature $T\sim 10,000$ K, provides a sensitive probe of the tenuous CGM \citep[e.g.,][]{OsterbrockFerland2006,Draine2011}.  At $z\approx 2$--7, the \lya\ line at 1215 \AA\ is conveniently redshifted into the atmospheric transmission window and becomes accessible on the ground.  In the past two decades, narrow-band imaging and deep long-slit spectroscopic observations have successfully revealed extended line-emitting gas around galaxies \citep[e.g.,][]{Adelberger2006, Rauch2008,Rauch2011,Steidel2011,Xue2017} and QSOs \citep[e.g.,][]{Hennawi2009,Cantalupo2012,Cantalupo2014Natur}. Those observations have shed light on several important physical processes in the CGM, such as the ubiquity of large-scale gas flows on 10--100 physical kpc (pkpc) scales at high redshifts \citep[e.g.,][]{Rauch2016} and the non-trivial contribution of star-forming galaxies to reionization \citep[e.g.,][]{Dijkstra2014, Matthee2018}. 

The recent advent of high-throughput, wide-field optical integral field spectrographs (IFSs) on large ground-based telescopes, such as the Multi Unit Spectroscopic Explorer (MUSE) on the Very Large Telescopes (VLT) \citep{Bacon2010} and the Keck Cosmic Web Imager (KCWI) on the Keck Telescopes \citep{Morrissey2018} has brought a significant breakthrough in systematically uncovering extended \lya\ emission in typical, low-mass galaxies as well as QSOs at $z\approx 2-7$ \citep[e.g.,][]{Wisotzki2016,Wisotzki2018,Borisova2016,Leclercq2017,Cai2017,Cai2019,Battaia2019}.  These sensitive IFS observations have uncovered extended \lya\ emission out to $>20$ times the spatial extent of the stellar continuum, and revealed key insights into the physical nature of these extended \lya\ sources. For example, significant spatial variations of \lya\ line profiles are directly observed within a single line-emitting nebula \citep[e.g.,][]{Rauch2013,Vanzella2017,Erb2018}.  In addition, there exists a positive correlation between the full-width-at-half-maximum (FWHM) of the \lya\ line and the continuum UV brightness of the associated star-forming galaxies \citep[e.g.,][]{Wisotzki2018,Leclercq2020}, indicating an intimate connection between the origin of the \lya\ photons and star-forming activities \citep[e.g.,][]{Dijkstra2012, Cantalupo2017review}. 
 
Multiple processes can lead to \lya\ emission in the CGM, such as fluorescence powered by ionizing photons from star-forming regions or active galactic neuclei (AGN), cooling radiation, and scattering of \lya\ photons by mostly neutral hydrogen gas \citep[e.g.,][]{HoganWeymann1987,GouldWeinberg1996,Cantalupo2005,Kollmeier2010,FaucherGiguere2010,Hennawi2013}. Disentangling different processes that contribute to the observed \lya\ signal is challenging due to the resonant scattering nature of \lya\ photons, especially when \lya\ is the only observable line feature from the emission regions. At the same time, the detailed double-peak profiles of spectrally-resolved \lya\ lines provide a sensitive probe of the underlying gas kinematics. It is expected that \lya\ emission originating in infalling and outflowing medium will result in blue-enhanced and red-enhanced peak, respectively \citep[e.g.,][and references therein]{Dijkstra2017}. This has motivated increasingly sophisticated Monte Carlo radiative transfer models that incorporate different gas geometry and kinematics to accurately track \lya\ photon scattering and infer the physical properties of the gaseous clouds \citep[e.g.,][]{Dijkstra2006,Verhamme2006,HansenOh2006,Laursen2009,Schaerer2011,Gronke2015}. 

These Monte Carlo \lya\ radiative transfer codes can generally reproduce the observed \lya\ line width based on a combination of thermal broadening and bulk motions, but significant discrepancies are also seen between observations and model predictions \citep{Verhamme2008,Kulas2012,Orlitova2018}.  Such discrepancies have both theoretical and observational implications. Theoretically, there is a lot of room for better capturing the physical processes in radiative transfer simulations, such as a realistic treatment of dust attenuation and gas clumpiness \citep[e.g.,][]{Laursen2009,Dijkstra2012,Gronke2016}.  Observationally, as \lya\ photons are scattered both in spectral and spacial dimensions, it is critical to obtain observations with not only high spectral resolution, but also high spatial resolution to provide the best constraints on the source environment.

Strong gravitational lensing provides sharpened images of the high-redshift Universe via an enhanced spatial resolution of highly magnified images of distant galaxies \citep[e.g.,][]{Coe2013} and recently individual, luminous high redshift stars \citep{Kelly2018}.
Massive galaxy and cluster lenses have revealed detailed properties of lensed background sources down to sub-kpc or even as detailed as tens of pc scales \citep[e.g.,][]{Livermore2012,Bordoloi2016,Johnson2017,Berg2018,Florian2020}.  Multiply-lensed QSOs and extended, lensed arcs of bright background sources have been used to spatially resolve the diffuse CGM in absorption spectroscopy \citep{Rauch2002,Chen2014,Zahedy2016, Rubin2018,Lopez2018,Mortensen2020}.
Several gravitationally-lensed \lya\ emitting nebulae have also been reported, in which the enhanced spatial resolution has aided to reveal the underlying physical environment of the source in greater details \citep{Swinbank2007,Patricio2016,Caminha2017,Claeyssens2019,Erb2019}.

\begin{figure*}
	\includegraphics[width=\linewidth]{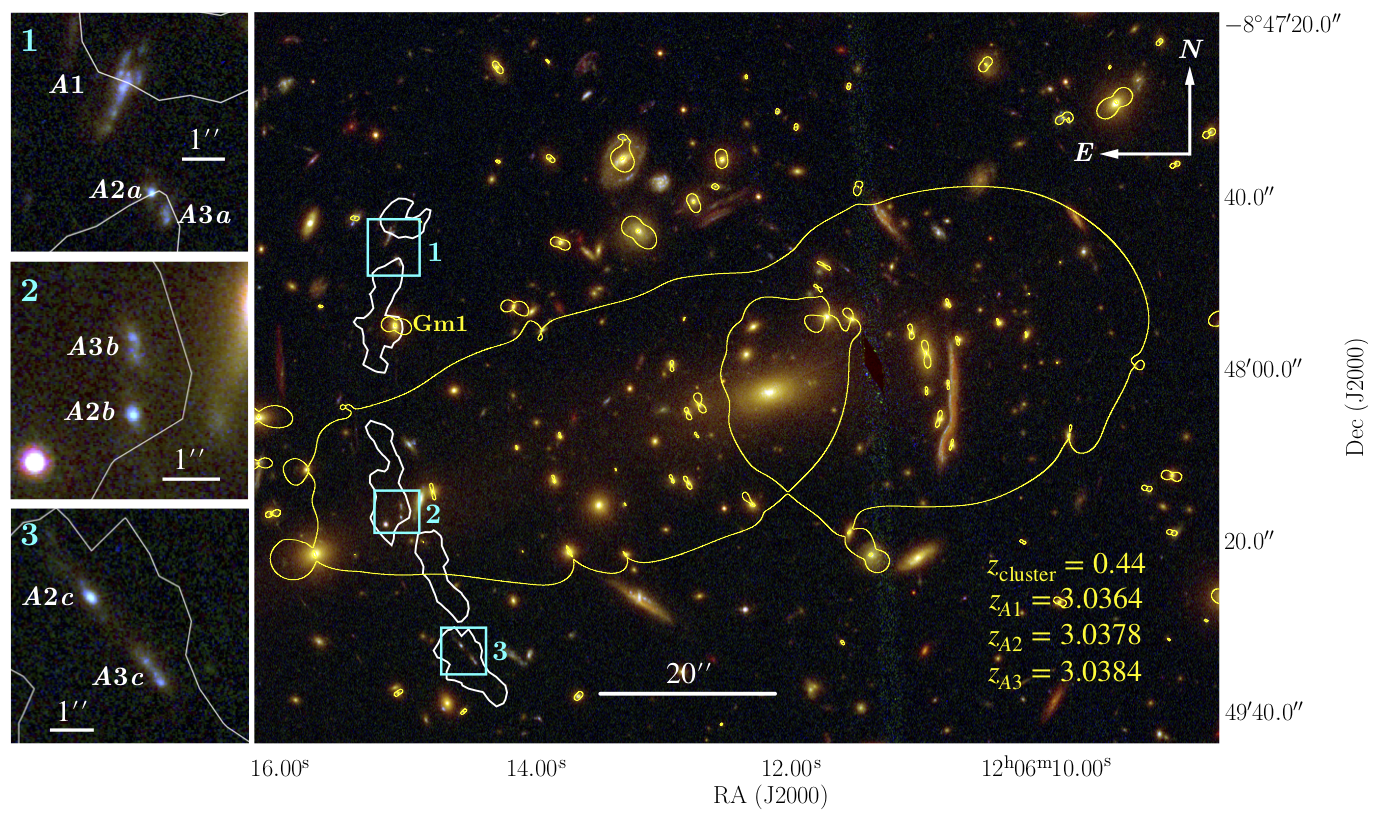}
     \caption{Composite image of the core region of MACS\,1206, produced using {\it HST} F475W (blue), F814W (green) and F160W (red) images.  White contours indicate the Ly$\alpha$ emission associated with System {\it A} at a surface brightness of ${\rm SB}_{{\rm Ly}\alpha}=3.7\times10^{-18} \, \mathrm{erg \, s^{-1} \, cm^{-2} \, arcsec^{-2}}$, integrated over the spectral window of 4890-4930 \AA\ (see \S\ \ref{sec:Lya_NB} below).  The surface brightness limit corresponds to a 3-$\sigma$ limiting flux over a circular aperture of $1\arcsec$ in diameter, roughly the size of the PSF measured in the MUSE data. Yellow contours show the critical curve of the cluster lens for a source at $z=3.038$. Left panels show zoomed-in regions around lensed images of galaxies {\it A}1, {\it A}2, and {\it A}3, along with the Ly$\alpha$ contours. Note that the galaxy {\it A}1 at $z=3.0364$ is magnified but not multiply-lensed. Cluster member galaxy Gm1 is located close to lensed images of System {\it A} and is individually optimised in the lens modeling process as described in \S\ref{sec:lens_model}. After correcting for the lensing magnification, the total \lya\ luminosity from the nebula is $L_{{\rm Ly}\alpha}=(5.2\pm 0.1)\times 10^{42} \, {\rm erg\,s^{-1}}$ (see \S\ \ref{sec:Lya_NB}).}
    \label{fig:hst_with_lya_contours_A}
\end{figure*}

\begin{figure*}
    \centering
	\includegraphics[width=\linewidth]{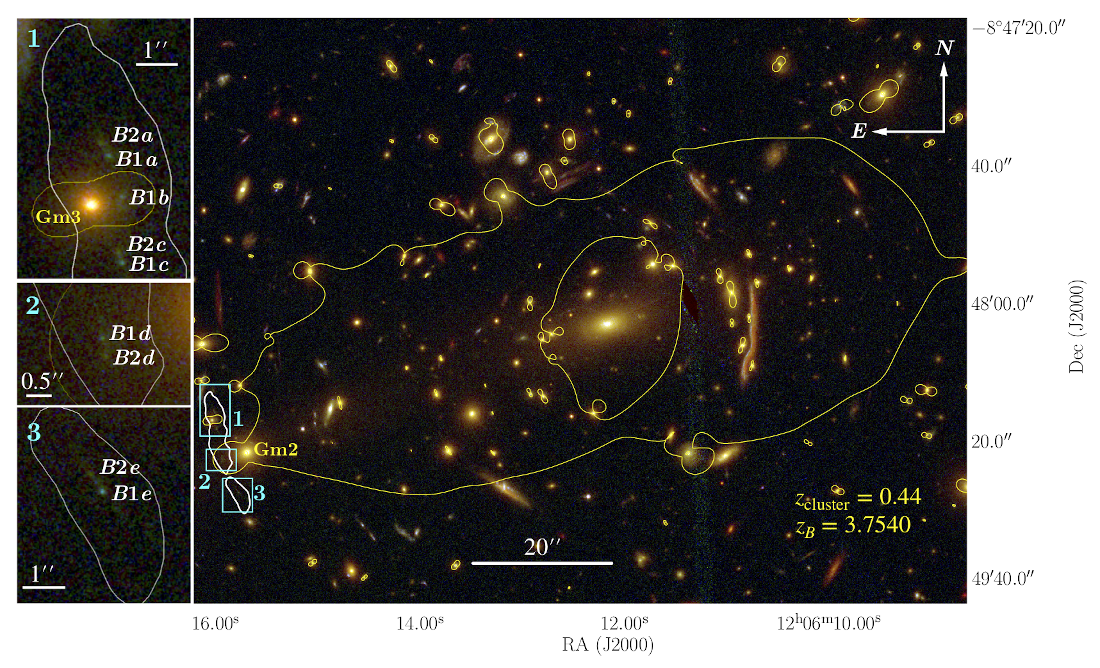}
    \caption{Same as Figure~\ref{fig:hst_with_lya_contours_A}, while highlighting the configuration of System {\it B}. White contours indicate the Ly$\alpha$ emission associated with System {\it B} at ${\rm SB}_{{\rm Ly}\alpha}=2.8\times10^{-18} \, \mathrm{erg \, s^{-1} \, cm^{-2} \, arcsec^{-2}}$, integrated over the spectral window of 5766-5796 \AA\ (see \S\ \ref{sec:Lya_NB} below).  The surface brightness limit corresponds to a 3-$\sigma$ limiting flux over a circular aperture of $1\arcsec$ in diameter, roughly the size of the PSF measured in the MUSE data. Yellow contours show the critical curve of the cluster lens for a source at $z=3.754$.  Left panels show zoomed-in regions around the lensed images of galaxy {\it B} consisting of components {\it B}1 and {\it B}2, along with the Ly$\alpha$ contours. Cluster member galaxies Gm2 and Gm3 are located close to lensed images of System {\it B} and are individually optimised in the lens modeling process as described in \S\ref{sec:lens_model}. After correcting for the lensing magnification, the total \lya\ luminosity from the nebula  is $L_{{\rm Ly}\alpha}=(9.8\pm0.2)\times 10^{41} \, {\rm erg\,s^{-1}}$ (see \S\ref{sec:Lya_NB}).}
    \label{fig:hst_with_lya_contours_B}
\end{figure*}

Here we present a detailed analysis of two gravitationally-lensed \lya\ emitting nebulae, System {\it A} at $z=3.038$ (Figure \ref{fig:hst_with_lya_contours_A}) and System {\it B} at $z=3.754$ (Figure \ref{fig:hst_with_lya_contours_B}), detected in deep MUSE observations of the field around the strong lensing cluster, MACS\,1206$-$0847 at $z=0.44$ (hereafter MACS\,1206).  Both nebulae are multiply-lensed to form giant tangential arcs in the image plane around the Einstein radius of the foreground cluster, and both exhibit a double-peaked \lya\ profile.  In particular, the serendipitous alignment of the nebula in System {\it A} results in an extended low surface brightness arc of ${\rm SB}_{{\rm Ly}\alpha}\approx 3\times 10^{-18}\, \mathrm{erg \, s^{-1} \, cm^{-2} \, arcsec^{-2}}$ and $\approx 1\arcmin$ in length, comprising three contiguous lensed images \citep{Caminha2017}, while System {\it B} forms an arc of $\approx 15''$ in length with high surface brightness peaks exceeding ${\rm SB}_{{\rm Ly}\alpha}\approx 2\times 10^{-17}\, \mathrm{erg \, s^{-1} \, cm^{-2} \, arcsec^{-2}}$.  In addition, the \lya\ emitting region in System {\it A} consists of two separate nebulae, detached from a group of three continuum sources with one being an $\approx 1.6\,L_*$ galaxy and the other two being sub-$L_*$ galaxies.  All three of these galaxies exhibit prominent interstellar absorption lines, including hydrogen damped \lya\ absorption (DLA) in their spectra.  One of the sub-$L_*$ galaxies ({\it A}3 in Figure~\ref{fig:hst_with_lya_contours_A} below) is further resolved into two high-intensity peaks.  In contrast, the \lya\ nebula in System {\it B} exhibits a symmetric morphology in the source plane, centered approximately at two compact continuum sources separated by $\approx 0\farcs1$ ($\approx 0\farcs3$--$0\farcs5$ in the image plane), both of which are low-luminosity $\approx 0.03\,L_*$ \lya\ emitters (LAE) with a %strong \lya\ emission line and a faint UV continuum 
rest-frame \lya\ equivalent width of $W ({\rm Ly\alpha})\approx 30\, \ang$.  

In this study, we examine the underlying gas flows by combining  spatially-resolved \lya\ emission profiles from MUSE and known star formation properties of the neighboring galaxies from available {\it Hubble Space Telescope} ({\it HST}) broadband photometry.
This paper is organized as follows.  First, the archival data included in our analysis are presented in Section~\ref{sec:data}, including broadband imaging data by {\it HST} and IFS data by VLT/MUSE.  The lens models fine-tuned to best reproduce multiple images from Systems {\it A} and {\it B} are presented in Section~\ref{sec:lens_model}. In Sections~\ref{sec:analysis_gal} and \ref{sec:analysis_gas}, we present detailed analysis of UV continuum galaxies and the \lya\ line-emitting gas, respectively.  We discuss our results in Section~\ref{sec:discussion}, and conclude in Section~\ref{sec:summary_conclusion}.  Throughout this paper, we adopt a Hubble constant of $H_0=70$ km/s/Mpc, $\Omega_\mathrm{M}=0.3$ and $\Omega_\Lambda = 0.7$ when deriving distances, masses and luminosities.  All magnitudes quoted are in the AB system.

\section{Observational Data}\label{sec:data}

MACS\,1206 is a well studied cluster, which was first identified as a luminous X-ray source in the ROSAT All Sky Survey \citep{Voges1999,Bohringer2001} and later confirmed to be a massive, strong-lensing cluster by the Massive Cluster Survey \citep{Ebeling2001,Ebeling2009}.  It was also selected as one of the 25 clusters in the {\it Cluster Lensing And Supernova Survey with Hubble} (CLASH) program \citep{CLASH2012}.  Exquisite imaging and spectroscopic data of this cluster field are available in public data archives, including high-quality multi-band imaging data from the {\it HST}, follow-up galaxy spectroscopic survey data from the CLASH-VLT redshift survey \citep{Biviano2013,Rosati2014}, and wide-field IFS data obtained using VLT/MUSE \citep{Bacon2010,Caminha2017}. High-level science products are retrieved from these public data archives for our study. In this section, we provide a summary of these data products.

\subsection{{\it HST} images}

\label{sec:data_hst_photometry} 
High spatial resolution, UV, optical and near-infrared imaging data of the field around MACS\,1206 obtained using the {\it HST} were retrieved from the Mikulski Archive for Space Telescopes (MAST) archive\footnote{\url{https://archive.stsci.edu/pub/hlsp/clash/macs1206/data/hst/scale_30mas}} (PI: M.\ Postman).  These include images taken using the Advanced Camera for Surveys (ACS), the Wide Field Camera 3 (WFC3), and a suite of UV, optical, and near-infrared filters (see Table~\ref{tab:photometryA} below).  Figures~\ref{fig:hst_with_lya_contours_A} and \ref{fig:hst_with_lya_contours_B} show composite images of the central region of MACS\,1206 from combining F475W (blue), F814W (green), and F160W (red) images, highlighting the lensing configurations of System {\it A} and System {\it B}, respectively. Detailed photometric properties of each system derived from these {\it HST} data are described in \S\ref{sec:analysis_gal}.

Given the close proximity of {\it B}1 and {\it B}2 in the source plane (see \S\ref{sec:Lya_NB} below), it is possible that they correspond to distinct star-forming regions in the same galaxy at $z=3.754$. However, without high-resolution infrared data, we cannot determine confidently whether or not {\it B}1 and {\it B}2 originate in the same galaxy. We therefore proceed with referring to {\it B}1 and {\it B}2 as individual galaxies for simplicity. 

\subsection{MUSE IFS Data}\label{sec:data_muse} 

Wide-field IFS data of MACS\,1206 were obtained using MUSE on the VLT UT4.  The observations were carried out under Program ID's 095.A-0181(A) and 097.A-0269(A) to cover an effective area of 2.63 arcmin$^2$ around the cluster in three pointings (PI: J.\ Richard), which are part of a systematic survey of 12 strong lensing clusters using MUSE \citep[][]{Richard2020}.   In the region where lensed images of Systems {\it A} and {\it B} are found, a total exposure time of $\approx 4$ hours were collected.  Pipeline-processed and flux-calibrated data cubes were retrieved from the ESO Phase 3 Archive, covering a wavelength range of 4750-9300 \AA\ with a spectral resolution of ${\rm FWHM}\approx 170$ (110) \kms\ at $\approx 5000$ (7000) \AA\ and a pixel scale of $0\farcs2\times0\farcs2$.  The mean point spread function (PSF) in the final combined data cube was determined using a bright star, and found to be $\approx 1\arcsec$ at $7000$ \AA.  Astrometry of the combined MUSE data cube was re-calibrated to match the world coordinate system of available {\it HST} images.  The pipeline generated combined data cube contains non-negligible sky residuals that affected the detection of faint emission features.  Additional sky subtraction was therefore performed using a median sky residual spectrum generated from object-free spaxels in the data cube. Detailed spectroscopic properties of both continuum sources and \lya\ emitting nebulae are described in \S\ref{sec:analysis_gal} and \S\ref{sec:analysis_gas}, respectively.  Finally, the wavelength array is converted to vacuum to facilitate accurate velocity calculations based on known rest-frame UV wavelengths.

\section{Cluster lens modeling}\label{sec:lens_model}

To determine the intrinsic properties of both Systems {\it A} and {\it B}, it is necessary to construct a cluster lens model to correct for the gravitational lensing effect. Here we employ the software {\it LENSTOOL} (version 6.5) \citep{lenstool} to construct a parametric cluster lens model of MACS\,1206 by incorporating known multiply-lensed galaxies identified in the MUSE data \citep{Caminha2017} and those reported in the literature \citep[e.g.,][]{Zitrin2012,Umetsu2012,Eichner2013}.  As both Systems {\it A} and {\it B} are in the core region of the cluster, we only include the strong lensing constraints and do not consider weak lensing effect in our lens modeling process.  We first obtain a fiducial cluster lens model that gives a good fit to a total of 72 multiple images from 21 background sources.  Those images cover a field of view (FOV) of $\approx 2\arcmin$ relatively evenly, providing robust constraints for the projected cluster mass distribution within this FOV.  We then fine-tune the lens model by considering only multiple images of Systems {\it A} and {\it B}, optimising the mass distribution projected close to those particular images as well as the multiply-lensed extended Ly$\alpha$ emitting nebulae.  Details regarding the lens modeling process are described below.

\subsection{Fiducial cluster lens model}
Following \cite{Caminha2017}, we adopt a parametric model based on a pseudo-isothermal elliptical mass distribution (PIEMD) \citep{PIEMD1993} of ellipticity $\epsilon$ and include two additional isothermal halo components to represent the cluster-scale diffuse mass.  This three-halo configuration is found to minimize the dispersion between predicted and observed image positions for all multiply-lensed sources \cite[see][for detailed discussions]{Caminha2017}.  The convergence of PIEMD is given by
\begin{equation}
\kappa_c = \frac{\sigma_{v}^2}{2\,G\,\Sigma_{cr}\,\sqrt{R_{\epsilon}^{2}+r_{c}^{2}}} , \label{eq_PIEMD} 
\end{equation}
where $R_{\epsilon}$ is the distance from the center of the cluster, defined as
\begin{equation}
R_{\epsilon}^2 = \frac{x^2}{(1+\epsilon)^2}+\frac{y^2}{(1-\epsilon)^2}, \label{eq_elliptical radius} 
\end{equation}
$r_{c}$ is the core radius and $\Sigma_{cr}$ is the projected critical mass density.  Given the angular diameter distances between the observer and the lens ($D_l$), the lens and the source ($D_{ls}$),
and the observer and the source ($D_s$), the projected critical mass density is defined as 
\begin{equation}
    \Sigma_{cr} = \frac{c^2}{4\pi G}\frac{D_s}{D_lD_{ls}}.
\end{equation} 
All six parameter of the three PIEMD halos (x, y, $r_c$, $\epsilon$, position angle, velocity dispersion $\sigma_v$) are free to vary. We also include external shear (parameterized by the intensity $\gamma_{\rm shear}$ and position angle $\theta_{\rm shear}$) to account for possible massive structures in regions further away from the cluster core. 

In addition to the cluster-scale diffuse mass distribution, we account for local perturbations in the vicinity of individual galaxies by including 128 cluster member galaxies in the lens model.  These member galaxies are selected based on their redshifts in the catalog of \cite{Molino2017}, which is downloaded from the MAST archive \footnote{\url{https://archive.stsci.edu/pub/hlsp/clash/macs1206/catalogs/molino/}}. We first eliminate galaxies fainter than $AB=24$ mag in the F160W band. For galaxies with spectroscopic redshifts, we select those with $0.425<z_{\rm spec}<0.453$.  For galaxies without $z_{\rm spec}$, we apply the same criterion based on available photometric redshifts.  A total of 128 cluster members are selected from this exercise.  Note that in general, the cluster lensing potential is dominated by the large-scale diffuse mass distribution, which is primarily in the form of dark matter. Member galaxies only introduce perturbations local to the location of individual galaxies. Therefore, in cases where lensed images do not appear close to individual member galaxies ($\leq5\arcsec$, corresponding to typical Einstein radius of individual galaxies), the variation in the selection of member galaxies does not introduce significant uncertainties to the cluster lensing potential. However, in cases where lensed images form close to individual galaxies, careful modeling of those individual galaxies is required to accurately reproduce the positions of nearby images. As our goal here is to obtain a good cluster-scale lens model instead of optimising individual galaxy mass distributions, we exclude image systems 2, 7, 13, 21, 24 and 27 in \cite{Caminha2017} (see their Fig. 1), whose multiple images fall very close to massive cluster member galaxies. This way we do not need to fine-tune every member galaxy with lensed images nearby and still maintain the accuracy of the large-scale cluster lens model. 

We include cluster member galaxies as 128 dual pseudo-isothermal elliptical mass distributions (dPIE) \citep{dPIE2007} located at their detected light centroids, with the ellipticity and position angle fixed to their observed values obtained from the \cite{Molino2017} catalog.  The convergence of the dPIE profile is given by
\begin{equation}
\kappa_g = \frac{\sigma_{g,v}^2}{2\,G\,\Sigma_{cr}}\left (\frac{1}{R_{g,\epsilon}}-\frac{1}{\sqrt{R_{g,\epsilon}^2+r_{g,t}^2}}\right) , \label{eq_dPIE} 
\end{equation}
where $r_{g,t}$ is the truncation radius. To reduce the total number of free parameters, we scale all 128 member galaxies with a constant mass-to-light ratio through
\begin{equation}
    \sigma_{g,v} = \sigma_{g,v}^0(\frac{L}{L_0})^{\frac{1}{4}}, \, r_{g,t} = r_{g,t}^0(\frac{L}{L_0})^{\frac{1}{4}},
\end{equation}
where $L_0$ is the reference luminosity with magnitude $m_{\rm F814W}=19.6$. Hence there are only two free parameters for member galaxies: $\sigma_{g,v}^0$ and $r_{g,t}^0$.

\begin{table}
    \centering
    \caption{Mean lensing magnification of multiple images of Systems {\it A} and {\it B}. Calculated based on the fine-tuned lens model as described in Section \ref{sec:fine_tune_lens_model}.}
    \begin{tabular}{lrclr}
    \hline
     \multicolumn{1}{c}{Image} & \multicolumn{1}{c}{$\Bar{\mu}$} & | & \multicolumn{1}{c}{Image} & \multicolumn{1}{c}{$\Bar{\mu}$} \\
    \hline
    {\it A}1 & 3.8 & | & {\it B}1{\it a} & 15.2 \\
    {\it A}2{\it a} & 4.3 & | & {\it B}1{\it c} & 10.4\\
    {\it A}2{\it b} & 5.4 & | & {\it B}1{\it d} & 12.1 \\
    {\it A}2{\it c} & 7.5 & | & {\it B}1{\it e} & 7.5 \\
    {\it A}3{\it a} & 4.5 & | & {\it B}2{\it a} & 8.2 \\  
    {\it A}3{\it b} & 4.4 & | & {\it B}2{\it c} & 13.0 \\
    {\it A}3{\it c} & 6.2 & | & {\it B}2{\it d} & 12.0 \\
        &     & | & {\it B}2{\it e} & 8.4 \\
    \hline
    \end{tabular}
%    \caption{Mean lensing magnification of multiple images of Systems {\it A} and {\it B}. Calculated based on the fine-tuned lens model as described in Section \ref{sec:fine_tune_lens_model}.}
    \label{tab:magnification}
\end{table}

Constraints of this fiducial cluster lens model are positions of 72 multiple images from 21 background sources identified by \cite{Caminha2017}, excluding image systems 2, 7, 13, 21, 24 and 27 for reasons described above. The optimization is performed based on object positions in the source plane.  We obtain similar best-fit parameters as \cite{Caminha2017}. The root-mean-square positional offset between observed and predicted images is $\langle\,{\rm rms}\,\rangle_{\rm im} =0\farcs76$ in the image plane, averaged over all 72 images of 21 sources.  The rms position offsets for Systems {\it A} and {\it B} are found to be $\langle\,{\rm rms}\,\rangle_{\rm im}=0\farcs38$ and $\langle\,{\rm rms}\,\rangle_{\rm im}=0\farcs73$, respectively. In the Appendix, we list the coordinates and redshifts of all 72 images used as constraints, as well as the best-fit model parameters. 

\subsection{Fine-tuned lens model for Systems {\it A} and {\it B}}\label{sec:fine_tune_lens_model}
Based on the fiducial cluster lens model described above, we now optimize the lens model for Systems {\it A} and {\it B} separately to ensure the highest accuracy in matching the observed locations of multiply-lensed images in these two systems. In the fiducial model, the respective centers of the three cluster-scale PIEMD halos are located at $\approx 2\arcsec$ from the brightest cluster galaxy (BCG), $\approx 13\arcsec$ northwest and $\approx 30\arcsec$ southeast of the BCG (see Table~\ref{tab:lenstool_parameters_fiducial} for a summary). As the southeast cluster-scale PIEMD halo occurs close to the lensed images of Systems {\it A} and {\it B}, we obtain a refined lens model, leaving all parameters of this PIEMD halo free while fixing the other two cluster-scale PIEMD halos to their best-fit parameters in the fiducial model. We also notice that three of the cluster member galaxies (marked as Gm1, Gm2 and Gm3 in Figures \ref{fig:hst_with_lya_contours_A} and \ref{fig:hst_with_lya_contours_B}) are located close to some images of Systems {\it A} and {\it B}. We therefore allow the velocity dispersion $\sigma_{g,v}$ and truncation radius $r_{g,t}$ of these three cluster members to vary freely in the fine-tuned model optimization, instead of being scaled together with the rest of member galaxies. Finally, the external shear parameters are fixed to their best-fit values in the fiducial model. 

Because we are particularly interested in accurately producing the lensing effect for Systems {\it A} and {\it B}, we also include constraints from the two substructures of {\it A}3 (designated {\it A}31 and {\it A}32 in Table \ref{tab:multiple_image_coordinates1}), and the fainter galaxy {\it B}2 in System {\it B}, which are not used in \cite{Caminha2017}. With a total of 18 multiple images of {\it A} and {\it B} as constraints (the first 18 entries in Table~\ref{tab:multiple_image_coordinates1}), we then run {\it LENSTOOL} again with the above set-up, and obtain a fine-tuned model. This model places significantly more weight on the local perturbers (Gm1, Gm2, and Gm3) and provides much improved root-mean-square positional offsets for the systems of interest in this study.  The rms position offsets for Systems {\it A} and {\it B} are reduced to ${\rm rms}_{\rm im}=0\farcs1$ and ${\rm rms_{im}}=0\farcs21$, respectively.  The best-fit parameters are listed Table~\ref{tab:lenstool_parameters_finetune} in the Appendix. In Figures~\ref{fig:hst_with_lya_contours_A} and \ref{fig:hst_with_lya_contours_B}, we show the predicted critical curves by this fine-tuned model for sources at the redshifts of Systems {\it A} and {\it B}.  Mean lensing magnification factors of multiple images of Systems {\it A} and {\it B} based on the fine-tuned lens model are presented in Table \ref{tab:magnification}.  Wherever required in subsequent analyses, we use this fine-tuned model to derive image position deflections and magnifications. 
 
\section{Analysis: galaxy properties}\label{sec:analysis_gal}

Both Systems {\it A} and {\it B} comprise two distinct components: (1) the continuum sources detected in the broadband {\it HST} images and (2) the \lya\ emitting nebulae that are more spatially extended than the continuum sources and are only visible in the MUSE IFS data. Available broadband photometry and spectra of the galaxies provide important constraints for the star formation histories and the underlying stellar populations. In this section, we investigate the properties of the galaxies by analysing the photometric and spectroscopic data of the continuum sources.  We will present the analysis of the associated \lya\ emitting nebulae in \S\ref{sec:analysis_gas}.

\begin{table*}
	\centering
	\caption{Summary of galaxy photometry for System {\it A}$^a$.}
	
	\label{tab:photometryA}
	\begin{tabular}{c | c c c c c c c c} 
 		\hline
 		& redshift & $M_{\rm 1500}^b$ & F300W$^c$ & F390W & F435W & F475W & F606W & F625W  \\ %& F850LP & F105W & F110W & F125W & F140W & F160W & Spitzer 3.6$\mu$m \\
 		\hline
{\it A}1 & 3.0364 & $-21.52$ & $>26.14$ & $25.85 \pm 0.08$ & $25.02 \pm 0.03$ & $24.59 \pm 0.02$ & $24.03 \pm 0.07$ & $23.90 \pm 0.01$    \\
{\it A}2 & 3.0378 & $-19.87$ & $>28.37$ & $26.97 \pm 0.12$ & $26.29 \pm 0.05$ & $26.10 \pm 0.03$ & $25.67 \pm 0.04$ & $25.43 \pm 0.02$    \\
{\it A}3 & 3.0384 & $-19.63$ & $>28.57$ & $27.53 \pm 0.12$ & $26.82 \pm 0.05$ & $26.46 \pm 0.03$ & $25.91 \pm 0.04$ & $25.76 \pm 0.02$   \\
        \hline
         & F775W & F814W & F850LP & F105W & F110W & F125W & F140W & F160W     \\
 		\hline
{\it A}1 & $23.76 \pm 0.01$ & $23.76 \pm 0.01$ & $23.73 \pm 0.02$ & $23.73 \pm 0.01$ & $23.71 \pm 0.01$ & $23.71 \pm 0.01$ & $23.52 \pm 0.01$ & $23.34 \pm 0.01$  \\
{\it A}2 & $25.37 \pm 0.02$ & $25.32 \pm 0.01$ & $25.31 \pm 0.03$ & $25.47 \pm 0.02$ & $25.50 \pm 0.01$ & $25.54 \pm 0.02$ & $25.36 \pm 0.01$ & $25.32 \pm 0.01$    \\
{\it A}3 & $25.66 \pm 0.02$ & $25.62 \pm 0.01$ & $25.61 \pm 0.03$ & $25.72 \pm 0.02$ & $25.70 \pm 0.01$ & $25.75 \pm 0.02$ & $25.51 \pm 0.01$ & $25.34 \pm 0.01$   \\
		\hline
		\multicolumn{9}{l}{$^a$All magnitudes are de-magnified  based on the lens model described in \S\ref{sec:lens_model}, and averaged among images {\it a} and {\it c}.} \\
		\multicolumn{9}{l}{$^b$At $z=3$, typical star-forming galaxies have $M_{\rm 1500}*=-21.1\pm 0.2$ \citep[e.g.,][]{Reddy2008}}\\
		\multicolumn{9}{l}{$^c$$2\sigma$ UV flux upper limit, averaged among the F225W, F275W and F336W bandpasses.}
\end{tabular}
\end{table*}

\begin{table*}
	\centering
	\caption{Summary of galaxy photometry for System {\it B}$^a$.}
	
	\label{tab:photometryB}
	\begin{tabular}{c | c c c c c c c c} 
 		\hline
 		& redshift & $M_{\rm 1500}^b$ & F450W$^c$ & F475W & F606W & F625W & F775W & F814W  \\ %& F110W & F125W & F140W & F160W & Spitzer 3.6$\mu$m \\
 		\hline
{\it B}1  & 3.7540 & $-17.23$ & $>30.08$ & $ 29.99\pm 0.20 $ & $ 29.15\pm 0.06 $ & $ 28.94\pm 0.08 $ & $ 28.69\pm 0.08 $ & $ 28.81\pm 0.05 $  \\
{\it B}2  & 3.7540 & $-17.01$ & $>29.99$ & $>30.68^d$ & $ 29.59\pm 0.11 $ & $ 29.57\pm 0.17 $ & $ 28.91\pm 0.11 $ & $ 28.95\pm 0.07 $  \\
        \hline
        & F850LP & F105W & F110W & F125W & F140W & F160W &  &  \\
 		\hline
{\it B}1 & $ 28.80\pm 0.11 $ & $ 29.29\pm 0.09 $ & $ 29.13\pm 0.05 $ & $ 29.19\pm 0.09 $ & $ 29.28\pm 0.08 $ & $ 29.22\pm 0.08 $ & &  \\
{\it B}2 & $ 28.87\pm 0.15 $ & $ 29.15\pm 0.10 $ & $ 29.11\pm 0.06 $ & $ 29.12\pm 0.10 $ & $ 28.98\pm 0.07 $ & $ 28.83\pm 0.07 $ & & \\ 
        \hline
		\multicolumn{9}{l}{$^a$All magnitudes are de-magnified  based on the lens model described in \S\ref{sec:lens_model}., and averaged among images {\it a}, {\it c} and {\it d}.} \\
		\multicolumn{9}{l}{$^b$At $z=4$, typical star-forming galaxies have $M_{\rm 1500}*=-21.1\pm 0.1$ \citep[e.g.,][]{Bouwens2007}}\\
		\multicolumn{9}{l}{$^c$$2\sigma$ UV flux upper limit, averaged among the F225W, F275W, F336W, F390W and F435W bandpasses.}\\
		\multicolumn{9}{l}{$^d$$2\sigma$ flux upper limit.}
\end{tabular}
\end{table*}

\begin{table*}
    \centering
    \caption{SED fitting results, showing $16\%$--$84\%$ confidence interval for each parameter. }
    \begin{tabular}{c c c c c c c c}
    \hline
     galaxy & redshift & ${\rm log} (\mstar/\msun)$ & SFR ($\msun \, \mathrm{yr^{-1}}$) & Age (Gyr) & $\tau$ (Gyr) & $A_V$  \\
    \hline
    {\it A}1 & 3.0364 & $[9.93, 9.98]$ & $ [89.84, 101.85] $ & $[0.11,0.14]$ & $[1.37,4.35]$ & $[0.72,0.77]$ \\[0.15cm]
    {\it A}2 & 3.0378 & $[8.95, 8.98]$ & $ [10.71, 11.45] $ & $[0.05,0.06]$ & $[1.30,4.38]$ & $[0.62,0.65]$ \\[0.15cm]
    {\it A}3 & 3.0384 & $[9.23,9.27]$ & $ [13.02,15.81] $ & $[0.14,0.19]$ & $[1.44,4.37]$ & $[0.64,0.71]$ \\[0.15cm]
    {\it B}1 & 3.7540 & $[7.59,7.96]$ & $ [0.23,0.40] $ & $[0.13,0.53]$ & $[1.26,4.25]$ & $[0.05,0.25]$ \\[0.15cm]
    {\it B}2 & 3.7540 & $[8.43,8.72]$ & $ [0.50,0.91] $ & $[0.43,1.31]$ & $[1.31,4.35]$ & $[0.47,0.74]$ \\
    \hline
    \end{tabular}
    \label{tab:Bagpipes_sedfitting}
\end{table*}

\subsection{Photometric properties}\label{sec:galaxy_photometry}

Accurate photometric measurements of galaxies in Systems {\it A} and {\it B} are challenging due to the crowding of members of the lensing cluster and non-negligible intracluster light (e.g., Figures \ref{fig:hst_with_lya_contours_A} \& \ref{fig:hst_with_lya_contours_B}).  We first measure broadband magnitudes of individual lensed images of each galaxy in different bandpasses using a combination of circular and isophotal apertures determined by SExtractor \citep[v.2.19.5;][]{SExtractor1996}.  These measurements (presented in the Appendix) are then corrected for lensing magnifications based on the fine-tuned lens model (see Table \ref{tab:magnification} presented in \S\ref{sec:lens_model}).  
%A mean de-magnified magnitude per galaxy in each bandpass is then calculated based on an average across multiple images.  

For galaxies {\it A}2 and {\it A}3 in System {\it A}, their {\it b} images occur between two bright foreground galaxies, resulting in uncertain background subtraction in the photometric measurements.  The de-magnified apparent magnitudes of {\it A}2 and {\it A}3 are therefore determined based on an average of images a and c.  The de-magnified magnitudes of {\it A}2 and {\it A}3 in image {\it a} are $\approx 0.2$ magnitudes fainter than that in image c, suggesting that the true magnification factor for image {\it a} relative to image {\it c} is smaller than what is predicted by the lens model. In \S\ref{sec:analysis_gas} below, we also show that the apparent \lya\ surface brightness in the extended nebulae from image {\it a} is fainter than what is seen in images {\it b} and {\it c}, supporting a smaller relative magnification factor at the location of image {\it a}. Such a discrepancy in image brightnesses is commonly seen in strongly-lensed galaxies and quasars, and is often due to the limited accuracy of lens models and/or the presence of small-scale substructures in the lens \citep[e.g.][]{McKean2007,Hezaveh2016}.  The discrepancy of $\approx 0.2$ magnitudes seen here is within the typical scatter of $\gtrsim 25\%$ between de-lensed magnitudes of multiply-lensed galaxies in cluster lenses \citep[e.g.][]{Lam2014,Caminha2016_b}. By averaging the de-lensed magnitudes between images {\it a} and {\it c}, we therefore mitigate the effect of lensing uncertainty on the magnification of these two galaxies.

Similarly, the {\it b} images of galaxies {\it B}1 and {\it B}2 are excluded due to the contamination from the nearby cluster member galaxy Gm3.  In addition, image {\it e} of {\it B}1 is unusually bright compared with its counter part in images {\it a}, {\it c} and {\it d}, which are between 0.8 and 1.2 magnitudes fainter than image {\it e} across different bandpasses after the lensing correction.  Such an enhancement in brightness is not observed in image {\it e} of {\it B}2. This brightness anomaly of {\it B}1{\it e} can also be seen in the color image in Figure~\ref{fig:hst_with_lya_contours_B}, and may be attributed to magnification perturbation caused by unseen substructures local to {\it B}1{\it e} \citep[e.g.][]{McKean2007,Hezaveh2016}. Consequently, the de-magnified apparent magnitudes of {\it B}1 and {\it B}2 are determined by averaging measurements of images {\it a}, {\it c} and {\it d}.  Finally, Galactic extinction corrections are calculated using the NED Galactic Extinction Calculator\footnote{\url{https://ned.ipac.caltech.edu/extinction_calculator}} and applied to the observed magnitudes in individual bandpasses following the \cite{Schlafly2011} extinction map.   

For galaxies in System {\it A} ({\it B}), the bandpasses bluer of F390W (F475W) correspond to rest-frame wavelengths $\lambda_{\rm rest}<912\text{\AA}$, and no fluxes are detected above the background noise.  We therefore place a 2-$\sigma$ upper limit of the observed flux in each of these bandpasses. Unfortunately, these images are not sufficiently sensitive to provide meaningful constraints for the escape fraction of ionizing photons from these galaxies. 
The final de-magnified apparent magnitudes of galaxies {\it A} and {\it B} in different bandpasses are presented in Tables \ref{tab:photometryA} and \ref{tab:photometryB}, while the direct measurements of individual images are presented in Table \ref{tab:photometry_all} for reference. 

To characterize the intrinsic luminosities of these galaxies, we also estimate the rest-frame UV absolute magnitudes at 1500 \AA, $M_{\rm 1500}$, using the observed F606W (F775W) brightness for galaxies in System {\it A} ({\it B}).  At the respective redshifts of Systems {\it A} and {\it B}, these bandpasses correspond roughly to the rest-frame 1500 \AA, and provide a robust estimate of the intrinsic UV luminosity.  The absolute magnitudes of {\it A}1, {\it A}2, {\it A}3, {\it B}1 and {\it B}2, at rest-frame $1500\ang$ are found to be $M_{\rm 1500}=-21.52$, $-19.87$, $-19.63$, $-17.23$ and $-17.01$, corresponding to 1.61, 0.35, 0.28, 0.03, $0.03\,L_*$, respectively, for a characteristic rest-frame absolute magnitude of $M_* = -21$ \citep[e.g.,][]{Bouwens2007,Reddy2008}.

\subsection{Stellar population parameters}\label{sec:SED_fitting}

The observed broadband spectral energy distributions (SEDs) of galaxies in Systems {\it A} and {\it B} based on the photometric measurements presented in Tables~\ref{tab:photometryA} and \ref{tab:photometryB} are typical of star-forming galaxies at $z=3$--4 \citep[e.g.,][]{Bouwens2007}.  To quantify the star formation histories, we perform a stellar population synthesis analysis 
using {\it Bayesian Analysis of Galaxies for Physical Inference and Parameter EStimation} \citep[\textsc{Bagpipes},][]{Bagpipes2018}, which employs the 2016 version of the \cite{BC03} stellar synthesis models.  We assume an exponentially declining star formation model, SFR$(t)\propto e^{-t/\tau}$, where $\tau$ represents the e-folding time and is a free parameter, and infer the stellar mass ($\mstar$), star formation rate (SFR), age and dust extinction ($A_V$) of the continuum sources in both systems based on the observed SEDs from F606W to F160W.  Because of a strong degeneracy between stellar age, metallicity, and dust attenuation \citep[e.g.,][]{Conroy2013}, we impose a metallicity prior based on the mass-metallicity relation for high-redshift galaxies \citep[e.g.,][]{Ma2016} and fix the metallicity of {\it A}1 to $20\%$ of the solar value, 10\% for {\it A}2 and {\it A}3, and 5\% for {\it B}1 and {\it B}2.

\begin{figure*}
    \centering
    \includegraphics[width=17cm]{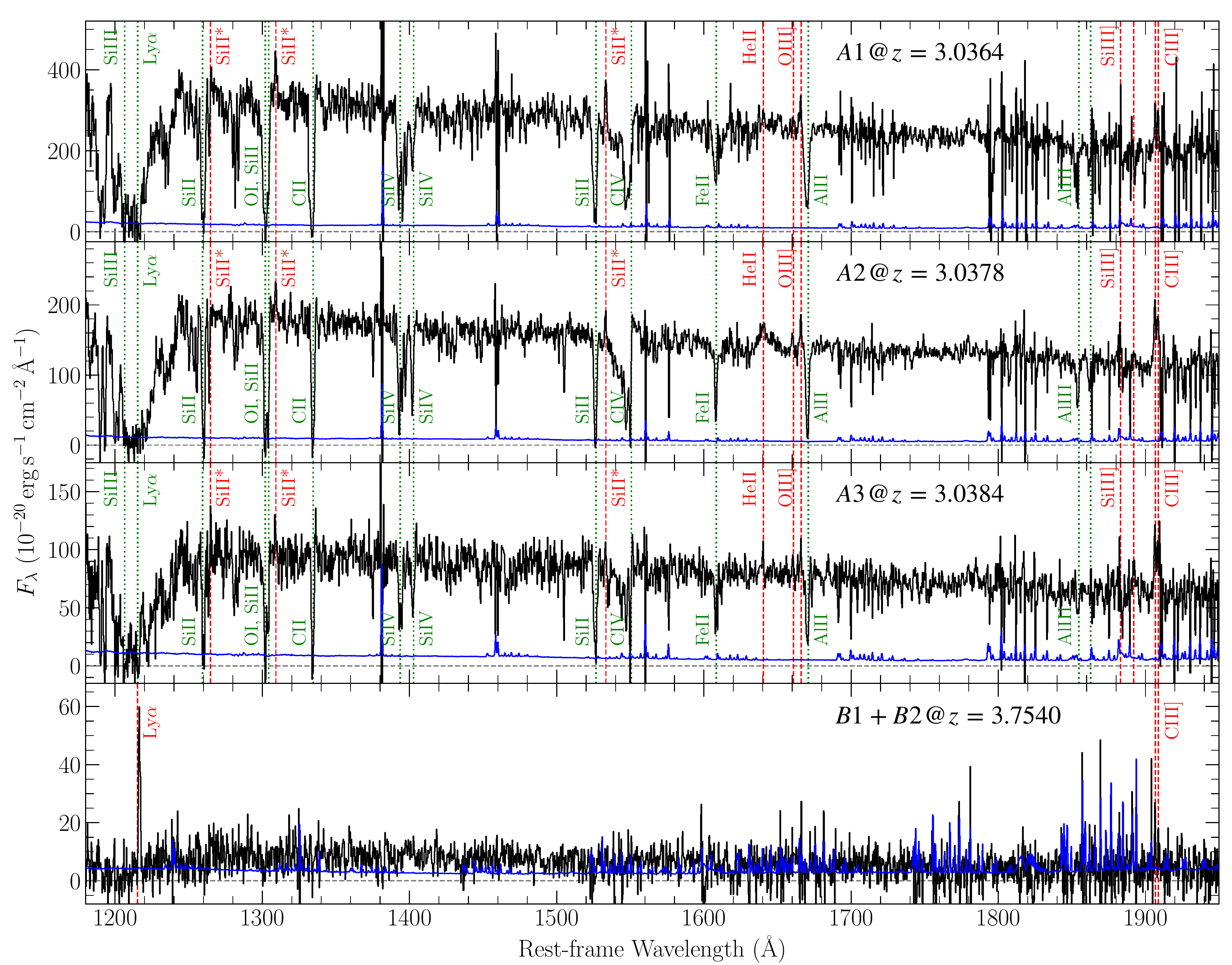}
    \caption{Sky-subtracted spectra of Systems {\it A} and {\it B} without lensing magnification corrections. For {\it A}2 and {\it A}3, multiple images {\it a} and {\it c} are stacked. For {\it B}1+{\it B}2 combined spectrum, images {\it a}, {\it c}, {\it d} and {\it e} are stacked. Rest-frame wavelength is calculated according to the best-fit redshift of each galaxy, as marked in their individual panels. The corresponding 1-$\sigma$ error spectrum is shown in blue in each panel.  Red dashed lines indicate major emission features while green dotted lines indicate major absorption features. } 
    \label{fig:galaxy_fullspec_A}
\end{figure*}

The $16\%$--$84\%$ confidence intervals in $\mstar$, SFR, age, and $A_V$ are presented in Table~\ref{tab:Bagpipes_sedfitting}.  All five galaxies are best characterized by a long star formation e-folding time that exceeds $\tau=1$ Gyr, along with a relatively young, best-fit stellar age.  In particular, the best-fit stellar ages for galaxies in System {\it A} are less than 200 Myr, making the adopted exponentially declining star formation model equivalent to a constant SFR scenario. This makes the inferred stellar age and SFR insensitive to the adopted star formation history, either exponentially declining or rising \citep[][]{Reddy2012}. As discussed below, a constant star formation history is also consistent with the spectral features uncovered in the MUSE data.  The inferred SFR for galaxies {\it A}1, {\it A}2, and {\it A}3 range between 10 and 100 $\msun\,{\rm yr}^{-1}$ and $\mstar$ between $10^9$ and $10^{10}\,\msun$, typical of UV luminous star-forming galaxies at $z\approx 3$ \citep[e.g.][]{Shapley2011}.  In contrast, galaxies {\it B}1 and {\it B}2 have significantly lower SFR and stellar mass with $\mstar\approx 10^8\,\msun$, more typical of \lya\ emitters (LAE) at $z\approx 3$ with a characteristic star formation time scale of $\lesssim 1$ Gyr \citep[e.g.][]{Feltre2020}.  

\subsection{Spectroscopic properties}\label{sec:galaxy_spec}

At $z=3$--4, available MUSE data cover the rest-frame wavelength range from $\lambda_{\rm rest}>1200$ \AA\ to $\lambda_{\rm rest}<1920$ \AA, and provide additional constraints for the star-forming interstellar medium (ISM) and the stellar populations in Systems {\it A} and {\it B}.  We extract individual galaxy spectra using spherical apertures centered on the location of the continuum sources, with varying sizes for different images depending on the intrinsic image size and magnification.  Because galaxies {\it B}1 and {\it B}2 are blended in the ground-based MUSE data, we are only able to extract a single spectrum for these two galaxies.  The extracted spectra (without lensing correction) are presented in Figure \ref{fig:galaxy_fullspec_A}, along with the corresponding 1-$\sigma$ error spectra. For galaxies {\it A}2 and {\it A}3, the spectra shown are combined from images {\it a} and {\it c}, while image {\it b} is excluded due to possible contamination from nearby cluster member galaxies.  Similarly for {\it B}1 and {\it B}2, image b is excluded from the combined spectrum due to possible contamination from the nearby elliptical galaxy.  Note that the brightness anomaly of {\it B}1e described in \S\ref{sec:galaxy_photometry} does not affect the spectral features due to the achromatic nature of lensing.  Image {\it e} is therefore included in the combined spectrum. 

\begin{figure*}
    \centering
    \includegraphics[width=16cm]{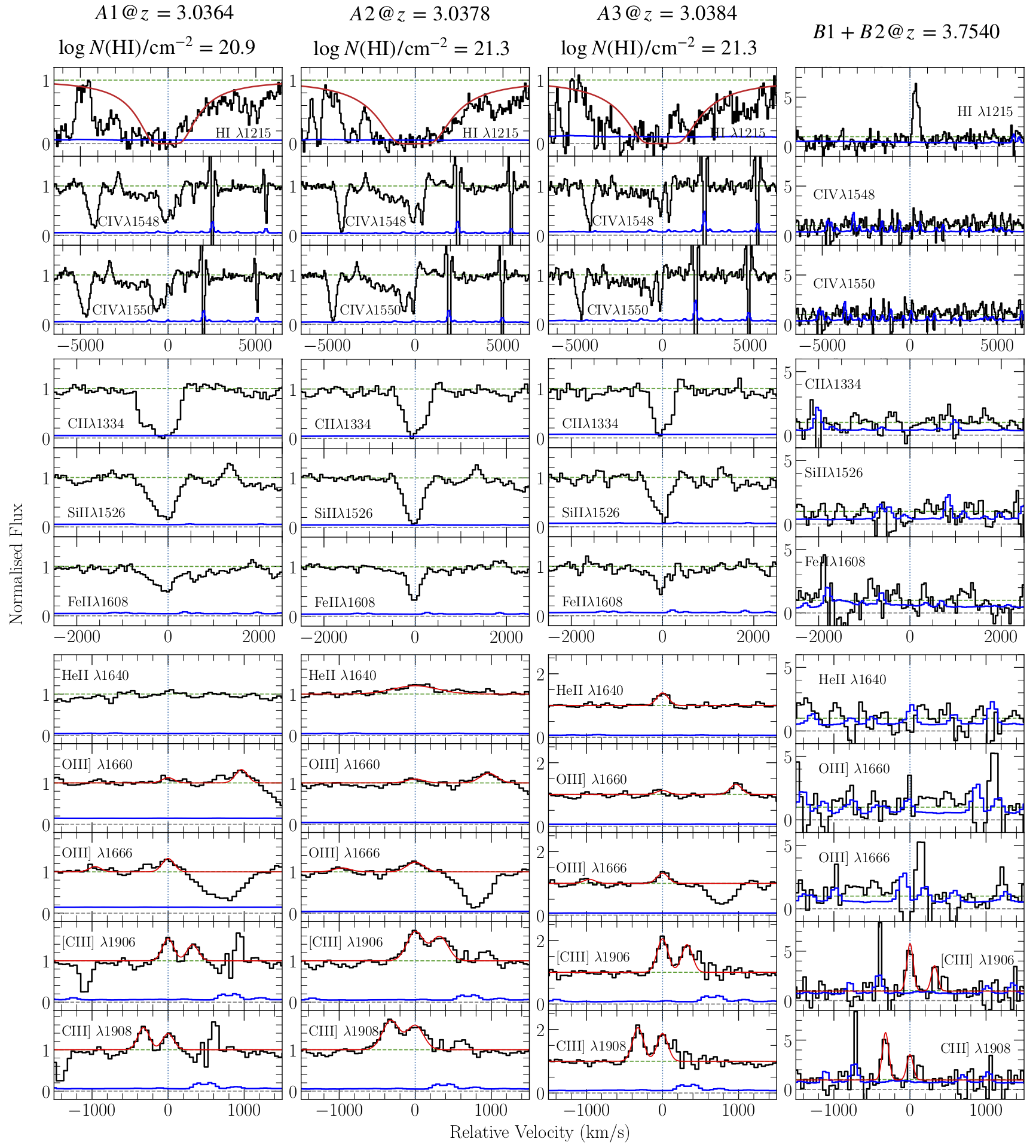}
    \caption{Summary of the ISM absorption and emission features of Systems {\it A} (left three columns) and {\it B} (right column).  Zero velocity corresponds to the systemic redshift determined from nebular emission lines of each galaxy (see Table \ref{tab:line_fitting_HeII_CIII}).  The red curve in the \ion{H}{I} panels shows the best-fit DLA profile with the estimated $\nhone$ displayed at the top of each column.  At negative velocities, the DLA profiles are contaminated by the forest of \lya\ absorption lines in the foreground.  The $\ion{C}{IV}\lambda\lambda\,1548,1550$ absorption profiles are presented in the second and third rows, showing blue absorption tail extending beyond $-2500$ \kms.  The 4th--6th rows show the low-ionization lines $\ion{C}{II}\lambda\,1334$, $\ion{Si}{II}\lambda\,1526$ and $\ion{Fe}{II}1608$, which exhibit asymmetric absorption profiles with extended blue tails, indicating the presence of gas outflows in the ISM. The best-fit Gaussian models of $\ion{He}{II}\lambda1640$, $\ion{O}{III]}\lambda\lambda1660,1666$ and $\ion{C}{III]}\lambda\lambda1906,1908$ emission lines are shown in red curves in bottom five rows. $\ion{He}{II}\lambda1640$ is fitted with a single Gaussian. The doublets are all fitted with a double Gaussian, and the separation between two Gaussian components are fixed by their rest-frame wavelength separation. We fix the flux ratio $\ion{O}{III]}\lambda1666$/$\ion{O}{III]}\lambda1660$ = 2.5. The redshift is tied to be consistent among all lines in each galaxy, and the best-fit values (shown at the top of each column) sets the zero velocity marked by the vertical dotted line.  Data spectrum (continuum normalised) is shown in black, 1-$\sigma$ error spectrum in blue and best-fit models in red. Flux and rest-frame equivalent width measured from the best-fit models for each emission line is listed in Table \ref{tab:line_fitting_HeII_CIII}.  While galaxies {\it B}1/{\it B}2 display a strong \lya\ and modest $\ion{C}{III]}\lambda\lambda1906,1908$ emission features, the data quality is not sufficient to place meaningful constraints on \ion{He}{II}, or \ion{O}{III]}}.
    %The reduced $\chi^2$ of the fitting is 3.6, 2.5, and 2.0 for {\it A}1, {\it A}2, and {\it A}3, respectively.
    \label{fig:HeII_CIII_fitting}
\end{figure*}
%$\ion{Si}{III]}\lambda\lambd{\it A}1883,1892$

The spectra of galaxies {\it A}1, {\it A}2, and {\it A}3 are characterized by three distinct features: (1) a blue continuum consistent with the broadband photometry presented in Table~\ref{tab:photometryA}; (2) strong interstellar absorption due to neutral hydrogen and heavy ions (marked in green, dotted line) that are commonly seen in $z\approx 3$ galaxies \citep[e.g.,][]{Shapley2003,Erb2014}, and (3) nebular emission lines due to 
$\ion{He}{II}\lambda\,1640$,  $\ion{O}{III]}\lambda\lambda\,1660,1666$, and $\ion{C}{III]}\lambda\lambda\,1906,1908$, as well as excited \ion{Si}{II}*$\lambda$\,1264, 1309, and 1530 lines. The strong DLA features observed in the spectra of galaxies {\it A}1, {\it A}2, and {\it A}3 reveal the presence of a significant amount of neutral gas in the ISM of these galaxies.  A Voigt profile analysis of the red damping wing at the systemic redshifts of these galaxies (see below) yields best-fit \ion{H}{I} column densities of $\log\,N({\rm HI})/{\rm cm}^{-2}=20.9\pm 0.1$, $21.3\pm 0.1$, and $21.3\pm 0.1$ for galaxies {\it A}1, {\it A}2, and {\it A}3, respectively, indicating a minimum surface neutral gas mass density of $\Sigma_{\rm gas}=8$--$20\,M_\odot\,{\rm pc}^{-2}$.  The best-fit DLA profiles are presented in the top row of Figure \ref{fig:HeII_CIII_fitting}.  Note that the blue-side of the observed DLA profiles are contaminated by the forest of \lya\ absorption lines in the foreground and therefore excluded from the fit.  

Apart from the strong DLA features, the prominent $\ion{C}{IV}\lambda\lambda\,1548,1550$ absorption profiles in galaxies {\it A}1--3 show a blue tail extending beyond $-2500$ \kms, indicating the presence of stellar winds produced by massive young stars. The $\ion{C}{IV}\lambda\lambda\,1548,1550$ lines are shown in the second and third rows of Figure \ref{fig:HeII_CIII_fitting}.  In addition, low-ionization absorption lines, such as $\ion{C}{II}\lambda\,1334$, \ion{Si}{II}$\lambda\,1526$ and $\ion{Fe}{II}\lambda\,1608$ as presented in the 4th--6th rows of Figure~\ref{fig:HeII_CIII_fitting}, indicate the presence of outflowing gas in the ISM of these galaxies. These absorption lines are clearly asymmetric with an extended blue wing in {\it A}1.  A weak, extended blue wing is also visible in $\ion{Fe}{II}\lambda\,1608$ for {\it A}2 and {\it A}3, while the core is more symmetrically distributed around the systemic velocity.  

To quantify outflow velocities, we measure the absorption velocity centroid, $v_{\rm center}$, and the maximum velocity of the absorption, $v_{\rm max}$. Both quantities are measured with respect to the galaxy systemic velocities derived from nebular emission lines (see below).  $v_{\rm center}$ is determined to be at the location of the deepest absorption trough. $v_{\rm max}$ measures the blueshifted velocity at which the absorption is consistent with the continuum to within 1-$\sigma$ level.  It is determined through the relation $f(v_{\rm max}) = 1.0 - \delta (v_{\rm max})$, where $f$ is the continuum-normalised flux and $\delta$ is the associated flux error \citep[e.g.][]{Martin2012}. To estimate the uncertainties of both $v_{\rm max}$ and $v_{\rm center}$, we repeat the measurements on 1000 random Gaussian generations of the spectra based on the observed intensities and error arrays, and report the mean and standard deviation of the 1000 measured values for both quantities. In addition, to quantify the internal velocity width of the absorption features, we fit a Gaussian profile to the red side of the absorption feature that is redward of the measured $v_{\rm center}$, and obtain a FWHM$_{\rm red}$ that is not affected by the extended blue wing. The model Gaussian profile is convolved with the instrument line spread function before fitting with the data. 

Because both $\ion{C}{II}\lambda\,1334$ and \ion{Si}{II}$\lambda\,1526$ absorption lines are saturated, we make the measurements using $\ion{Fe}{II}\lambda\,1608$ line. We find [$v_{\rm center}$, $v_{\rm max}$, FWHM$_{\rm red}$] of [$-42\pm58$, $-757\pm 72$, $474\pm52$], [$-22\pm30$, $-614 \pm 147$, $301\pm26$] and [$-38\pm28$, $-480\pm 140$, $211\pm56$] \kms\ for {\it A}1, {\it A}2 and {\it A}3, respectively. 
%The velocity centroids of all three galaxies show a blueshifting trend with respect to the systemic galaxy velocities, although 
Both {\it A}1 and {\it A}2 exhibit an absorption velocity centroid  consistent with $v=0$ to within measurement uncertainties, while {\it A}3 displays a slightly more significant blueshift.  At the same time, the maximum velocity $v_{\rm max}$ of $\approx 500$--$760$ \kms\ observed in $\ion{Fe}{II}\lambda\,1608$ exceeds the respective FWHM$_{\rm red}$ measured for these galaxies,
%clearly shows that $\ion{Fe}{II}\lambda\,1608$ in all three galaxies exhibits an extended blue wing, %(the MUSE instrument line spread function is approximately 110 \kms\ at this observed wavelength). The fact that $v_{\rm max}$ is
%significantly larger than the respective FWHM$_{\rm red}$ measured for these galaxies.  The observed line also 
clearly indicating the presence of high-velocity outflows. We also note that the measured FWHM$_{\rm red}$ is broader than the FWHM measured for nebular emission lines in all three galaxies (see below). In particular, for galaxy {\it A}1, the absorption line width is $\approx 3$ times the width inferred from nebular emission lines (see Figure~\ref{fig:HeII_CIII_fitting}), suggesting the presence of turbulence ISM local to the star forming regions.

All three galaxies show significantly smaller outflow velocities in the line centroids in comparison to typical \lya\ emitting galaxies, which is $\sim 200$ km/s as reported in \citep[][]{Shibuya2014}.  If the outflows originate in a biconical structure, the small outflow velocities in System {\it A} may suggest a large inclination angle of the cones of $\sim 80^{\circ}$, assuming that the mean $v_{\rm center}$ among the three galaxies of $\sim 34$ km/s is the projected velocity from an inclined cone flowing out with 200 km/s.  Such a large inclination angle is also consistent with the side-lobe like morphology of the \lya\ nebulae (i.e., the two clouds bracketing the continuum galaxies) and the elongated gap between the two clouds, as described below in \S\ref{sec:Lya_NB}.  In addition, the fact that all three galaxies show similar uncharacteristically small outflow velocities might suggest that they all reside in the same outflow bubble likely originating from galaxy {\it A}1.

%For low-ionization absorption lines $\ion{C}{II}\lambda\,1334$, \ion{Si}{II}$\lambda\,1526$ and $\ion{Fe}{II}\lambda\,1608$, we measure the velocity centroid $\bar{v}$, the blue velocity extent $v_{90}$, and the FWHM of each line in all three galaxies. These three lines are chosen for measurements because they are not blended with other emission/absorption features, and are not affected by strong sky contamination. We adopt the column density-weighted velocity centroid, $\bar{v} = \frac{\sum {\rm ln}(1/f_i)v_i}{\sum{\rm ln}(1/f_i)}$, where $f_i$ is the continuum normalised flux within each spectral pixel. $v_{90}$ is determined such that 90\% of total column density-weighted absorption blue-ward of $\bar{v}$ is enclosed between $v_{90}$ and $\bar{v}$. All three quantities are measured within the velocity limits of [-800, +400] \kms, [-500, +450] \kms and [-500, +360] \kms for galaxies {\it A}1, {\it A}2 and {\it A}3, respectively (marked as vertical dashed lines in the middle three rows of Figure~\ref{fig:HeII_CIII_fitting}). The uncertainties are estimated through 1000 random Gaussian generations of the spectra based on the error arrays, and the measured values are listed in Table~\ref{tab:absorption_lines}. 

Different from galaxies in System {\it A}, galaxies {\it B}1 and {\it B}2 exhibit a strong \lya\ emission with no apparent DLA trough, and resolved $\ion{C}{III]}\lambda\lambda\,1906,1908$ doublet features on top of a faint UV continuum.  No strong absorption features are detected, but the spectrum does not have sufficient sensitivities for placing strong constraints.  We measure a rest-frame equivalent width (${\rm EW}_{\rm rest}$) of the \lya\ emission line of galaxies {\it B}1 and {\it B}2 over the observed wavelength window from $\lambda_1 = 5760$ \AA\ to $\lambda_2=5796$ \AA, and obtain ${\rm EW}_{\rm rest}({\rm Ly\alpha})=33.3\pm 1.5 \ang$.  

For all galaxies, we are able to determine an accurate systemic redshift for each of these galaxies by simultaneously fitting multiple emission lines with a Gaussian function, convolved with an appropriate instrument line spread function, which shares a common velocity centroid. Specifically for galaxies in System {\it A}, we adopt a single Gaussian model for $\ion{He}{II}\lambda\,1640$ and a double Gaussian model for both $\ion{O}{III]}\lambda\lambda\,1660,1666$ and $\ion{C}{III]}\lambda\lambda\,1906,1908$ doublets.  In addition, the flux ratio of $\ion{O}{III]}\lambda\,1666$/$\ion{O}{III]}\lambda\,1660$ is fixed at 2.5 as expected from their radiative transition probabilities. For galaxy {\it A}1, $\ion{He}{II}\lambda\, 1640$ is excluded from the fitting due to the lack of detection, and $\ion{O}{III]}$ and $\ion{C}{III]}$ doublets are fitted with a common line width.  For galaxy {\it A}2, $\ion{He}{II}\lambda\, 1640$ is visibly broader than both $\ion{O}{III]}$ and $\ion{C}{III]}$ doublets.  We therefore allow the width of $\ion{He}{II}\lambda\, 1640$ to be a free parameter while the the doublets share a common line width in the fit for {\it A}2. The difference in line width between $\ion{He}{II}\lambda\, 1640$ and $\ion{O}{III]}$/$\ion{C}{III]}$ doublets is not surprising, as $\ion{He}{II}\lambda\, 1640$ emission is expected to have both stellar and nebular contributions which can sometimes lead to complex line structures \citep[e.g.][]{Berg2018,Kehrig2018,Nanayakkara2019,Feltre2020}. For galaxy {\it A}3, fittings with or without $\ion{He}{II}\lambda\, 1640$ line width being a free parameter return consistent results within uncertainties. Therefore, we assign a common line width to all lines fitted for {\it A}3 for simplicity. 

For galaxies {\it B}1 and {\it B}2, we fit a double Gaussian model with a fixed doublet separation to the $\ion{C}{III]}\lambda\lambda\,1906,1908$ intercombination lines.  The best-fit redshifts, line widths, integrated line fluxes, and ${\rm EW}_{\rm rest}$, along with associated errors of individual galaxies are presented in Table \ref{tab:line_fitting_HeII_CIII}.  The best-fit line profiles of the emission features are also presented in Figure \ref{fig:HeII_CIII_fitting}.

\begin{table}
	\centering
	\caption{ Emission line fitting results, with lensing magnification corrected in all flux measurements based on mean magnification values listed in Table~\ref{tab:magnification}. }
	\label{tab:line_fitting_HeII_CIII}
	\begin{tabular}{l c c c } % four columns, alignment for each
	\hline
		\multicolumn{4}{c}{{\it A}1 at $z=3.0364\pm0.0001^a$} \\
 		\hline
 		 & FWHM & Flux & ${\rm EW}_{\rm  rest}^b$ \\ 
 		 & (\kms) &  ($\mathrm{10^{-20} \, erg \, s^{-1} \, cm^{-2}}$) & (\AA) \\ 
 		\hline
        $\ion{He}{II} \, \lambda 1640$ &  & $<45^c$ & $<0.18^d$ \\
        $\ion{O}{III]} \, \lambda 1660$ & $147\pm34$ & $35 \pm 9$ & $0.14 \pm 0.04$ \\
        $\ion{O}{III]} \, \lambda 1666$ & ... & $86 \pm 22$ & $0.36 \pm 0.11$ \\
        $\ion{[C}{III]} \, \lambda 1906$ & ... & $120 \pm 30$ & $0.64 \pm 0.19$ \\
        $\ion{C}{III]} \, \lambda 1908$ & ... & $87 \pm 25$ & $0.47 \pm 0.16$ \\
        \hline
	    \multicolumn{4}{c}{{\it A}2 at $z=3.0378\pm0.0001^e$} \\
 		\hline
        $\ion{He}{II} \, \lambda 1640$ & $673\pm83^f$ & $49 \pm 7$ & $0.79 \pm 0.13$ \\
        $\ion{O}{III]} \, \lambda 1660$ & $237\pm14$ & $9 \pm 1$ & $0.5\pm 0.02$  \\
        $\ion{O}{III]} \, \lambda 1666$ & ... & $22\pm2$ & $0.37\pm 0.05$ \\
        $\ion{[C}{III]} \, \lambda 1906$ & ... & $58\pm4$ & $1.23\pm 0.10$   \\
        $\ion{C}{III]} \, \lambda 1908$ & ... & $48\pm3$ &  $1.02\pm0.09$ \\
        \hline
	    \multicolumn{4}{c}{{\it A}3 at $z=3.0384\pm0.0001$} \\
 		\hline
        $\ion{He}{II} \, \lambda 1640$ & $136\pm9$ & $13\pm2$ & $0.41\pm 0.10$ \\
        $\ion{O}{III]} \, \lambda 1660$ & $...$ & $4\pm1$ & $0.13\pm 0.02$\\
        $\ion{O}{III]} \, \lambda 1666$ & ... & $10\pm1$ & $0.34\pm 0.04$\\
        $\ion{[C}{III]} \, \lambda 1906$ & ... & $32\pm2$ &  $1.20\pm 0.10$ \\
        $\ion{C}{III]} \, \lambda 1908$ & ... & $26\pm1$ & $0.99\pm 0.08$ \\
        \hline
        \multicolumn{4}{c}{{\it B}1/{\it B}2 at $z=3.7540\pm0.0001$} \\
 		\hline
        $\ion{[C}{III]} \, \lambda 1906$ & $43\pm20$ & $0.9\pm0.2$ & $3.36\pm 0.54$\\
        $\ion{C}{III]} \, \lambda 1908$ & ... & $0.5\pm0.2$ & $1.80\pm 0.53$  \\
        \hline
        \multicolumn{4}{l}{$^a$ Obtained from a simultaneous fit of $\ion{O}{III]}\lambda\lambda\,1660,1666$ and } \\
        \multicolumn{4}{l}{\, \, $\ion{C}{III]}\lambda\lambda\,1906,1908$} \\
        \multicolumn{4}{l}{$^b$ Rest-frame equivalent width} \\
        \multicolumn{4}{l}{$^c$ 2-$\sigma$ upper limit} \\
        \multicolumn{4}{l}{$^d$ 2-$\sigma$ upper limit} \\
        \multicolumn{4}{l}{$^e$ Obtained from a simultaneous fit of all lines listed; same for}\\
        \multicolumn{4}{l}{\, \, {\it A}3 and {\it B}1/{\it B}2} \\
        \multicolumn{4}{l}{$^f$ FWHM of $\ion{He}{II} \, \lambda 1640$ in {\it A}2 is not tied with other lines due}\\
        \multicolumn{4}{l}{\, \, to its wide line width}\\
        \\
	\end{tabular}
\end{table}

\subsection{Emission line diagnostics}\label{sec:cloudy}

The UV emission line properties presented in Table \ref{tab:line_fitting_HeII_CIII} are typical of star-forming galaxies at $z\approx 3$ \citep[e.g.,][]{Maseda2017,Nanayakkara2019,Feltre2020}, and reveal a turbulent and high-density nature in the ISM with a radiation field dominated by massive young stars in these galaxies. 
%In particular, galaxy {\it A}2 with $\mstar\approx 10^9\,\msun$ is an order of magnitude less massive than typical UV bright galaxies at $z\approx 3$ \citep[e.g.,][]{Erb2006b,Kulas2012}, but exhibits an emission line width that is 60\% broader.  
The best-fit FWHMs of the emission lines correspond to velocity dispersions of $\approx 60$ \kms\ in A1 and A3, and $\approx 100$ \kms\ in A2, which are within the typical range measured for $z=2$--3 galaxies (e.g., $108\pm86$ \kms\ reported in \citealt{Erb2006b}, and $\approx 50$--150 \kms\ reported in \citealt{Kulas2012}).
The ratio between the \ion{C}{III]} intercombination lines serves as an important UV diagnostic of the electron density, $n_{\rm e}$, in the ISM, although it saturates at density below $n_{\rm e}\approx 10^3\,{\rm cm}^{-3}$ \citep[e.g.,][]{Kewley2019}. The observed
$\ion{[C}{III]}\lambda\,1906$/$\ion{C}{III]}\lambda\,1908$ ratios of these galaxies range from $1.2\pm 0.2$ for {\it A}2 and {\it A}3 to $1.9\pm 0.6$ for {\it B}1 and {\it B}2, constraining the ISM electron density in both Systems {\it A} and {\it B} to be $\lesssim 2\times 10^4\,{\rm cm}^{-3}$ for a gas temperature of $10,000$ K \citep{OsterbrockFerland2006}.  The high-density limits are also comparable to what is seen in \ion{C}{III]} emitters at $z\approx 3$ \citep[e.g.,][]{Maseda2017}.
In addition, the detection of \ion{He}{II}$\lambda\,1640$ emission, along with the presence of a prominent P-Cygni profile in \ion{C}{IV}$\lambda\lambda\,1548,1550$ with blue absorption tail extending beyond $|\Delta\,v|\approx 2000$ \kms\ (second and third rows in Figure \ref{fig:HeII_CIII_fitting}), indicate the presence of massive young stars with $M\gtrsim 30\,M_\odot$ \citep[e.g.,][]{Leitherer1999,Pettini2000,Crowther2007,Brinchmann2008,Cabanac2008}.  The presence of broad \ion{He}{II}$\lambda\,1640$ emission line in {\it A}2 is also a sign of Wolf-Rayet stars that have a short lifetime of $\sim 5$ Myr \citep[e.g.][]{Schaerer1998,Crowther2007,Cabanac2008}, in agreement with the constant SFR scenario suggested by photometric SED analysis (see Table \ref{tab:Bagpipes_sedfitting} and discussion in \S\ref{sec:SED_fitting}). 

%young stellar age obtained from the stellar population synthesis analysis summarized in Table \ref{tab:Bagpipes_sedfitting}.  

\begin{figure}
	\includegraphics[width=7cm]{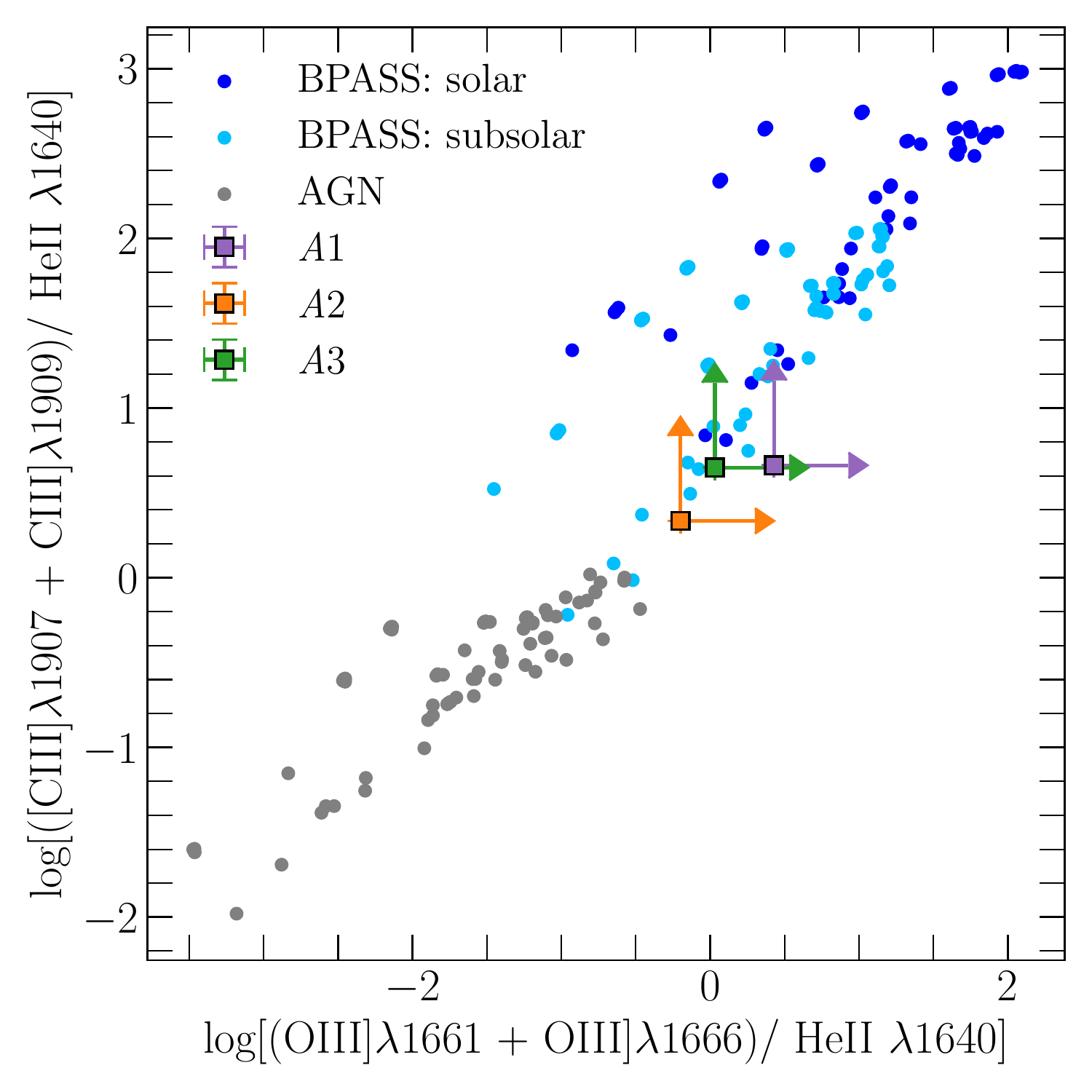}
    \caption{UV diagnostic diagram between AGN- (grey points) and stellar-dominated (blue and cyan points) radiation fields described in Section \ref{sec:cloudy}. Gas metallicities range from subsolar $Z = 0.002$ to supersolar $Z = 0.04$.  The observed line ratios for galaxies {\it A}1, {\it A}2, and {\it A}3 are shown in squares.  All three galaxies exhibit line ratios that are consistent with young stars (rather than AGN) dominating the ISM radiation field.} 
    \label{fig:UV_BPT}
\end{figure}

Finally, we investigate the possibility of these galaxies hosting an active galactic nucleus (AGN) using emission line diagnostics in the UV.  Specifically, \cite{Feltre2016} shows that the combination of collisionally excited nebular lines $\ion{O}{III]}\lambda\lambda1660,1666$, $\ion{C}{III]}\lambda\lambda1906,1908$ and the $\ion{He}{II}\lambda 1640$ recombination line can serve as a good indicator of the ISM ionization state.
We compute the expected line ratios of $\ion{O}{III]}\lambda\lambda1660,1666/\ion{He}{II}\lambda1640$ and $\ion{C}{III]}\lambda\lambda1906,1908/\ion{He}{II}\lambda1640$ under different AGN and star formation (SF) ionization radiation fields, using the \texttt{CLOUDY} code \citep[version 17.01;][]{cloudy2017}.  For the AGN spectrum, we adopt the model continuum specified in \texttt{CLOUDY} with an effective temperature of $10^6$ K, an X-ray to UV ratio of $\alpha_{\rm ox} = -1.4$, a UV slope of $\alpha_{\rm uv} = -0.5$ and an X-ray slope of $\alpha_{\rm x} = -1$.  For the SF model, we consider two stellar populations with sub-solar ($Z=0.001$) and solar ($Z=0.02$) metallicity, respectively. We use the FSPS code \citep[v3.1;][]{Conroy2009,Conroy2010} to generate a composite SF spectrum at the age of 250 Myr with BPASS models, which assumes a Salpeter stellar initial mass function with an upper mass cutoff at $100 \mathrm{M_\odot}$ \citep{BPASS2017}.
For each adopted AGN or SF spectrum, we generate a grid of models for the expected line ratios with the following parameters: gas metallicity $Z = [0.002, 0.02, 0.04]$, hydrogen density $n_{\rm H}/ \mathrm{cm^{-3}}=[10, 10^2, 10^3, 10^4]$ and ionization parameter $U = [-4, -3.5, -3, -2.5, -2, -1.5]$. We set a temperature floor of 10,000 K.  The predicted line flux ratios are shown in Figure \ref{fig:UV_BPT}, along with the observed values for galaxies {\it A}1, {\it A}2, and {\it A}3. The non-detection of $\ion{He}{II}\lambda\,1640$ in galaxy {\it A}1 naturally leads to lower limits of both line flux ratios. We also treat the line flux ratios in both {\it A}2 and {\it A}3 as lower limits because of a possible stellar contribution to the measured $\ion{He}{II}\lambda1640$ flux. 
%Here to determine whether there are active galactic nuclei (AGN) in {\it A}1, {\it A}2 and {\it A}3 contributing as ionizing sources, 
%In the case of AGN radiation fields, we use the command ``AGN 6.00 -1.40 -0.50 -1.0" to specify an AGN continuum  

Figure \ref{fig:UV_BPT} shows that all three galaxies in System {\it A} are consistent with an ISM radiation field being dominated by massive young stars and that there is no evidence of an AGN dominating the radiation field in these galaxies. Furthermore, the lack of $\ion{C}{IV}\lambda\,1548,1550$ in emission also suggests the absence of AGN as  $\ion{C}{IV}\lambda\,1548,1550$ emission is expected to be prominent with a hard ionization background \citep[e.g.][]{Gutkin2016}. %contributing to the ionization of the ISM and Ly$\alpha$ nebula. 
Due to the lack of relevant emission lines in the combined spectrum of {\it B}1 and {\it B}2, we cannot conduct the same exercise for System {\it B}. 
A close examination of available deep X-ray data taken by {\it Chandra} also shows that there is no apparent excess of X-ray signal at the locations of Systems {\it A} and {\it B}. %\textcolor{red}{not sure about sys B, seems like there is a trace of X-ray signal in an arc shape, will check with Irina}

\section{Analysis: Ly$\alpha$ nebula properties}\label{sec:analysis_gas}

The observed broadband photometric and spectroscopic properties of galaxies in System {\it A} indicate that these galaxies are typical of UV continuum selected star-forming galaxies at $z\approx 3$ with an ISM radiation field dominated by massive young stars, while galaxies in System {\it B} display properties that resemble low-mass LAEs in the early epoch.  The large amount of ISM gas revealed in the spectra of galaxies {\it A}1, {\it A}2, and {\it A}3, coupled with a wide-spread \lya\ nebula revealed in the MUSE data shows that this is a particularly gas-rich system.  Here we present an analysis of the morphologies and line profiles of the extended \lya\ nebulae in these two systems.

\subsection{Pseudo narrow-band \lya\ image and source plane reconstruction}\label{sec:Lya_NB}

\begin{figure*}
	\includegraphics[width=0.8\linewidth]{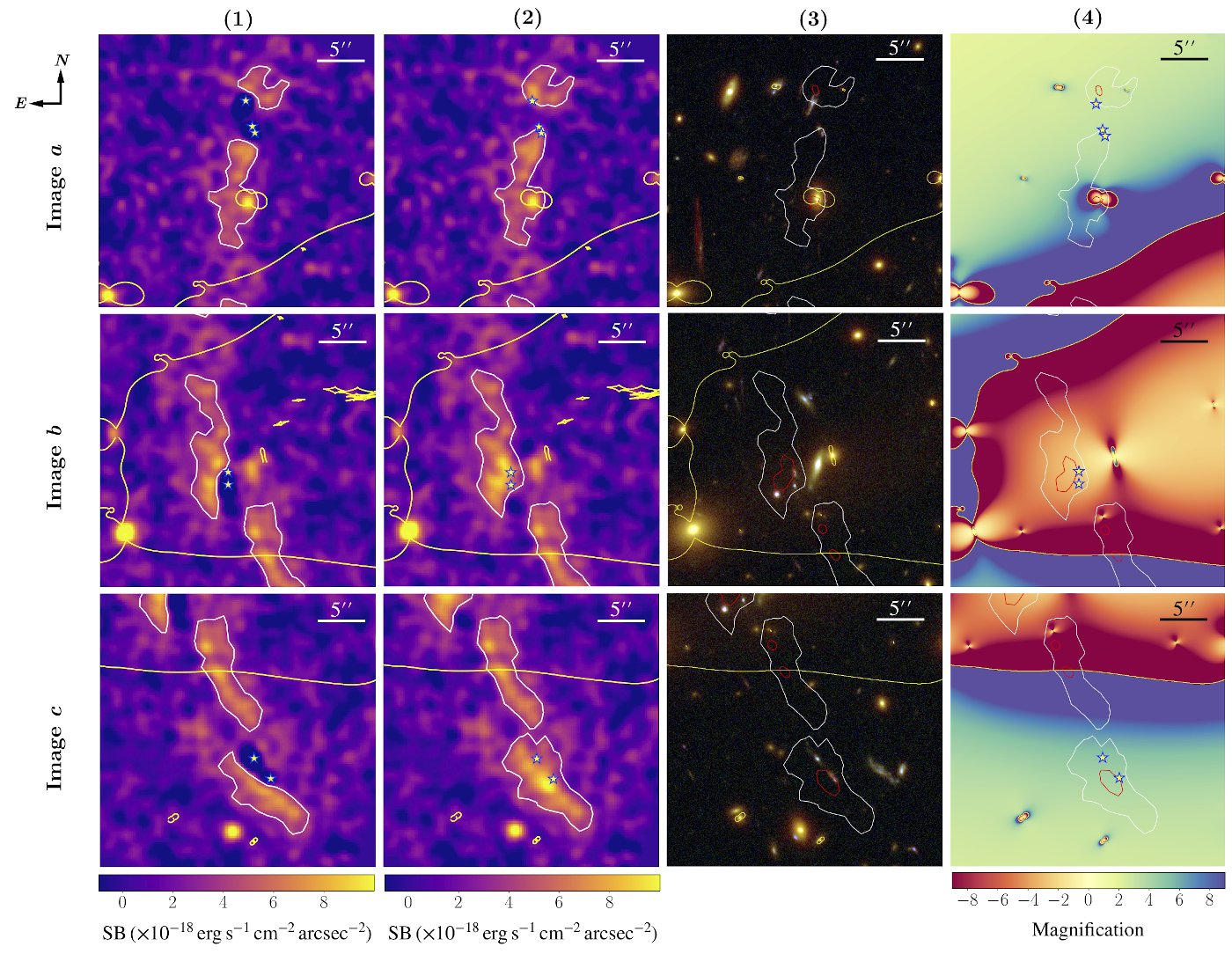}
	\caption{Summary of the lensing configuration of the observed \lya\ arc in System {\it A}.  Column ({\it 1}): pseudo narrow-band images without the emergent \lya\ flux within the DLA troughs at the locations of galaxy continuum.  The images have been smoothed using a Gaussian kernel of ${\rm FWHM}_{\rm smooth}=1''$.   The contour marks constant surface brightness of $3.7\times 10^{-18}\sbunit$, which is detected at the 3-$\sigma$ level of significance in the smoothed image.  Star symbols mark the positions of the associated star-forming galaxies identified in {\it HST} images.
    Column ({\it 2}): same as Column ({\it 1}) but includes \lya\ flux  from the DLA troughs at the locations of galaxy continuum (see text). 
    Column ({\it 3}): contours of multiply-lensed \lya\ nebulae determined from Column ({\it 2}) overlaid on individual galaxy images in the {\it HST} data to illustrate the relative alignment between star-forming regions and the line-emitting gas (see also Figure \ref{fig:hst_with_lya_contours_A}).  \lya\ surface brightness contours of of $3.7\times10^{-18} \, \mathrm{erg \, s^{-1} \, cm^{-2} \, arcsec^{-2}}$ and $7.3\times10^{-18} \, \mathrm{erg \, s^{-1} \, cm^{-2} \, arcsec^{-2}}$ are shown in white and red, respectively, and the yellow contours mark the critical curves of the cluster lens for sources at $z=3.038$. 
    Column ({\it 4}): the magnification map overlaid with the same \lya\ contours to illustrate the spatial variation of lensing magnification across the nebulae. Negative magnification factors indicate flipped parity of the image. 
    }
    \label{fig:img_plane_contours_A}
\end{figure*}

\begin{figure}
    \centering
    \includegraphics[width=\linewidth]{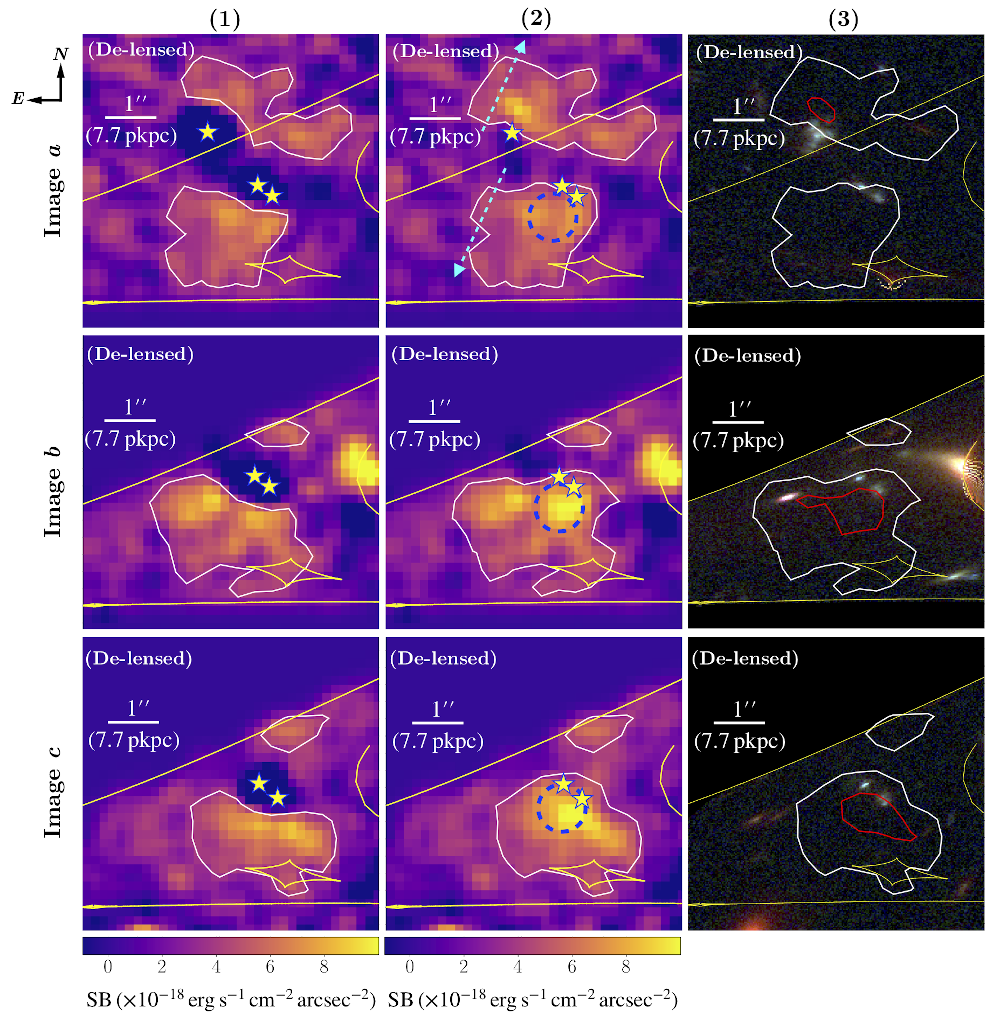}
    \caption{Column ({\it 1}): de-lensed narrow-band image without \lya\ flux from the DLA troughs at the locations of galaxy continuum. The images are smoothed using a Gaussian kernel of ${\rm FWHM}_{\rm smooth}=0.5''$ in the source plane. The contour marks constant surface brightness of $3.7\times 10^{-18}\sbunit$, corresponding to the 3-$\sigma$ level of significance detected in the image plane, same as contours shown in Column({\it 1}) of Figure~\ref{fig:img_plane_contours_A}. Star symbols mark the de-lensed positions of the associated star-forming galaxies identified in {\it HST} images.
    Column ({\it 2}): same as Column ({\it 1}) but includes \lya\ flux  from the DLA troughs at the locations of galaxy continuum. 
    Column ({\it 3}): de-lensed \lya\ contours overlaid on de-lensed {\it HST} data, with the yellow contours showing the caustics in the source plane.  The blue dashed circles in Column ({\it 2}) mark the apertures for the template spectrum extraction, which we use for the shell model analysis (see \S\ref{sec:velocity_shell_fitting}). The cyan dashed arrows show the directions along which we extract the one-dimensional surface brightness profile (see Figure~\ref{fig:1D_SB_profiles} below)}
    \label{fig:src_plane_contours_A}
\end{figure}

To characterize the extended \lya\ nebulae in both systems, we first form a pseudo narrow-band \lya\ image for each system.  We first note that all three galaxies in System {\it A} exhibit asymmetric \lya\ emission feature within the DLA trough, with an enhanced red peak at $\Delta\,v\approx +500$ \kms\ (e.g., top row of Figure \ref{fig:HeII_CIII_fitting}) from the respective systemic redshifts.  The observed asymmetric profile of these emergent \lya\ photons is similar to what is seen in the extended nebulae (see \S\ref{sec:variation} below) and is characteristic of large-scale outflows that have been commonly identified in high-redshift galaxies \citep[e.g.][]{Franx1997,Frye1998,Pettini2000,Frye2002,Cabanac2008}.  
An origin of the emergent \lya\ photons in outflows is qualitatively consistent with the presence of gas outflows seen in absorption lines in galaxy spectra and the presence of massive young stars inferred from the UV spectral properties of the galaxies described in \S\ref{sec:cloudy} \citep[see also][for examples]{Pettini2000,Cabanac2008}. Given the uncertainty of the lint-of-sight distance between the galaxies and the \lya\ emission location, whether these photons originate in the star-forming ISM of the galaxies or in the extended nebulae that are blended with the galaxy image by projection remains uncertain.  Therefore, we construct two versions of the pseudo narrow-band \lya\ image for System {\it A}: one without including the emergent \lya\ photons in the DLA trough of the continuum sources, and a second one which incorporates both the \lya\ photons in the DLA troughs and those in the extended nebulae.  As discussed below and shown in Figures \ref{fig:img_plane_contours_A} and \ref{fig:src_plane_contours_A}, this exercise enables a clearer understanding of the differences in the observed surface brightness profiles between multiple images, as well as establishing a direct connection between the galaxies and the line-emitting gas at large distances.

To construct a pseudo narrow-band \lya\ image for System {\it A} without including the \lya\ photons from the DLA troughs, we perform a local continuum subtraction per spaxel within the \lya\ line.  We determine a wavelength-dependent continuum level based on a linear interpolation of the continuum fluxes observed on the blue and red sides of the \lya\ line.  Specifically, we determine a medium flux over a spectral window of 4830-4863 \AA\ on the blue side and a median flux over 4961-4994 \AA\ on the red side.  At $z\approx 3.04$, these correspond roughly to [$-5000$,$-3000$] and [$+3000$, $+5000$] km/s from the Ly$\alpha$ centroid (see Figure \ref{fig:lya_profile_and_tlac_models} below). The interpolated value is then subtracted from the observed flux at each spaxel. A pseudo narrow-band image of the \lya\ emission is then created by integrating the flux of each spaxel over the wavelength range from 4890 \AA\ to 4930\AA, where \lya\ flux is detected at a high level of significance (see Figure \ref{fig:lya_profile_and_tlac_models} below).  A smoothed pseudo narrow-band \lya\ image, using a Gaussian kernel of ${\rm FWHM}_{\rm smooth}=1''$, is presented in Column ({\it 1}) of Figure \ref{fig:img_plane_contours_A}, which shows two nebulae separated roughly by $\approx 2''$ in the image plane and bracketing galaxies {\it A}1, {\it A}2, and {\it A}3 from the north and south. Furthermore, at the locations of galaxy continuum, there is a net absorption in this pseudo narrow-band image due to the presence of DLAs. 
%(see also Figure \ref{fig:hst_with_lya_contours_A}). 

Next, we construct a pseudo narrow-band image that includes the emergent \lya\ photons in the DLA troughs.   This is accomplished by 
first identifying the spaxels within the continuum emitting regions of galaxies {\it A}1, {\it A}2, and {\it A}3 as defined by SExtractor (see \S\ref{sec:galaxy_photometry}).  We then adopt the best-fit DLA model profile for each galaxy presented in Figure \ref{fig:HeII_CIII_fitting}, and multiply the model by the best-fit continuum obtained using a low-order polynomial fit to line-free regions in the integrated continuum spectrum presented in Figure \ref{fig:galaxy_fullspec_A}.  Next, the combined DLA-continuum model spectrum is scaled to match the continuum level of the spectrum in each spaxel and subtracted from the data.  The amplitude of the continuum model for each spaxel is determined using the spectrum in the wavelength window from 5430 \AA\ to 5560 \AA, corresponding to rest-frame wavelengths from 1345 \AA\ to 1375 \AA, where no narrow-line features are present.  The resulting difference data cube is combined with the previous continuum-subtracted data cube in the extended nebula region.  A pseudo narrow-band image is then created by integrating over the wavelength range from 4890 \AA\ to 4930 \AA.  Similarly, we smooth the image using a Gaussian kernel of FWHM$_{\rm smooth}=1\arcsec$, and present the smoothed pseudo narrow-band image in Column ({\it 2}) of Figure~\ref{fig:img_plane_contours_A}. 

In both versions of the pseudo narrow-band image presented in Columns ({\it 1}) and ({\it 2}) of Figure~\ref{fig:img_plane_contours_A}, the white contours mark a constant \lya\ surface brightness of $3.7\times10^{-18} \, \mathrm{erg \, s^{-1} \, cm^{-2} \, arcsec^{-2}}$, which is detected at the 3-$\sigma$ level of significance.  A strong variation in \lya\ surface brightness is clearly seen across both the northern and southern nebulae, suggesting large spatial fluctuations in the underlying gas properties. While there exists a clear gap between the northern and southern nebulae, after including the \lya\ signal inside the DLA troughs, the overlap between the constant \lya\ surface brightness contours and these galaxies supports a continuous flow of dense gas from star-forming regions into a low-density halo environment. Furthermore, the surface brightness of the southern nebula in the vicinity of the galaxy continuum in images {\it b} and {\it c} is relatively more enhanced than that in image {\it a} after incorporating the \lya\ signal in the DLA troughs (also see Figures~\ref{fig:src_plane_contours_A} and \ref{fig:1D_SB_profiles} below). Specifically, in Column ({\it 2}) of Figure~\ref{fig:img_plane_contours_A}, the \lya\ surface brightness in image {\it a} in the vicinity of galaxies {\it A}2 and {\it A}3 is fainter by $\approx 25\%$ compared with images {\it b} and {\it c}. The reduced \lya\ surface brightness in images {\it a} suggests that the magnification factor of image {\it a} relative to images {\it b} and {\it c} is smaller than what is predicted by the lens model. Such a difference in surface brightness of lensed \lya\ nebulae is also seen in \cite{Caminha2016}, and is consistent with the discrepancy in de-lensed continuum brightnesses of {\it A}2 and {\it A}3, for which image {\it a} is fainter by $\approx 0.2$ magnitude (see discussion in \S\ref{sec:galaxy_photometry}). 

In Column ({\it 3}) of Figure \ref{fig:img_plane_contours_A}, \lya\ surface brightness contours showing $3.7\times10^{-18} \, \mathrm{erg \, s^{-1} \, cm^{-2} \, arcsec^{-2}}$ and $7.3\times10^{-18} \, \mathrm{erg \, s^{-1} \, cm^{-2} \, arcsec^{-2}}$ (i.e., 3-$\sigma$ and 6-$\sigma$ determined from the pseudo narrow-band image shown in Column {\it 2}) are overlaid on top of the {\it HST} composite image from Figure \ref{fig:hst_with_lya_contours_A}. Note that image {\it b} is north-south flipped from images {\it a} and {\it c} in this lensing configuration.  As a guide, we include the magnification map in Column ({\it 4}) of Figure \ref{fig:img_plane_contours_A} (negative magnification factors indicate flipped parity of the image), overlaid with the same \lya\ contours.

Through the deflection field calculated using the fine-tuned lens model (see \S\ \ref{sec:fine_tune_lens_model}), we de-lens both the pseudo narrow-band image and the {\it HST} images back to the source plane.  The de-lensed pseudo narrow-band image smoothed with a Gaussian kernel of ${\rm FWHM}_{\rm smooth}=0.5''$ in the source plane is presented in Columns ({\it 1}) and ({\it 2}) of Figure~\ref{fig:src_plane_contours_A} for before and after including \lya\ emission in the DLA troughs, respectively.  The reconstructed source-plane images clearly show that most of the northern nebula is merely singly-lensed like galaxy {\it A}1, while the southern nebula stretches across the lensing field with rapidly changing magnification factors.  Image {\it a}, covering the full extent of the nebulae in the source plane, constrains the projected size of the \lya\ nebulae to approximately 30 pkpc from north to south.  The de-lensed {\it HST} broadband images, as shown in Columns ({\it 3}), are in excellent agreement among three multiple images, consistent with the low image position dispersion of ${\rm rms_{\rm im}}=0\farcs1$ predicted by the fine-tuned lens model (see \S\ref{sec:lens_model}).  The de-lensed pseudo narrow-band images show the same surface brightness discrepancy between multiple images as seen in the image plane (see Figure~\ref{fig:img_plane_contours_A}), where the surface brightness near the galaxy continuum regions is fainter in image {\it a} as discussed above. The white and red contours in Figure~\ref{fig:src_plane_contours_A} correspond to surface brightnesses of $3.7\times10^{-18} \, \mathrm{erg \, s^{-1} \, cm^{-2} \, arcsec^{-2}}$ and $7.3\times10^{-18} \, \mathrm{erg \, s^{-1} \, cm^{-2} \, arcsec^{-2}}$, same as the contours shown in Figure~\ref{fig:img_plane_contours_A}.

\begin{figure*}
	\includegraphics[width=0.85\linewidth]{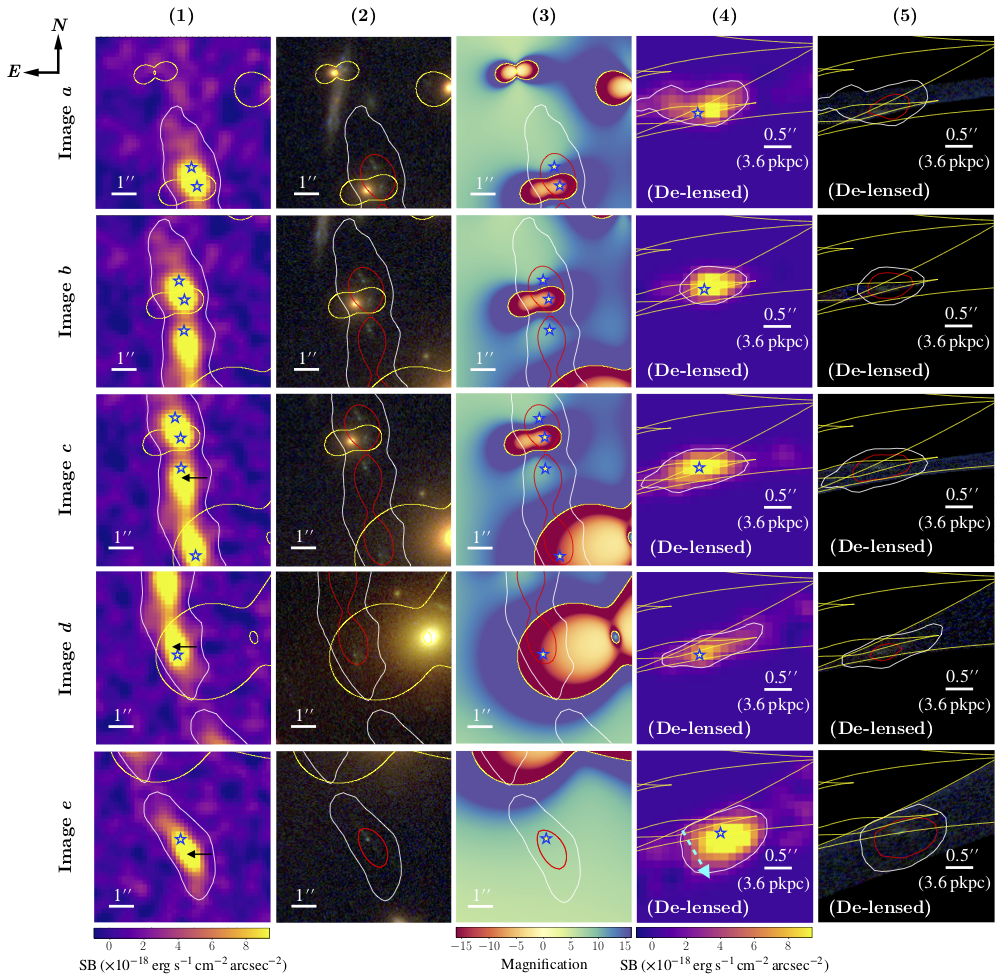}
    \caption{Summary of the lensing configuration of the observed \lya\ arc in System {\it B}.  Column ({\it 1}): pseudo narrow-band image of the \lya\ emission, smoothed with a Gaussian kernel of ${\rm FWHM}_{\rm smooth}=1''$.   The contour marks constant surface brightness of $2.8\times 10^{-18}\sbunit$, which is detected at the 3-$\sigma$ level of significance.  Star symbols mark the positions of the associated star-forming galaxies identified in {\it HST} images, and the yellow contours show the critical curve for a source at $z_{\rm sys}=3.754$.  Black arrows in the bottom three panels indicate the location of the brightest pixels in images c, d and e (one pixel from each image) that are included in the template \lya\ spectrum used for shell model analysis (see \S\ref{sec:velocity_shell_fitting}). 
    Column ({\it 2}): contours of multiply-lensed \lya\ nebulae overlaid on individual galaxy images in the {\it HST} data.  \lya\ surface brightness contours of $2.8\times10^{-18} \, \mathrm{erg \, s^{-1} \, cm^{-2} \, arcsec^{-2}}$ and $7.5\times10^{-18} \, \mathrm{erg \, s^{-1} \, cm^{-2} \, arcsec^{-2}}$ are shown in white and red, respectively. 
    Column ({\it 3}): the magnification map overlaid with the same \lya\ contours to illustrate the spatial variation of lensing magnification across the nebulae. 
    Column ({\it 4}): de-lensed narrow-band image, smoothed with a Gaussian kernel of ${\rm FWHM}_{\rm smooth}=0.5''$ in the source plane. White contours mark constant surface brightness of $2.8\times 10^{-18}\sbunit$, corresponding to the 3-$\sigma$ level of significance detected in the image plane, same as contours shown in Column({\it 1}). 
    Column ({\it 5}): de-lensed \lya\ contours overlaid on de-lensed {\it HST} data, with the yellow contours showing the caustics in the source plane.  White and red contours mark the 3- and 8-$\sigma$ \lya\ surface brightness, same as the contours in Column({\it 2}). The cyan dashed arrow shows the direction along which we extract the one-dimensional surface brightness profile (see Figure~\ref{fig:1D_SB_profiles} below). Compared with System {\it A}, the lensing configuration of System {\it B} is much more complicated, with images {\it a}-{\it d} being partial images with different levels of completeness. Image {\it e} is the only complete image of \lya\ emission above 3-$\sigma$ limiting surface brightness. }
    \label{fig:img_src_plane_contours_B}
\end{figure*}

When computing the total \lya\ flux in the nebulae, we consider image {\it a} for the northern nebula to avoid the confusion of partially lensed multiple images, and average images {\it a} and {\it c} for the southern nebula. Due to the contamination from a nearby galaxy at the east side of the southern nebula in image {\it b}, we leave out image {\it b} in the average. In contrast with the continuum sources, the \lya\ nebulae span a much larger area in the image plane, within which the magnification factor can vary significantly (see Column({\it 4}) of Figure~\ref{fig:img_plane_contours_A}). Therefore, instead of using a mean magnification factor, we correct the lensing magnification for each spaxel within the extended nebulae before summing over all spaxels within the 3-$\sigma$ contour for these images. We then integrate the flux over the wavelength range of 4890-4930 \AA\ (the same wavelength window for constructing the narrow-band image described above). The total de-lensed \lya\ flux of the southern nebula obtained from image {\it a} is $\approx 5\%$ ($25\%$) lower than that from image {\it c} before (after) including the \lya\ flux from the DLA troughs.  This difference of \lya\ flux between images {\it a} and {\it c} is in agreement with what is observed in the \lya\ surface brightness and de-lensed magnitudes of galaxies {\it A}2 and {\it A}3 (see \S\ref{sec:analysis_gal}), suggesting again that the magnification factor near the continuum regions in image {\it a} is smaller than what is predicted by the lens model.
%Based on the lens model described in \S\ref{sec:lens_model}, the mean magnification of the northern nebula from image {\it a} is $\Bar{\mu}\approx 3.2$, and $\Bar{\mu}\approx 4.6$ for image {\it c} of the southern nebula. Summing over all spaxels within the 3-$\sigma$ contour for these two images and 

After excluding the \lya\ flux from within the DLA troughs and correcting the lensing magnification, we obtain a total flux of $f_{\rm Ly\alpha}(A_{\rm north})=(2.0\pm0.1)\times 10^{-17}\,{\rm erg\, s^{-1}\, cm^{-2}}$ for the northern nebula, and $f_{\rm Ly\alpha}(A_{\rm south})=(2.9\pm0.1)\times 10^{-17}\,{\rm erg\, s^{-1}\, cm^{-2}}$ for the southern nebula.  Including the \lya\ flux from the DLA troughs, the total flux is increased to $f_{\rm Ly\alpha}^{\rm tot}(A_{\rm north})=(2.7\pm0.1)\times 10^{-17}\,{\rm erg\, s^{-1}\, cm^{-2}}$ for the northern nebula, and $f_{\rm Ly\alpha}^{\rm tot}(A_{\rm south})=(3.8\pm0.1)\times 10^{-17}\,{\rm erg\, s^{-1}\, cm^{-2}}$ for the southern nebula. The \lya\ signal inside the DLA troughs therefore accounts for $\approx 25\%$ of the total \lya\ emission from both the northern and southern nebulae. At $z\approx 3.038$, these flux measurements (including the \lya\ flux in the DLA troughs) correspond to a \lya\ luminosity of $L_{{\rm Ly}\alpha}(A_{\rm north})=(2.15\pm0.07)\times 10^{42} \, {\rm erg\,s^{-1}}$ for the northern nebula, and $L_{{\rm Ly}\alpha}(A_{\rm south})=(3.03\pm0.08)\times 10^{42} \, {\rm erg\,s^{-1}}$ for the southern nebula. Combining both northern and southern nebulae together leads to a total \lya\ luminosity of $L_{{\rm Ly}\alpha}(A)=(5.2\pm0.1)\times 10^{42} \, {\rm erg\,s^{-1}}$.
%$L(\lya)=(21.1\pm0.8)\times 10^{41} \, {\rm erg\,s^{-1}}$ and $(40.2\pm0.9)\times 10^{41} \, {\rm erg\,s^{-1}}$, respectively.  

%\begin{figure}
%    \centering
%    \includegraphics[width=\linewidth]{fig_SysA_lya_variations.pdf}
%    \caption{{\it Left:} Pseudo NB image of the \lya\ emitting nebulae in System {\it A}, smoothed with a Guassian kernel with $\sigma=0\farcs5$. The contours %correspond to $3\sigma$ surface brightness limit (i.e., $2.4\times 10^{-18}\sbunit$), with the blue (red) contours indicate images of the northern %(southern) nebula. {\it Right: } Top panel shows spectra of the northern and southern nebula, summed within the $3\sigma$ contour over all multiple %images, and bottom panel shows spectra from two individual apertures indicated in the pseudo NB image on the left. Both apertures have a radius of %$0\farcs8$. Velocity of 0 corresponds to the systemic redshift of {\it A}1 ($z=3.0367$. )Spectra shown are smoothed with a boxcar filter with a width of 3 %pixels (i.e., $3.75\ang$).  Note that the smoothing is only for display purposes, and all analyses are conducted with unsmoothed spectra.}
%    \label{fig:LyaNB_variation_sysA}
%\end{figure}

For System {\it B}, no apparent DLA or strong ISM absorption features are detected in the MUSE spectra of the star-forming regions, but the low $S/N$ as a result of a faint continuum makes gas column density estimates highly uncertain.  The apparent discontinuity in the continuum blueward and redward of the \lya\ emission line is consistent with the expectation from the \lya\ forest in the intergalactic medium at $z\approx 3.75$ \citep[e.g.][]{Becker2007}.  To construct a pseudo narrow-band image for this system, we first subtract the expected continuum at the observed \lya\ line.  Following the approach described above for System {\it A}, we determine the continuum level in each spaxel of the star-forming regions by matching the low-order polynomial fit of the UV continuum presented in Figure~\ref{fig:HeII_CIII_fitting} to the observed spectrum.  For spaxels outside of the continuum emitting regions, the continuum level at the \lya\ emission line is determined based on a linear interpolation between blue and red continuum fluxes observed within 5730--5760 \AA\ and 5800--5830 \AA, respectively.  A pseudo narrow-band image is then constructed by integrating the flux in the wavelength range from 5766 \AA\ to 5796 \AA.  A smoothed version using a Gaussian kernel of ${\rm FWHM}=1''$ is presented in Column ({\it 1}) of Figure~\ref{fig:img_src_plane_contours_B}, overlaid with constant surface brightness contours of $2.8\times10^{-18}\, \mathrm{erg \, s^{-1} \, cm^{-2} \, arcsec^{-2}}$ detected at 3-$\sigma$.  In Column ({\it 2}) of Figure~\ref{fig:img_src_plane_contours_B}, contours of 2.8 and $7.5\times10^{-18} \, \mathrm{erg \, s^{-1} \, cm^{-2} \, arcsec^{-2}}$ (i.e., 3-$\sigma$ and 8-$\sigma$) are presented along with the {\it HST} composite image to illustrate the relative alignment between the \lya\ nebulae and the associated galaxies.  The magnification map presented in Column ({\it 3}) shows the fast changing magnification factors across all five lensed images in System {\it B}, as the lensed \lya\ emitting regions straddles multiple critical curves in the image plane.  De-lensed \lya\ pseudo narrow-band images (smoothed using a a Gaussian kernel of ${\rm FWHM}=0\farcs5$) and the {\it HST} images in the source plane, based on the fine-tuned lens model described in \S\ \ref{sec:fine_tune_lens_model}, are also presented in Columns ({\it 4}) and ({\it 5}) of Figure~\ref{fig:img_src_plane_contours_B}, respectively.  Compared with System {\it A}, the lensing configuration of System {\it B} is more complicated, with images {\it a}--{\it d} being partial images of different completeness levels. Image {\it e} is the only complete lensed image of the \lya\ nebula defined at the 3-$\sigma$ level of significance in surface brightness.  The source plane reconstruction from image {\it e} reveals a relatively symmetric \lya\ emission morphology, roughly centered near the UV continuum sources.  Using image {\it e}, we estimate the projected size of the \lya\ emitting nebula to be approximately 10 pkpc in diameter.  A small spatial offset, $\approx 0\farcs1$, is seen between UV continuum sources and the peak of \lya\ emission, corresponding to $\approx 0.7$ pkpc at $z=3.754$.  It is commonly observed among LAEs that the \lya\ emission signals can have an offset from the UV continuum, with a median 1D projected offset of $\approx 0.6$ pkpc in previous slit spectroscopic data \citep[e.g.,][]{Hoag2019,Ribeiro2020,Lemaux2020}.  Larger offsets have also been found in narrow-band imaging data \cite[e.g.,][]{Shibuya2014}.  However, we note that the continuum fluxes of galaxies {\it B}1 and {\it B}2 are much fainter than the LAEs considered in those studies.

We use image {\it e}, the most complete image among all five multiple images of System {\it B}, to compute the total flux of the \lya\ emission.  Despite of the flux anomaly observed in image {\it e} of galaxy {\it B}1 as discussed in \S\ref{sec:galaxy_photometry}, the effect is likely localised (since image {\it e} of {\it B}2 does not show the same brightness enhancement) and therefore will not significantly bias the total \lya\ flux from the extended nebula.  After correcting the lensing magnification for each spaxel,
%With a mean magnification of $\Bar{\mu}=8.3$ for the nebula in image {\it e}, 
we obtain a total flux of $f_{\rm Ly\alpha}(B)=(7.4\pm0.2)\times 10^{-18}\, {\rm erg\, s^{-1}\, cm^{-2}}$, integrated across the wavelength range of 5766-5796 \AA\ (the same wavelength window for constructing the narrow-band image described above) and summed over all spaxels within the 3-$\sigma$ contour in image {\it e}.  At $z=3.754$, the observed \lya\ flux translates to a total luminosity of $L_{{\rm Ly}\alpha}(B)=(9.8\pm0.2)\times 10^{41} \, {\rm erg\,s^{-1}}$. 

\begin{figure*}
    \centering
    \includegraphics[width=\linewidth]{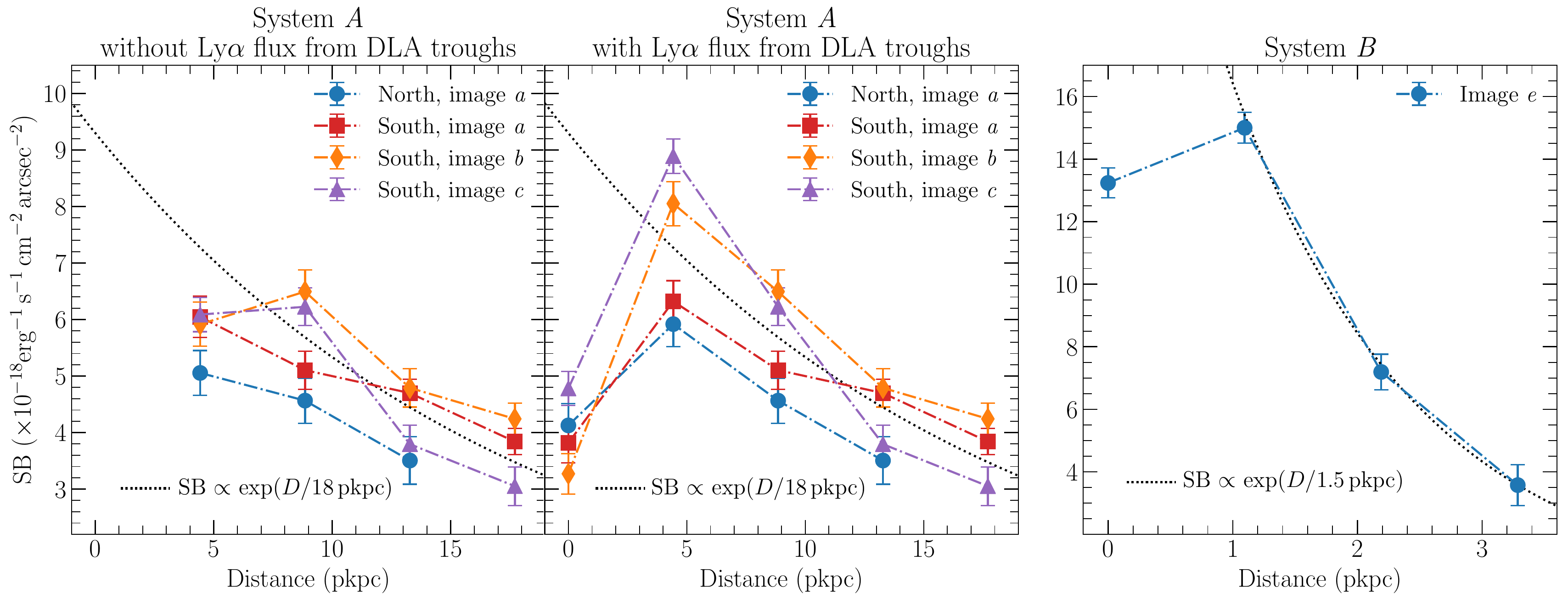}
    \caption{De-lensed \lya\ surface brightness profile, extracted along the directions indicated in Figures~\ref{fig:src_plane_contours_A} and \ref{fig:img_src_plane_contours_B}. For System {\it A}, we present the surface brightness profile both before and after including the \lya\ flux from the DLA troughs at the locations of galaxy continuum. Distance is measured from the location of {\it A}1 ({\it A}2 and {\it A}3) for the northern (southern) nebula. For System {\it B}, zero distance corresponds to the location of {\it B}1. Note that we use rectangle apertures to extract the surface brightness profile as guided by the velocity gradient within the nebulae, instead of circular annuli (see text). In both systems, there is a decrease in surface brightness at small distances. The suppression may be attributed to either attenuation by the observed high neutral gas column density and possibly high dust content in System {\it A} or by a reduced total gas column as a result of galactic scale outflows in System {\it B}.}
    \label{fig:1D_SB_profiles}
\end{figure*}

For both systems, we also extract the de-lensed one-dimensional \lya\ surface brightness profile in the source plane starting from the galaxy continuum regions to the edge of each nebula (near the 3-$\sigma$ surface brightness contours), as shown in Figure~\ref{fig:1D_SB_profiles}. In \S\ref{sec:velocity_shell_fitting} below, we derive the velocity gradient within the nebulae in both systems. As the velocity gradient suggests non-spherical gas flows in both systems, we therefore use a series of $2\arcsec \times 0\farcs6$ ($1\arcsec \times 0\farcs15$) pseudo slits for System {\it A} (System {\it B}), instead of circular annuli. 
We then extract the surface brightness profiles along directions guided by the velocity gradient within the nebulae (cyan dashed arrows in Figures~\ref{fig:src_plane_contours_A} and \ref{fig:img_src_plane_contours_B}; also see \S\ref{sec:velocity_shell_fitting} below). The position angle of the pseudo slit is $25^\circ$ north through east for System {\it A} and $60^\circ$ north through west for System {\it B}.  
%,, which is perpendicular to the gradient of the velocity field observed in both northern and southern nebulae.
%, which is at an angle of $25^\circ$ from north to west (south to east) for the northern (southern) nebula in System {\it A} and
%The position angle of the pseudo slit is $60^\circ$ north through west, perpendicular to the observed velocity gradient %from south to west for the nebula in System {\it B}, as 
%indicated by cyan dashed arrows in Figures~\ref{fig:src_plane_contours_A} and \ref{fig:img_src_plane_contours_B}. 
The first aperture (distance of zero) is centered on the de-lensed locations of galaxy {\it A}1 ({\it A}2 and {\it A}3) for the northern (southern) nebula in System {\it A}, and the distance of the subsequent apertures are measured from these corresponding continuum regions. For System {\it B}, the distance is measured from the location of {\it B}1, where we put the first aperture. We show the surface brightness profiles for System {\it A} both before and after including the \lya\ flux inside the DLA troughs from star forming regions (see Figure~\ref{fig:HeII_CIII_fitting}). 

As discussed above, \lya\ surface brightness from the southern nebula agrees well across all three multiple images before including \lya\ flux from DLA troughs, while image {\it a} becomes dimmer than images {\it b} and {\it c} after including the \lya\ flux from the DLA troughs, suggesting a relatively smaller magnification factor in image {\it a} than what is predicted by the lens model at the locations of {\it A}2{\it a} and {\it A}3{\it a}.  Figure~\ref{fig:1D_SB_profiles} shows that both Systems {\it A} and {\it B} exhibit a general decline in \lya\ surface brightness with increasing projected distance.  Applying a simple exponential profile to characterize the observed surface brightness, ${\rm SB}({\rm Ly\alpha})\propto {\rm exp}(-D/D_s)$, we find a best-fit scale radius of $D_s\approx 18$ pkpc for System {\it A} and $D_s\approx 1.5$ kpc for System {\it B} (see Figure \ref{fig:1D_SB_profiles}), corresponding to a half-light radius of $r_e\approx 30$ and 2.5 pkpc for Systems {\it A} and {\it B}, respectively.  These are consistent with the typical size found for Lyman break galaxies \citep[e.g.][]{Steidel2011} and LAEs \citep[e.g.][]{Wisotzki2016,Leclercq2017}. 

At the same time, we also see a suppressed \lya\ surface brightness at the locations of the galaxies.  The suppression resembles what is seen in the ``net absorption'' sub-sample of stacked \lya\ surface brightness profiles of \citep[][]{Steidel2011}.  We propose that the suppression may be attributed to attenuation by dusty outflows, which is supported by the observed high neutral gas column density and blueshifted low-ionization ISM absorption line in System {\it A}.  Under the dusty outflow scenario, the radial extent of the  observed dip in the center of the \lya\ surface brightness profile is then a direct measure of the projected radius of the dusty outflows, which in the present cases amounts to  $\lesssim 5$ pkpc for System {\it A} and  $\lesssim 1$ pkpc for System {\it B}. Dust in the ISM could also contribute to the suppression of the \lya\ signal in the gap, which would imply an anisotropic distribution of the dusty material in the ISM given the presence of extended \lya\ nebulae at larger distances away from the line-of-sight. Alternatively, the suppression may be attributed to a reduced $N({\rm HI})$ as a result of galactic scale outflows or galaxy interactions \citep[e.g.,][]{Johnson2014}. 

%- A is much shallower, with rs 20 and 5 pkc, while B has rs around 1.2 kpc.
 
% \begin{figure}
%    \centering
%    \includegraphics[width=8cm]{fig_SysB_lya_variations.pdf}
%    \caption{{\it Left:} Pseudo NB image of the \lya\ emitting nebula in System {\it B}, smoothed with a Guassian filter with $\sigma=0\farcs3$. The blue %(red) contour corresponds to $3\sigma$ ($20\sigma$) surface brightness limit. {\it Right: } Spectra extracted within $3\sigma$ and $6\sigma$ contours, %as well as the spectrum extracted from the three brightest pixels from images c, d and e. Velocity of 0 corresponds to the systemic redshift of {\it B}1/{\it B}2 %($z=3.7540$). Spectra shown are smoothed with a boxcar filter with a width of 3 pixels (i.e., $3.75\ang$).  Note that the smoothing is only for display %purposes, and all analyses are conducted with unsmoothed spectra.}
%    \label{fig:LyaNB_variation_sysB}
%\end{figure}

%\subsection{Ly$\alpha$ line profile}

\subsection{Spatial variation of line profiles}\label{sec:variation}

\begin{figure*}
    \centering
    \includegraphics[width=0.9\linewidth]{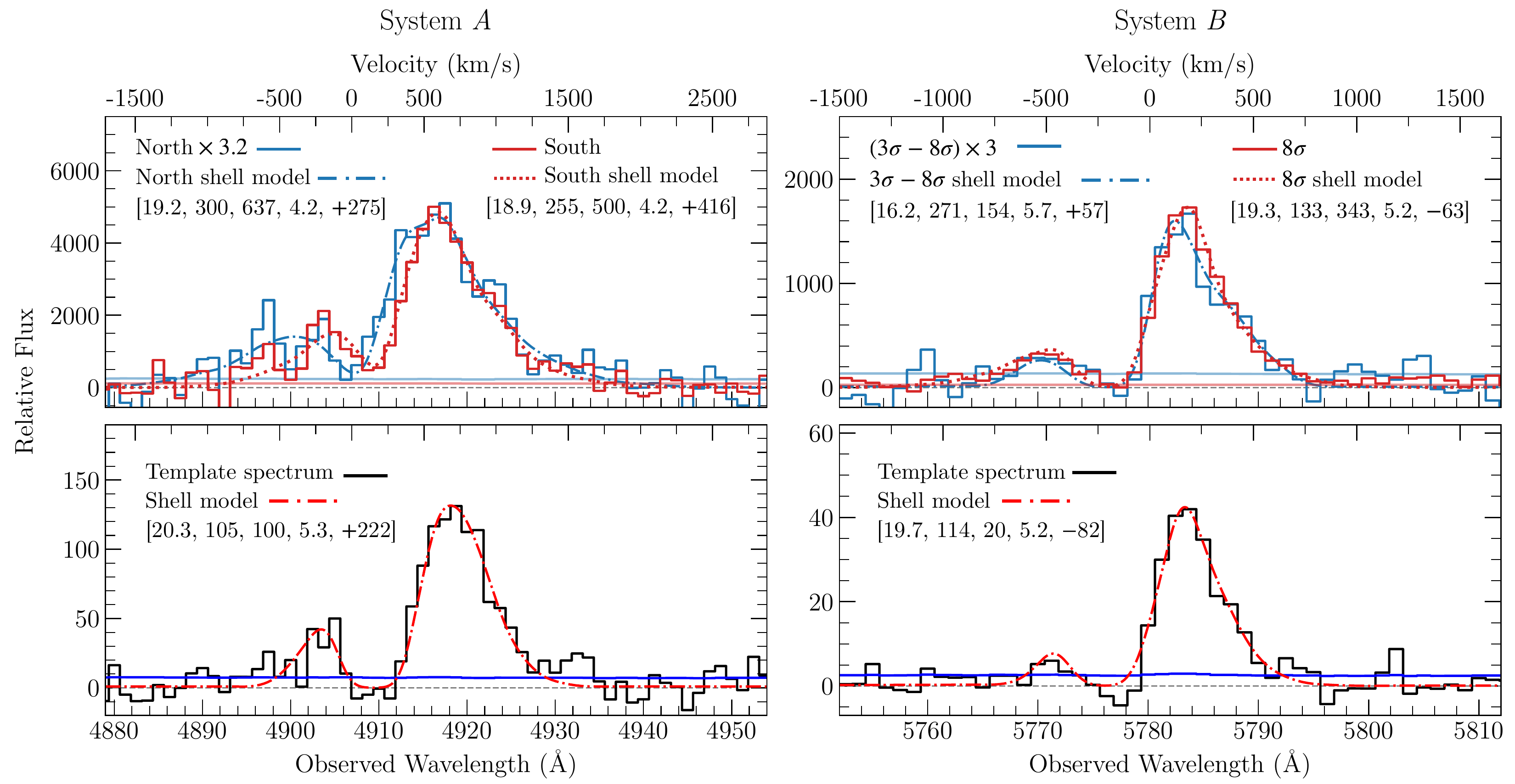}
    \caption{Spatial variation of the observed \lya\ profiles and its impact on the best-fit shell model for Systems {\it A} ({\it left}) and {\it B} ({\it right}).  {\it Top} panels display the summed \lya\ line profiles over a large area, while {\it bottom} panels display the template \lya\ profiles extracted from localized, small apertures indicated in Figure \ref{fig:src_plane_contours_A} for System {\it A} and stacked brightest pixels from multiple images {\it c}, {\it d} and {\it e} for System {\it B} (see text) .  The corresponding best-fit model profiles from the expanding shell model described in the text are included as dotted and dash-dotted curves with the best-fit parameters summarized in the legend, following the order of [$\log\,N({\rm HI})/{\rm cm}^{-2}$, $v_{\rm exp}\,{\rm (km\,s^{-1})}$, $\sigma_i \,{\rm (km\,s^{-1})}$, $\log\,T_{\rm eff}/{\rm K}$, $\Delta\,v\,{\rm (km\,s^{-1})}$].  For the large area sums in the {\it top} panels, System {\it A} is naturally divided into northern and southern nebulae, while System {\it B} is broadly divided by low- and high-surface brightness regions, between within and outside of the 8-$\sigma$ contours.  Zero velocity corresponds to $z_{\rm sys}=3.0364$ for System {\it A}, which is the systemic redshift of {\it A}1, and $z_{\rm sys}=3.7540$ for System {\it B}, which is the systemic redshift of {\it B}1 and {\it B}2.  The largest distinction between large and small aperture stacks is seen in System {\it A}, both in terms of the flux level in the valley between the blue and red peak as well as the profile line width, which is captured by a combination of neutral hydrogen column density $\log\,N({\rm HI})/{\rm cm}^{-2}$, intrinsic line width ($\sigma_i$) and effective temperature ($\log\,T_{\rm eff}/{\rm K}$).  The line profiles are significantly broader in the stacked spectra obtained over a larger area.  In contrast, such distinction is much less visible in System {\it B}.
    %To account for possible presence of spatial variation in the \lya\ profile that may lead to systematic bias in the model analysis, a template spectrum is constructed for each nebula from a restricted, high $S/N$ area (see text).  The template spectra of Systems {\it A} and {\it B} are presented in the bottom panels, along with the best-fit model profiles.
    }
    \label{fig:lya_profile_and_tlac_models}
\end{figure*}

In addition to the surface brightness variation in the narrow-band images, the \lya\ nebulae in both systems exhibit a double-peak profile with a significantly enhanced red peak that indicates expansion/outflowing motions.  In the {\it top-left} panel of Figure \ref{fig:lya_profile_and_tlac_models}, we present stacked \lya\ spectra from the northern and southern nebulae in System {\it A}.  The spectra are extracted separately from within the 3-$\sigma$ contours in Column ({\it 2}) of Figure \ref{fig:img_plane_contours_A}.  In the {\it top-right} panel of Figure \ref{fig:lya_profile_and_tlac_models}, we present stacked \lya\ spectra for System {\it B}, extracted from within the low-surface brightness (between 3-$\sigma$ and 8-$\sigma$ contours) and high-surface brightness (within the 8-$\sigma$ contours) regions shown in Figure \ref{fig:img_src_plane_contours_B}.  An overall shift in wavelength, both in the peak locations and the location of the valley of the \lya\ line, is clearly seen between the northern and southern nebulae in System {\it A}, with the northern nebula being blueshifted by $\approx 200$ \kms\ relative to the southern one, suggesting a large velocity gradient across the line-emitting region.  At the same time, no significant differences are seen between low- and high-surface brightness regions in System {\it B}.

%, suggesting both a velocity shear within the gas cloud as well as variation in physical properties of the gas. 
% Both spectra show the double-peaked feature with a more extended red wing, typical for expanding gas. Possible additional secondary peaks are also present in both red and blue sides. There is a velocity difference of $\approx 200$ km/s between the northern nebula and the southern nebula, with the northern nebula being relatively blue-shifted. 

To investigate in detail the velocity offset and possible spatial fluctuations in the \lya\ profiles across both nebulae, we need to employ smaller apertures for extracting \lya\ spectra.  Specifically, we consider two competing factors when determining the extraction apertures: (1) the $S/N$ necessary to obtain significant signal in both the blue and red peaks and (2) possible spatial smearing of the extracted \lya\ profile over a large aperture that may lead to erroneous characteristics of the \lya\ profile. Because of the low surface brightness nature across all regions in System {\it A}, the \lya\ line per spaxel does not have sufficiently high signals.  We therefore experiment with extracting \lya\ spectra from a range of aperture sizes to identify an appropriate aperture size for achieving a sufficiently high $S/N$ while limiting the smearing effect from combining different regions. We obtain the optimal extraction aperture from a localized, small area with a radius of $0\farcs5$ centered near the highest surface brightness peak in the reconstructed source-plane narrow-band image (blue dashed circles in Column ({\it 2}) of Figure~\ref{fig:src_plane_contours_A}). We then identify the spaxels located whitin this area in the image plane in all three multiple images {\it a}, {\it b} and {\it c}, and construct a template spectrum for System {\it A} by coadding the spectra from all identified spaxels, which contains the information of gas properties in the brightest region of the nebula. The template spectrum is displayed in the {\it bottom-left} panel of Figure \ref{fig:lya_profile_and_tlac_models}. 

Although the $S/N$ of the template spectrum is lower than what is seen in the large-area stacks ({\it upper-left} panel of Figure~\ref{fig:lya_profile_and_tlac_models}), the signal is strong enough to demonstrate the significant difference between the \lya\ profiles extracted from small and large areas.  Specifically, the template spectrum has a narrower width than the large-area stacks from both the northern and southern nebulae. In addition, the template spectrum exhibit a flux level that is consistent with zero at the bottom of the valley between the red and blue peaks.  The observed zero flux in the valley is consistently seen across the nebulae in all spectra extracted from small apertures, and differs from the distinctly non-zero flux observed in the stacked spectra over the larger nebulae (see also Figure 3 of \citealt{Caminha2017}). Such difference can be naturally explained by the presence of a large velocity gradient in the nebulae that results in smearing of the combined \lya\ profile. But because a non-zero flux in the valley of a double-peak \lya\ profile would lead to very different parameters constraints for the expanding shell model \citep[e.g.][also see below]{Dijkstra2006,Verhamme2006,HansenOh2006,Laursen2009,Schaerer2011,Gronke2015}, the ability to spatially resolve the velocity field is necessary for obtaining accurate constraints for the underlying gas properties. In our study, we leverage lensing magnifications to resolve spatial variations on scales as small as $\approx 2$ pkpc along both nebulae (Systems {\it A} and {\it B}) in ground-based, seeing-limited data, though we caution that variations on smaller scales may still be present in these clouds  \citep[e.g.][]{Cantalupo2019}.

For System {\it B}, because the nebula is significantly brighter than what is seen in System {\it A} and the distinction in the observed \lya\ profile is subtle between different locations, we construct a template spectrum using only the brightest pixels in images {\it c}, {\it d} and {\it e} (one pixel from each image) to better constrain possible velocity gradient and spatial variation over a small area. The locations of the three brightest pixels included in the template spectrum are indicated by the black arrows in Column ({\it 1}) of Figure~\ref{fig:img_src_plane_contours_B}.
The template spectrum for System {\it B} is displayed in the {\it bottom-right} panel of Figure \ref{fig:lya_profile_and_tlac_models}, and does not show significant differences from the stacked spectra from larger areas ({\it upper-right} panel of Figure~\ref{fig:lya_profile_and_tlac_models}).

\subsection{Physical properties of Ly$\alpha$ nebulae under an expanding shell model}\label{sec:velocity_shell_fitting}

\begin{table*}
    \centering
    \caption{Summary of the best-fit parameters (95\% confidence interval) for characterizing the observed \lya\ profile under an expanding shell model. }%See Figure~\ref{fig:lya_profile_and_tlac_models} for data and model spectra.}
    \begin{tabular}{l c c c c c c}
    \hline
    \multicolumn{7}{c}{System {\it A}, $z_{\rm sys} = 3.0364$} \\
    \hline
      & & $v_{\rm exp}$ & $\sigma_i^a$ &  & $\Delta\,v^b$ &  \\
     Spectrum & $\log\,N({\rm HI})/{\rm cm}^{-2}$ & (km/s) & (km/s) & $\log\,T_{\rm eff}/{\rm K}$ & (km/s) & $\chi^2_\nu$ \\
    \hline
    North & $19.2^{+0.2}_{-0.3}$ & $300^{+40}_{-30}$ & $637^{+28}_{-51}$ & $4.2^{+0.6}_{-1.2}$ & $275^{+60}_{-37}$ &6.6 \\[0.15cm]
    South & $18.9^{+0.2}_{-0.1}$ & $255^{+6}_{-36}$ & $500^{+49}_{-7}$ & $4.2^{+0.1}_{-0.5}$ & $416^{+7}_{-44}$ &10.6 \\[0.15cm]
    % Region {\it 1} & $20.3^{+0.1}_{-0.7}$ & $172^{+166}_{-31}$ & 125 & $5.8^{+0.1}_{-2.8}$ & $7^{+466}_{-22}$ &1.9 \\[0.15cm]
    % Region {\it 2} & $19.1^{+1.5}_{-1.8}$ & $293^{+29}_{-219}$ & 125 & $5.7^{+0.3}_{-1.2}$ & $437^{+44}_{-311}$ &3.6 \\[0.15cm]
    Template spectrum & $20.3^{+0.2}_{-0.2}$ & $105^{+27}_{-20}$ & 100 & $5.3^{+0.2}_{-0.2}$ & $222^{+22}_{-30}$ &2.4 \\
    \hline
    \multicolumn{7}{c}{System {\it B}, $z_{\rm sys} = 3.7540$} \\
    \hline
    $3\sigma-8\sigma$ & $16.2^{+3.4}_{-0.7}$ & $271^{+63}_{-174}$ & $154^{+218}_{-57}$ & $5.7^{+0.2}_{-2.4}$ & $57^{+57}_{-138}$ &0.9 \\[0.15cm]
    $8\sigma$ & $19.3^{+0.2}_{-0.1}$ & $133^{+10}_{-15}$ & $343^{+20}_{-22}$ & $5.2^{+0.0}_{-0.2}$ & $-63^{+13}_{-19}$ &5.3 \\[0.15cm]
    % Region {\it 1} & $19.9^{+0.4}_{-0.2}$ & $91^{+48}_{-48}$ & 13 & $5.1^{+0.2}_{-0.7}$ & $-69^{+38}_{-44}$ &1.6 \\[0.15cm]
    % Region {\it 2} & $19.7^{+0.2}_{-0.1}$ & $97^{+12}_{-24}$ & 13 & $4.9^{+0.2}_{-0.7}$ & $-57^{+13}_{-31}$ &2.0 \\[0.15cm]
    Template spectrum & $19.7^{+0.1}_{-0.2}$ & $114^{+17}_{-19}$ & 20 & $5.2^{+0.2}_{-0.2}$ & $-82^{+25}_{-19}$ &1.3 \\
    \hline
    \multicolumn{7}{l}{$^a$ Values without errors indicate a prior specified by the nebular emission lines (see Table \ref{tab:line_fitting_HeII_CIII}).}\\
    \multicolumn{7}{l}{$^b$ Relative velocity with respect to $z_{\rm sys}$.} 
    \end{tabular}
    \label{tab:parameters}
\end{table*}

We utilize the spatially and spectrally resolved \lya\ profiles from MUSE observations and a \lya\ Monte Carlo radiative transfer code \texttt{tlac} \citep{Gronke2014,Gronke2015} to model the physical properties of the line-emitting gas.  We adopt an expanding shell model that has successfully explained many observed \lya\ spectra across a wide range of redshifts based on a finite set of parameters, including the neutral hydrogen column density, $N(\ion{H}{I})$, the expansion velocity, $v_{\rm exp}$, intrinsic line width, $\sigma_i$, effective temperature, $T_{\rm eff}$, and systemic velocity, $\Delta\,v$ \citep[e.g.][]{Verhamme2006,Yang2017,Gronke2017}.  

As illustrated in \cite{Verhamme2006}, while there are considerable degeneracies between different parameters of the shell model, the peak separation increases primarily with $N(\ion{H}{I})$, and the red-to-blue peak ratio increases with $v_{\rm exp}$, while $T_{\rm eff}$ and $\sigma_i$ set the overall line width \citep[see also][for a more detailed discussion on the effect of these parameters]{Gronke2015}.  In most cases this simple shell model provides a crude estimate of the underlying kinematic properties of the gas, but there are also known cases where the model failed to provide a good fit to data \citep[e.g.][]{Kulas2012,Orlitova2018}.  %Likely complications include the presence of clumpy medium, in which \lya\ photons travelling through low column density ``holes'', will contribute to enhanced blue emission \citep{Erb2014,Alexandroff2015} and non-zero flux in the valley between the peaks \citep[e.g.,][]{Gronke2016}.

We note that the shell models are developed for a spherical shell expanding radially outward, which may work better for unresolved Lya nebulae under the assumption that the emission sources are at the center of the gas. In applying these models to System {\it A}, for which the \lya\ photons may originate outside of the nebulae, we attribute the enhanced red peak in the observed \lya\ profile per spaxel to cloud expansion relative to a fiducial reference point interior to the cloud along the observer's line of sight.  In addition, we attribute the observed velocity shear to the motion of this reference point relative to the systemic redshift of the galaxies.  Although the source of photons likely lies outside of the nebulae (see \S\ref{sec:discussion} below for discussions on the origin of \lya\ photons), the problem is equivalent to extracting the \lya\ signal from one hemisphere of a spherical shell.  Because of spherical isotropy inherent to the shell model, we expect that considering one hemisphere would result in an overall reduction in the amplitude of the signal, instead of altering the line profile.  Guided by this understanding, we proceed with approximating the signal in each spaxel with expectations from an expanding shell model for constraining the systemic velocity at each location. 
%This enables us to map out the offset in the systemic velocity of individual spaxels across the nebulae. }

\begin{figure}
    \centering
    \includegraphics[width=\linewidth]{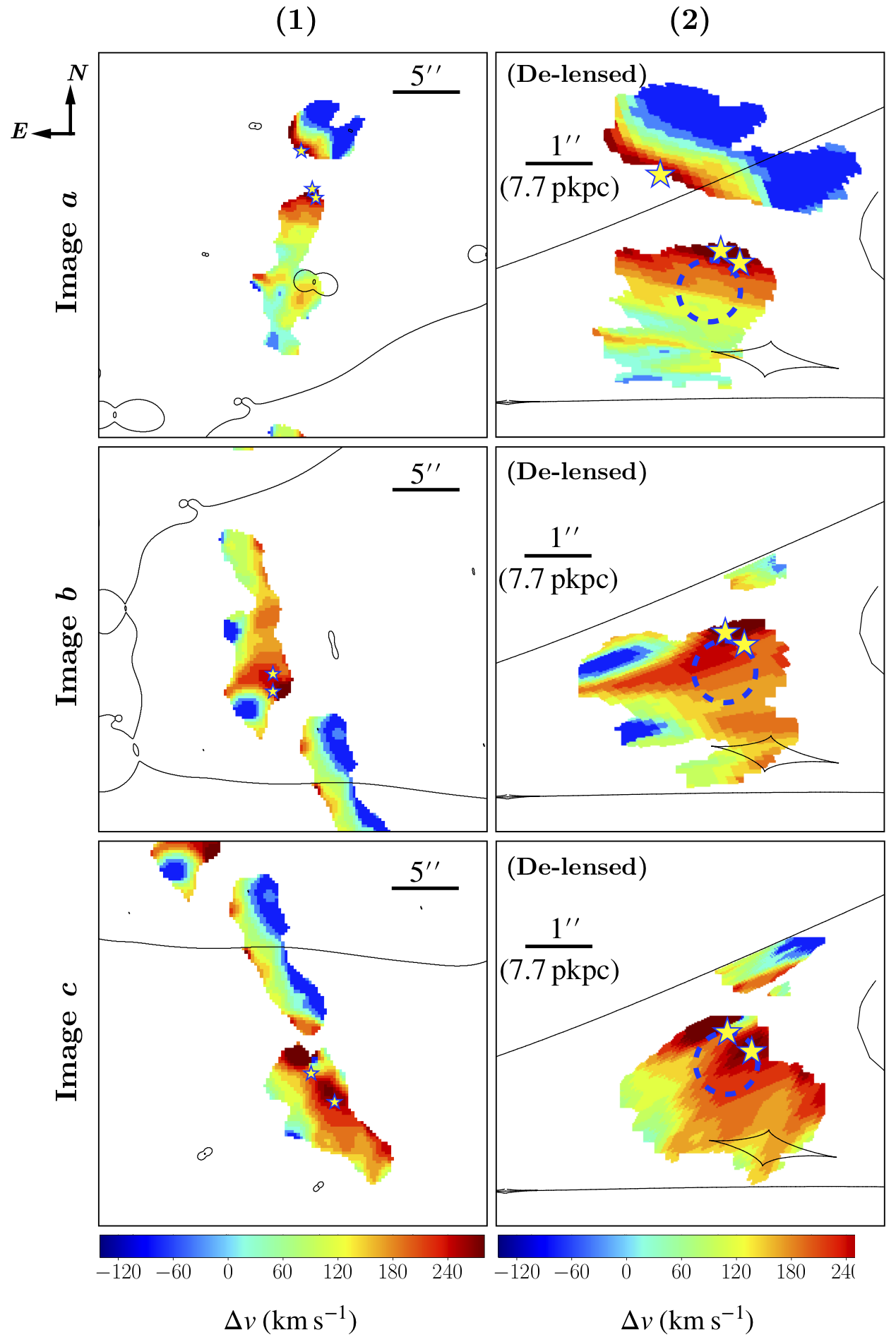}
    \caption{Column ({\it 1}): velocity map of multiple images {\it a}, {\it b} and {\it c} in System {\it A}, derived from cross-correlating the best-fit shell model for the template spectrum ({\it bottom-left} panel of Figure~\ref{fig:lya_profile_and_tlac_models}) and spectra extracted from spaxels within the 3-$\sigma$ contours. Zero velocity corresponds to $z_{\rm sys}=3.0364$, which is the systemic redshift of {\it A}1 derived from nebular emission lines.  
    Column ({\it 2}): de-lensed velocity map of individual images in the source plane. Star symbols mark the positions of the associated star-forming galaxies identified in {\it HST} images (see Figures~\ref{fig:img_plane_contours_A} and \ref{fig:src_plane_contours_A}). The blue dashed circles, same as in Column ({\it 2}) of Figure~\ref{fig:src_plane_contours_A}, mark the apertures for the template spectrum extraction, which we use for the shell model analysis.
    }
    \label{fig:sysA_vmap}
\end{figure}

\begin{figure}
    \centering
    \includegraphics[width=\linewidth]{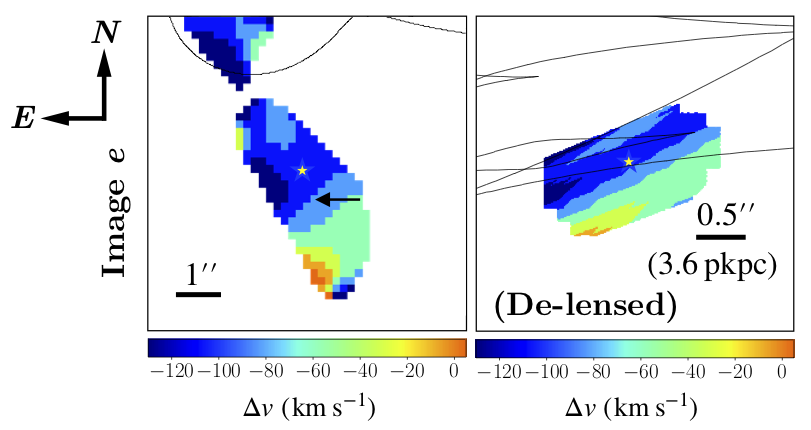}
    \caption{{\it Left}: velocity map of image {\it e} in System {\it B}, derived from cross-correlating the best-fit shell model for the template spectrum ({\it bottom-right} panel of Figure~\ref{fig:lya_profile_and_tlac_models}) and spectra extracted from spaxels within the 3-$\sigma$ contour. Zero velocity corresponds to $z_{\rm sys}=3.7540$, which is the systemic redshift of {\it B}1/{\it B}2 derived from nebular emission lines 
    {\it Right}: de-lensed velocity map of image {\it e} in the source plane. Star symbols mark the position of {\it B}1/{\it B}2 determined from {\it HST} images (see Figure~\ref{fig:img_src_plane_contours_B}). The black arrow, same as in Column ({\it 1}) of Figure~\ref{fig:img_src_plane_contours_B}, indicate the brightest pixel in image e that is included in the template spectrum.}
    
    \label{fig:sysB_vmap}
\end{figure}

For our analysis, we assume a homogeneous medium of constant gas density and compare the observed \lya\ profiles with predictions over a grid of model parameters.  To fully explore the allowed parameter space, we construct a model grid that covers $\log\,N(\ion{H}{I})/{\rm cm^{-2}}$ from 15.1 to 21.1, $v_{\rm exp}$ from 10 to 400 \kms, $\sigma_i$ from 25 to 700 \kms, $\log\,T_{\rm eff}/{\rm K}$ from 3.0 to 6.0, and $\Delta\,v$ from $-100$ to 550 \kms.  The velocity offset, $\Delta\,v$, is calculated with respect to the systemic redshift $z_{\rm sys}$ listed in Table \ref{tab:parameters}.  We use 10,000 photons and 100 frequency bins to generate each model profile.  Each model profile is also convolved with MUSE line spread function before compared to observations.  Given the uncertainty of the dust attenuation effect on \lya\ photons, we do not include dust in our models and it will be explored separately in the future. For each model, we compute a likelihood function ${\mathscr L}$ defined as
\begin{eqnarray}\label{eq:likelihood}
%    \chi^2 = \sum_j \frac{[S(\lambda_{j})-M(\lambda_{j}| N_{\ion{H}{I}}, v_{\rm exp}, \sigma_i, T_{\rm eff}, \Delta\,v)]^2}{dS^2(\lambda_{j})} \, .
    & &{\mathscr L}(N_{\ion{H}{I}}, v_{\rm exp}, \sigma_i, T_{\rm eff}, \Delta\,v) \nonumber \\
    & & = \prod_j \exp\left\{-\frac{[D(\lambda_{j})-M(\lambda_{j}| N_{\ion{H}{I}}, v_{\rm exp}, \sigma_i, T_{\rm eff}, \Delta\,v)]^2}{2\,S^2(\lambda_{j})}\right\}, \nonumber\\
    & &
\end{eqnarray}
where $D(\lambda_{j})$ and $M(\lambda_{j})$ are the observed and model spectra, respectively, and $S(\lambda_j)$ is the corresponding error spectrum.  The likelihood function is computed over the wavelength range of 4890-4930 \ang\ (5766-5796 \ang) for System {\it A} (System {\it B}), and can be translated to $\chi^2$ following $\chi^2=-2\ln\,{\mathscr L}$.  We then construct a marginalised likelihood function for each parameter by integrating ${\mathscr L}$ over all other parameters, and find the $95\%$ confidence interval centered around the best-fit value. Note that since we do not explicitly include turbulent broadening in the models here, the temperature inferred from the model represents an effective temperature that includes non-thermal motion.  For reference, for an intrinsic gas temperature of $T=10^4$ K, an inferred effective temperature of $T_{\rm eff}=10^5$ ($10^6$) K implies an underlying bulk flow of $\sigma_v^{\rm bulk}\approx 30$ (90) \kms.  

% To constrain the expanding shell model, we consider two competing factors in extracting the observed \lya\ profiles: (1) the $S/N$ necessary to obtain robust constraints for the model parameters and (2) spatial smearing of the empirical \lya\ profile over a large aperture that may lead to erroneous models (see \S\ \ref{sec:variation}).
%, when coadding the data from multiple spaxels to improve the $S/N$ of the \lya\ spectrum used for the fit.  
% Because of its low surface brightness nature, the \lya\ line per spaxel does not have sufficiently high signals for constraining the shell model at high confidence levels.  We therefore experiment with fitting \lya\ spectra extracted from a range of aperture sizes to identify an appropriate aperture size for achieving a sufficiently high $S/N$ while limiting the smearing effect from combining different regions. 
  
To illustrate the impact of velocity smearing on the \lya\ profile analysis, we first consider stacked spectra obtained over a large area %.  System {\it A} is naturally divided into northern and southern nebulae, while System {\it B} is broadly divided between low- and high-surface brightness regions, between within and outside of the 8-$\sigma$ contours.  The stacked spectra are presented in the 
along with the best-fit model profiles ({\it top} panels of Figure \ref{fig:lya_profile_and_tlac_models}).  The best-fit parameters and the associated 95\% confidence intervals are presented in Table \ref{tab:parameters}.  %A reduced $\chi^2$, $\chi_\nu^2$, is calculated for each best-fit model over the wavelength range from  4890-4930 \AA\ for System {\it A} and 5766-5796 \AA\ for System {\it B}, which encloses the full \lya\ profile and minimizes noise contributions from continuum regions.  
The large $\chi_\nu^2$ values in Table \ref{tab:parameters} show that an expanding shell model fails to provide a good fit for the high $S/N$ stacked \lya\ spectra for both systems.  A close examination of the profiles in the {\it top} panels of Figure \ref{fig:lya_profile_and_tlac_models} shows that the best-fit models with an uncharacteristically large intrinsic line width of $\sigma_i\approx 500$-650 \kms\ provide a poor fit to the blue peak of the northern and southern nebulae in System {\it A}.  The best-fit $\sigma_i$ is substantially broader than either what is seen in the nebular emission lines (see Table \ref{tab:line_fitting_HeII_CIII}) or what is expected for the velocity dispersion in halos of a comparable mass scale for the host galaxies \citep[e.g. $M_{\rm halo}< 10^{12}\,M_\odot$;][]{Trainor2012}. 
For System {\it B}, while the small $\chi_\nu^2$ for the stacked spectrum from low-surface brightness regions suggests a good fit to the data, the model remains poorly constrained with large associated uncertainties due to the low $S/N$ of the data.  At the same time, the best-fit shell model produces a relatively poor fit to the blue peak of high $S/N$, high-surface brightness regions, leading to a large $\chi_\nu^2$.  %At the same time, despite a smaller $\chi_\nu^2$ for the spectrum from low surface brightness regions, the model is poorly constrained due to a low $S/N$ of the spectrum (see Table \ref{tab:parameters}).

%on find that the \lya\ spectrum of northern nebula is best described by a simple shell model of $z_{\rm sys}=3.0374 $, $\log\,N(\ion{H}{I})=19.25\pm 0.20$, $v_{\rm exp}=270\pm30$ \kms, $T\approx 5000$ K, and $\sigma_i\approx 630$ \kms, while the southern nebula is best described by a shell model of $z_{\rm sys}=3.04$, $\log\,N(\ion{H}{I})=19.1\pm 0.15$, $v_{\rm exp}=283\pm30$ \kms, $T\approx 1000$ K, and $\sigma_i\approx 520$ \kms.  The best-fitting models are overplotted on top of the stacked spectra in Figure \ref{fig:lya_profile_and_tlac_models} for comparisons.  A velocity shear of $\Delta\,v\approx +180$ \kms\ from north to south across the entire \lya\ emitting gas is consistent with our initial visual inspection of the data.  

To improve the precision and accuracy of the model constraints, we perform the profile analysis for the template spectra extracted from localized, small apertures presented in the {\it bottom} panels of Figure \ref{fig:lya_profile_and_tlac_models}.  In addition, we adopt the observed nebular line width (see Table \ref{tab:line_fitting_HeII_CIII}) as a prior for modelling the \lya\ profiles.  This is justified  by the understanding that these \lya\ photons likely originate in the star-forming regions of the associated galaxies (see \S\ref{sec:discussion} below).  Specifically, we set $\sigma_i=100$ \kms\ for System {\it A} based on the observed FWHM of $\approx 240$ \kms\ in galaxy {\it A}2, and $\sigma_i=20$ \kms\ for System {\it B} based on the observed FWHM of $\approx 40$ \kms\ in galaxies {\it B}1/{\it B}2.  The best-fit model profiles are shown in dotted and dash-dotted curves and best-fit parameters are presented in Table \ref{tab:parameters}.  It is immediately clear that the resulting $\chi_\nu^2$ is reduced substantially and the parameters are well-constrained.  In contrast, setting the same prior on $\sigma_i$ when fitting the stacked spectra from larger apertures results in poor model fits with larger $\chi^2_\nu$ values (see Appendix). This again underscores the smearing effect on stacked \lya\ profiles extracted from larger areas, which can significantly bias the constrained gas properties with the presence of velocity gradient and spatial variations in the line emitting region. The best-fit models also suggest that both Systems {\it A} and {\it B} consistently require a large neutral hydrogen column density, $\log\,N(\ion{H}{I})/{\rm cm}^{-2}\gtrsim 19$ for explaining the observed \lya\ profiles from localized locations.  %In System {\it A}, dramatic changes are seen in $\sigma_i$ and $T_{\rm eff}$, from the initial $\sigma_i>500$ \kms\ and $T_{\rm eff}<10^4$ K to $<200$ \kms\ and $T_{\rm eff}>10^4$ K.  The best-fit $N(\ion{H}{I})$ also increases by an order of magnitude to $\log\,N(\ion{H}{1})/{\rm cm}^{-2}\apg 20$ for these small-aperture spectra, closer to the $N(\ion{H}{I})$ observed in the ISM of the associated galaxies.  Furthermore, the best-fit models also implies a substantial velocity offset of $\approx 200$ \kms\ between the two locations of $\approx 6$ pkpc apart.  The best-fit $\sigma_i$ derived for the \lya\ spectra from small apertures is consistent with the velocity dispersions observed in the ISM of galaxies {\it A}1-3 (see Table \ref{tab:line_fitting_HeII_CIII}), and the large velocity gradient underscores the need of spatially-resolving the velocity field in the nebula.

%In contrast, such distinction is less significant in System {\it B}.  While the resulting $\chi_\nu^2$ is reduced substantially, particularly for high intensity spaxels (10-$\sigma$ versus Region {\it 2}), little distinction is seen in the best-fit parameter values.  The only exception is the velocity offset, $\Delta\,v$, which changes from $\Delta\,v\approx -130$ \kms\ for the 10-$\sigma$ coadd to $\Delta\,v\approx -30$ \kms\ in Region {\it 2}. 
%The large line width is understood to be driven by the non-zero value at the bottom of the valley between blue and red peaks, which arises as a result of blending between line-emitting gas moving at different speeds.  We therefore repeat the model analysis using the spectra presented in the {\it middle-left} panel of Figure \ref{fig:lya_profile_and_tlac_models}, which are extracted from a fixed aperture of $0\farcs8$ in radius aperture and exhibit no detectable flux at the valley between peaks.  The best-fitting model profiles are presented along with the data, with the spectrum of Region {\it 1} characterized by $z_{\rm sys}=3.0xx$, $\log\,N(\ion{H}{I})=20.x\pm 0.20$, $v_{\rm exp}=1x0\pm x$ \kms, $T\approx x000$ K, and $\sigma_i\approx 1x0$ \kms\ and the spectrum of Region {\it 2} characterized by $z_{\rm sys}=3.0xx$, $\log\,N(\ion{H}{I})=20.x\pm 0.20$, $v_{\rm exp}=1x0\pm x$ \kms, $T\approx x000$ K, and $\sigma_i\approx 1x0$ \kms.

Because of the competing factors between spectral qualities (i.e., $S/N$) and velocity smearing, in addition to a strong degeneracy between different model parameters with modest $S/N$ data, we continue the analysis with a focus on constraining the velocity field, $\Delta\,v$, across the \lya\ nebulae. %rather than constraining highly degenerate parameters such as $\log\,N(\ion{H}{I})$ and $v_{\rm exp}$.  
This is achieved by cross-correlating the best-fit shell model for the template \lya\ spectra with each spaxel within the 3-$\sigma$ contours in both systems to measure velocity offsets at different locations. To optimize the cross-correlation analysis, we spatially smooth the data cube with a Gaussian filter of ${\rm FHWM}_{\rm smooth}=1''$ before extracting individual spectra. The resulting velocity maps are presented in Figure~\ref{fig:sysA_vmap} for System {\it A} and  Figure~\ref{fig:sysB_vmap} for System {\it B}. We also present de-lensed velocity maps of both nebulae in the source plane. We have also experimented with constraining the velocity gradient by fitting an asymmetric Gaussian function \citep[e.g., see Eq.1 of][]{Leclercq2020} to the red peak of the \lya\ profile from every spaxel within the 3-$\sigma$ contour in both systems, and we obtain a similar velocity gradient as shown in Figures~\ref{fig:sysA_vmap} and \ref{fig:sysB_vmap}. Our method utilising the best-fit model of the template spectra enables us to determine the systemic velocity offset $\Delta v$ of the line-emitting gas in the frame of the nearby star-forming region, thereby connecting the nebulae with the associated star-forming regions.

Our analysis of System {\it A} has uncovered a highly organized velocity field across the \lya\ emitting nebulae, with increasing velocity offset from $\Delta\,v\approx 0$ \kms\ at $\approx 11$ pkpc south of galaxies {\it A}2 and {\it A}3 to $\Delta\,v\approx +250$ \kms\ at the locations of these galaxies (Figure~\ref{fig:sysA_vmap}).  In the north, the velocity offset decreases from $\Delta\,v\gtrsim +200$ \kms\ at the location of galaxy {\it A}1 to $\Delta\,v\approx -150$ \kms\ at $\approx 13$ pkpc northwest of {\it A}1. The inferred velocity offset shows that the line-emitting gas closest to the galaxies is receding from the galaxies. Different from the extended blue wings seen in ISM absorption lines, this velocity offset seen in \lya\ emission places the gas behind the star-forming region. The observed steep velocity gradient, $|\Delta\,v/\Delta\,r_\perp| \approx 22-27\,\kms\,{\rm pkpc}^{-1}$, together with a large best-fit $N(\ion{H}{I})$ and an enhanced red peak in the \lya\ profile across the nebula supports a scenario in which high column density gas is driven outward from the galaxies to beyond 10 pkpc in projected distance into the low-density surroundings.  Due to a lack of AGN activities (see \S\ref{sec:cloudy}), the outflows are likely driven by star formation in these young galaxies. 

It is interesting that there exists an apparent gap in \lya\ signal between the northern and southern nebulae.  One possible explanation for this gap is a reduced $N({\rm HI})$ as a result of galaxy interactions.  A lack of strong \lya\ absorber has been seen at projected distances of $<20$ pkpc from an interacting galaxy pair at low redshift with an upper limit of $\log\,N({\rm HI})/{\rm cm}^{-2}\lesssim 13.7$ \citep[e.g.][]{Johnson2014}.  In the optically thin regime, we estimate a 2-$\sigma$ upper limit of $\log\,N(\ion{H}{I})/{\rm cm}^{-2}< 16.4$ at $\approx 5$ pkpc based on the observed 2-$\sigma$ upper limit  in \lya\ surface brightness and an assumption of 100\% escape fraction of ionizing photons from the galaxies.  While at this limit, the gas would still be optically thick to \lya\ photons, we cannot rule out the possibility of a significantly lower $N({\rm HI})$.
%the nearby ionizing radiation field \citep[e.g.][]{Hennawi2013}. The current non-detection of \lya\ emission in the gap constrains $N({\rm HI})$ at $\log\,N(\ion{H}{I})/{\rm cm}^{-2}< 16.4$ (a 2-$\sigma$ upper limit), with the ionizing radiation field estimated based on the \textsc{Bagpipes} models for galaxies {\it A}1-{\it A}3 (assuming 100\% escape fraction of ionizing photons and a distance of 5 pkpc between the galaxies and the center of the gap). 
Other plausible explanations for the gap also include a lack of illumination from young stars due to anisotropic leakage of \lya\ and ionizing photons, and attenuation of \lya\ signal due to highly neutral, dusty gas in-between these galaxies (also see discussion in \S\ref{sec:discussion} below).

In contrast, System {\it B} exhibits distinct properties from System {\it A}.  The \lya\ nebula appears to be distributed symmetrically around galaxies {\it B}1 and {\it B}2 with the peak intensity located close to star forming regions.  The inferred line-of-sight velocity offset of $\approx  -100$ \kms\ near the location of the galaxies, coupled with the observed \lya\ profile, again supports an outflow scenario from the galaxies.  The observed velocity gradient of $|\Delta\,v/\Delta\,r_\perp| \approx 20\,\kms\,{\rm kpc}^{-1}$ toward the outer edge of the nebula may be explained by a line-of-sight projection effect.

\section{Discussion}\label{sec:discussion}

We have shown that by accounting for spatial variations in the observed \lya\ line profiles, we are able to determine the velocity field and constrain gas flows across the nebulae.  
Given the proximity of the line-emitting gas to star-forming galaxies and the relatively small velocity offset between gas and galaxies, we argue that the gas is being driven out of the star-forming regions at a modest speed. 
%and that the velocity gradient uncovered in both nebulae can be interpreted as due to line-of-sight projection of outflowing gas from the adjacent star-forming galaxies.  
%\subsection{Connection between the nebulae and the galaxies}
Specifically, the \lya\ nebula of System {\it B} exhibits a relatively symmetrical morphology with the peak of the \lya\ emission located close to the star-forming regions.  This configuration is typical of low-mass LAEs at high redshifts \citep[e.g.][]{Wisotzki2016,Leclercq2017}, and suggests that gas flows outward from the star-forming regions into the low-density halo environment.  
At the same time, galaxies {\it A}1, {\it A}2 and {\it A}3 share similar spectral and photometric properties (see \S\ref{sec:analysis_gal}).  The close proximity of these three galaxies suggest that they may share an interactive group environment and are part of a common CGM.  The \lya\ nebulae are clearly offset to one side of the galaxies with the highest surface brightness regions bordering the continuum-emitting regions (see Figures~\ref{fig:img_plane_contours_A} and \ref{fig:src_plane_contours_A}).  %This proximity suggests that 
While the star-forming regions contribute significantly to the extended \lya\ emission, %likely through a combination of recombination radiation from photoionized gas and resonant scattering of \lya\ photons produced in the star-forming regions, as discussed in the previous two sections.  However, 
the connection between the star-forming regions and the large-scale outflows remains uncertain.

We consider two plausible scenarios for the origin of the outflows. First, the northern nebula originates in gas flowing out of {\it A}1, while the southern nebula originates in gas flowing out of galaxies {\it A}2 and {\it A}3.  This is plausible if all three galaxies are capable of driving galactic scale super winds.  Applying the conversion factor of \cite{MadauDickinson2014}, we estimate an unobscured SFR of $\approx 22$, 5 and 4 $M_\odot\,{\rm yr}^{-1}$ for galaxies {\it A}1, {\it A}2, and {\it A}3, respectively, based on the observed rest-frame UV absolute magnitudes $M_{1500}$ presented in Table~\ref{tab:photometryA}.  In the presence of dust, this observed $M_{1500}$ and inferred SFR are likely lower limits to the intrinsic values.  In addition, we estimate a total projected area based on the continuum regions determined by SExtractor (see \S\ref{sec:galaxy_photometry}) and apply lensing magnification corrections based on the fine-tuned lens model (see \S\ref{sec:lens_model}).  We find the intrinsic projected area of {\it A}1, {\it A}2, and {\it A}3 to be $\approx 50$, 11 and 11 pkpc$^2$, respectively.  For galaxies {\it A}2 and {\it A}3, these are based on an average over all three images, {\it a}, {\it b}, and {\it c} after lensing magnification corrections.  Combining the estimated unobscured SFR and projected area leads to an estimate of SFR per unit area of $\gtrsim 0.4\,M_\odot\,{\rm yr}^{-1}\,{\rm pkpc}^{-2}$ in these individual galaxies.  This exceeds the empirical threshold seen in driving galactic scale super winds in local starburst galaxies \citep[e.g.][]{Heckman:2015}.  

Alternatively, galaxy {\it A}1 may be the single dominant source driving the outflows seen in both the northern and southern nebulae.  Apart from being the most massive galaxy with the highest SFR in the group, {\it A}1 also shows more extended blue wings in the low-ionization ISM absorption lines (see Figure~\ref{fig:HeII_CIII_fitting} and Table~\ref{tab:absorption_lines}), suggesting the presence of galactic outflows that are more prominent than what is seen from the same line features in galaxies {\it A}2 and {\it A}3. In this scenario where galaxy {\it A}1 is the origin of the outflows on both sides, the gap in \lya\ emission between the northern and southern nebulae is likely due to dusty outflow materials from galaxy {\it A}1 that cover the gap area along the line-of-sight.

A remaining question of the observed line-emitting nebulae is the origin of \lya\ photons.  As described earlier, multiple mechanisms can lead to \lya\ emission in diffuse gas, including cooling radiation, fluorescence powered by ionizing photons from either star-forming regions or AGN, and resonant scattering by neutral hydrogen gas \citep[e.g.,][]{HoganWeymann1987,GouldWeinberg1996,Cantalupo2005,Kollmeier2010,FaucherGiguere2010,Hennawi2013}.  Disentangling between different mechanisms that are responsible for the observed \lya\ signals is challenging, especially when the \lya\ line is the only observable feature in the nebulae.  

For the two systems in our study, however, the observed spectral properties of the \lya\ line enable us to rule out cooling radiation and photo-ionization due to the cosmic UV background radiation as a dominant mechanism for powering the observed emission.  Specifically, radiatively cooled gas is expected to condense and sink through the hot ambient medium, resulting in infall, and the majority of the photons will travel through the infalling clouds before escaping the medium \citep[e.g.,][]{FaucherGiguere2010}.  The expectation of an enhanced blue-peak from infalling gas in inconsistent with the observations.  In addition, the expected \lya\ fluorescence signal from cosmic UV background alone is low with surface brightness of $\lesssim 10^{-19}\,\sbunit$ \citep[e.g.,][]{Kollmeier2010}. %, photoionization from star-forming galaxies in Systems {\it A} and {\it B} should have a significant contribution to the observed signal. 

We therefore proceed with considerations of the two remaining scenarios: (1) \lya\ photons arising as a result of fluorescence powered by ionizing photons from star-forming regions and (2) \lya\ photons produced in the galaxies and resonantly scattered by neutral hydrogen in the nebulae.  The first scenario requires a non-zero escape fraction of ionizing photons from the galaxies, while the second scenario corresponds to the shell model analysis described in \S\ref{sec:velocity_shell_fitting}.  Here we also discuss the implications of these scenarios.

\subsection{Ly$\alpha$ emission from recombination radiation and implications for the escape fraction of ionizing radiation from star-forming galaxies}\label{sec:escape}

We first consider the possibility that the observed \lya\ signals are powered by {\it in situ} star formation directly underneath the nebulae. Available deep {\it HST} F606W image places strong constraints on the rest-frame UV continuum flux at the location of the nebulae.  Using the integrated \lya\ luminosity of $L_{\rm Ly\alpha}=(2.15\pm0.07)\times 10^{42} \, {\rm erg\,s^{-1}}$ ($[3.49\pm0.08]\times 10^{42} \, {\rm erg\,s^{-1}}$) for the northern (southern) nebula of System {\it A}, we infer an SFR of $\approx1.1 \, (2.3) \, \msun \, \mathrm{yr^{-1}}$ based on a conversion factor of ${\rm Ly\alpha}/{\rm H}\alpha=8.7$ \citep[][and references therein]{Hayes2015} and the H$\alpha$-SFR relation of \cite{Kennicutt2012}.  From the inferred SFR, we derive the expected
apparent magnitude in the F606W bandpass (corresponding to 1500 \AA\ in the rest frame at $z\approx 3$) of $AB({\rm F606W})\approx27.3$ (26.5) using the FUV flux-SFR relation of \cite{MadauDickinson2014} for the underlying star-forming regions in the northern (southern) nebula.  The inferred F606W magnitude is roughly more than two magnitudes brighter than the 2-$\sigma$ detection limit in the F606W bandpass ($AB({\rm F606W})\approx29$ within an aperture of $0\farcs5$ in diameter), but no flux is detected at the location of the nebulae away from the galaxies.  While we consider {\it in situ} star formation an unlikely scenario for powering the \lya\ signals, we cannot rule out the possibility that dust obscurations may have played a role in blocking the FUV photons along the line of sight.  Deeper imaging data at submillimeter are needed for constraining the effect of dust.  In the following discussion, we proceed with considerations of external sources for powering the observed \lya\ signals.

For photo-ionization by the nearby galaxies, the observed \lya\ intensity is connected to the incident ionizing radiation field and the discussion often involves considerations of two different regimes, optically thin versus optically thick gas.  For the purpose of our study, both Systems {\it A} and {\it B} consistently require a large $N(\ion{H}{I})$, exceeding $\log\,N(\ion{H}{I})/{\rm cm}^{-2}\approx 19$ (Table \ref{tab:parameters}), for explaining the observed \lya\ profile.  We therefore consider only optically-thick regime in the subsequent discussion.

In optically-thick regime, ionization occurs in the surface of a cloud illuminated by the ionizing source and roughly 66\% of all ionizing photons are converted into \lya\ photons through recombination cascades in the surface layer (i.e., $\eta_{\rm B}=0.66$) \citep{OsterbrockFerland2006}. The surface brightness of \lya\ emission is connected to ionizing photon flux according to
\begin{eqnarray}\label{eq:caseB}
%    \begin{aligned}
     \mathrm{SB_{Ly\alpha}}  &=& f_g\,f_{\rm esc}\,\frac{\eta_{\mathrm{B}}\,h\,\nu_{\mathrm{Ly\alpha}}}{(1+z)^4}\frac{\Phi}{\pi} \nonumber \\
        & = & 3.2\times 10^{-18}  f_g\,f_{\rm esc}\left( \frac{1+z}{4.0}\right)^{-4} \left(\frac{D}{{\rm 10\, pkpc}}\right)^{-2} \nonumber \\ 
        & & \left( \frac{\Phi_0}{10^7\, {\rm s^{-1}\, cm^{-2}}}\right) \ \mathrm{erg \, s^{-1}\, cm^{-2}\,arcsec^{-2}} 
%    \end{aligned}
\end{eqnarray}
where $f_g$ is the geometric correction coefficient to account for partial illumination of the nebula and redistribution of \lya\ photons, $f_{\rm esc}$ is the fraction of ionizing photons that escape the galaxies, $D$ is the distance of the cloud from the ionizing source, and $\Phi_0$ is the ionizing photon flux at a distance of 1 kpc from the source.  In principle, comparing the observed \lya\ surface brightness with the expected ionizing radiation field from the SED analysis of the galaxies constrains $f_{\rm g}$ and $f_{\rm esc}$ based on Eq.~\ref{eq:caseB}. 
In practice, uncertainties in the inferred galaxy spectra are large.  Therefore, it is not trivial to obtain accurate constraints for $f_g$ and $f_{\rm esc}$. 

For System {\it A}, we estimate the total ionizing photon fluxes from {\it A}1, {\it A}2 and {\it A}3 using the best-fit \textsc{Bagpipes} model spectra and find respectively $\Phi_0\approx 3.4\times 10^8$, $8.1\times 10^7$, $5.1\times 10^7\,{\rm s^{-1}\, cm^{-2}}$ at $D=10$ pkpc.  Assuming $f_{\rm g}=0.5$ from numerical simulations \citep[e.g.,][]{Cantalupo2005,Kollmeier2010} and $f_{\rm esc}<10\%$ as a fiducial upper limit for ionizing photon escape fraction \citep[e.g.][]{Chen2007,Vanzella2010,Grazian2017}, the observed peak \lya\ surface brightness of $7.3\times 10^{-18} \, \mathrm{erg \, s^{-1}\, cm^{-2}\,arcsec^{-2}}$ (the 6-$\sigma$ contour in Figure \ref{fig:src_plane_contours_A}) implies a distance limit of $D<8.5$ pkpc from {\it A}1 and $D<3.3$ pkpc from {\it A}3. Adopting the low-intensity contour of $3.7\times 10^{-18} \, \mathrm{erg \, s^{-1}\, cm^{-2}\,arcsec^{-2}}$ would increase the distance limit by 40\% to $D<12$ pkpc from ${\it A}1$ and $D<4.6$ pkpc from {\it A}3. The observed extent of \lya\ emission of $\gtrsim 10$ pkpc (see Figure~\ref{fig:src_plane_contours_A}) therefore requires {\it A}1 to be the dominant source of ionizing photons with an escape fraction $\sim 10\%$. Current observations suggest that the escape fraction of ionizing photons from massive (>$L_*$) galaxies at $z\approx 3$ is much smaller than $10\%$ \citep[e.g.][]{Grazian2017}. The inferred large log$(\ion{H}{I}$) based on \lya\ line profiles also suggests that $f_{\rm esc}$ is likely to be small. In addition, in \S\ref{sec:dust} below, we show that resonant scattering of \lya\ photons produced in the star-forming regions can account for the full extent of the \lya\ nebulae. We therefore conclude that recombination radiation from photo-ionized gas alone is unlikely to dominate the observed \lya\ signal in System {\it A}.

We repeat the same exercise for System {\it B}.  Due to the smaller physical scale of System {\it B}, we estimate the ionizing photon flux at a distance of $D=1$ pkpc. Using the best-fit \textsc{Bagpipes} model spectra, we obtain the total ionizing photon flux from {\it B}1 and {\it B}2 combined to be $\Phi\approx 1.4\times 10^8\, {\rm s^{-1}\, cm^{-2}}$ at 1 pkpc.  The observed surface brightness of $7.5\times 10^{-18} \, \mathrm{erg \, s^{-1}\, cm^{-2}\,arcsec^{-2}}$ (the 8-$\sigma$ contour in Figure \ref{fig:img_src_plane_contours_B}) leads to $(f_{\rm g}f_{\rm esc})\approx 33\%$, or $f_{\rm esc}=66\%$ assuming $f_{\rm g}=0.5$. At the limit of $f_{\rm esc}<1$, we infer the distance limit of $D<1.2$ pkpc for the high-intensity contours. With the low-intensity surface brightness of $2.8\times 10^{-18} \, \mathrm{erg \, s^{-1}\, cm^{-2}\,arcsec^{-2}}$ (the 3-$\sigma$ contour in Figure \ref{fig:img_src_plane_contours_B}), the inferred distance limit is increased to $D<2$ pkpc. Because the observed \lya\ emission extends to $\gtrsim 4$ pkpc in the source plane (see Figure \ref{fig:img_src_plane_contours_B}), we conclude that recombination radiation from photo-ionized gas alone {\it cannot} explain all of the observed \lya\ photons away from the galaxies in System {\it B}.

\subsection{Ly$\alpha$ emission from scattering and implications for dust attenuation}\label{sec:dust}

Given the star-forming nature of the galaxies in both Systems {\it A} and {\it B}, we now consider the scenario in which the \lya\ photons are produced in the star-forming ISM of the galaxies and resonantly scattered through the spatially extended nebulae.  Using the estimated SFR in the 16\%--84\% confidence interval for galaxies {\it A}1, {\it A}2 and {\it A}3 (see Table~\ref{tab:Bagpipes_sedfitting}), we infer a total intrinsic \lya\ luminosity of $L^{\rm int}_{\rm Ly\alpha}/(10^{44}\,{\rm erg}\,{\rm s}^{-1})=1.46$--1.65, 0.17--0.19, and 0.21--0.26 for galaxies {\it A}1, {\it A}2, and {\it A}3, respectively, using the conversion factor of ${\rm Ly}\alpha/{\rm H}\alpha=8.7$ \citep[][and references therein]{Hayes2015} and the H$\alpha$-SFR relation of \cite{Kennicutt2012}.  For System {\it B}, the same exercise leads to an intrinsic \lya\ luminosity of $L^{\rm int}_{\rm Ly\alpha}=$(1.2--2.1)$\times 10^{42}\,{\rm erg\, s^{-1}}$ for galaxies {\it B}1 and {\it B}2 combined. %, which gives $f_{\rm esc}^{\rm Ly\alpha}\approx 43$-78\% given the observed $L({\rm Ly\alpha})=9.2\times 10^{41}\,{\rm erg\,s^{-1}}$ from the nebula in System {\it B}. 

While these star-forming galaxies may be intrinsically luminous in \lya, we expect that a large fraction of these \lya\ photons are unable to escape the ISM due to a substantial amount of dust attenuation. We obtain an empirical estimate of the attenuation factor $k_{\rm dust}=1 - L_{\rm Ly\alpha}^{\rm obs}/L_{\rm Ly\alpha}^{\rm int}$ based on the observed \lya\ luminosity of $2.15\times 10^{42}\,{\rm erg\,s^{-1}}$ for the northern nebula and $3.43\times10^{42}\,{\rm erg\,s^{-1}}$ for the southern nebula, and the intrinsic \lya\ luminosity from star-forming regions described above.  Attributing the \lya\ emission of the northern (southern) nebula to the scattering of \lya\ photons from galaxy {\it A}1 (galaxies {\it A}2 and {\it A}3), we estimate $k_{\rm dust}$ to be $\approx 98\%$ and $\approx 92\%$ for the northern and southern nebula, respectively.  

%up to $A_V\approx 0.7$ mag inferred from the stellar population synthesis analysis presented in \S\ \ref{sec:SED_fitting} (see also Table \ref{tab:Bagpipes_sedfitting}).  Adopting the \cite{Calzetti2000} extinction law for starburst galaxies, we estimate an extinction magnitude of $A_{1215}\approx 5.2$ mag for the \lya\ emission line based on the estimated stellar extinction of $A_V\approx 0.7$ mag.  This exercise leads to an estimated attenuation factor of $k_{\rm dust}\approx 99$\% for the \lya\ photons.
Following the optically-thick prescription from Equation \ref{eq:caseB} and replacing ionizing photon flux with \lya\ photon flux $\Phi_{{\rm Ly}\alpha}$, we can now connect the \lya\ scattering surface brightness to $L^{\rm int}_{\rm Ly\alpha}$ following
\begin{eqnarray}\label{eq:Lya_scatter}
%    \begin{aligned}
     \mathrm{SB_{Ly\alpha}}  &=& \frac{h\,\nu_{\mathrm{Ly\alpha}}}{(1+z)^4}\frac{\Phi_{{\rm Ly}\alpha}}{\pi} \nonumber \\
        &=& \frac{h\,\nu_{\mathrm{Ly\alpha}}}{(1+z)^4}\frac{(1-k_{\rm dust})\,L^{\rm int}_{{\rm Ly}\alpha}}{4\pi^2\,D^2\,h\,\nu_{{\rm Ly}\alpha}} \nonumber \\
        & = & 2.4\times 10^{-18} \left( \frac{1+z}{4.0}\right)^{-4} \left(\frac{D}{10\,{\rm pkpc}}\right)^{-2} \nonumber \\ 
        & & \frac{(1-k_{\rm dust})\,L^{\rm int}_{{\rm Ly}\alpha}}{10^{42}\,{\rm erg}\,{\rm s}^{-1}}\ \mathrm{erg \, s^{-1}\, cm^{-2}\,arcsec^{-2}}.
%    \end{aligned}
\end{eqnarray}
Eq.~\ref{eq:Lya_scatter} leads to a distance estimate of $D_{\rm north}\approx 14$ pkpc between the northern nebula and galaxy {\it A}1 for an observed \lya\ surface brightness of $3.7\times 10^{-18} \, \mathrm{erg \, s^{-1}\, cm^{-2}\,arcsec^{-2}}$ (the 3-$\sigma$ contour in Figure \ref{fig:src_plane_contours_A}), an intrinsic \lya\ luminosity of $L^{\rm int}_{\rm Ly\alpha}=(1.46$--$1.65)\times 10^{44}\,{\rm erg}\,{\rm s}^{-1}$ for {\it A}1, and an attenuation factor of 98\%.  At higher intensity of $7.3\times 10^{-18} \, \mathrm{erg \, s^{-1}\, cm^{-2}\,arcsec^{-2}}$ (the 6-$\sigma$ contour in Figure \ref{fig:src_plane_contours_A}), the distance is reduced to $\approx 10$ pkpc.  The inferred distance range is fully consistent with the extent of the northern nebula relative to {\it A}1.  In addition, the estimated amount of dust attenuation agrees with $A_V\approx 0.7$ mag inferred from the SED analysis presented in \S\ref{sec:SED_fitting} (see also Table \ref{tab:Bagpipes_sedfitting}). Based on the \cite{Calzetti2000} extinction law for starburst galaxies, the estimated stellar extinction of $A_V\approx 0.7$ mag corresponds to an extinction magnitude of $A_{1215}\approx 5.2$ mag for the \lya\ emission line, or $k_{\rm dust}\approx 99$\% for the \lya\ photons. It suggests that resonant scattering alone can fully account for the observed \lya\ brightness in the northern nebula.

For the southern nebula, galaxies {\it A}2 and {\it A}3 together contribute to a total intrinsic \lya\ luminosity of $L^{\rm int}_{\rm Ly\alpha}=3.8$-$4.5\times 10^{43}\,{\rm erg}\,{\rm s}^{-1}$.  Adopting $k_{\rm dust}=92\%$ leads to a distance estimate of $D_{\rm south}\approx 10$ pkpc for the high-intensity region and $D_{\rm south}\approx 15$ pkpc for the low-intensity region between galaxies {\it A}2/{\it A}3 and the southern nebula. 
%Similarly, reducing the attenuation factor slightly to 98\% would lead to $D_{\rm south}\approx 5$ (7) pkpc for the high (low) intensity regions. 
Similarly, the estimated size is consistent with the observed extent of the southern nebula (see \S\ref{sec:Lya_NB}).  Although the dust attenuation of $92\%$ is in tension with the estimated $k_{\rm dust}\approx 99\%$ based on the SED analysis, we argue that a possible contribution of \lya\ photons from galaxy {\it A}1, together with uncertainties in $f_{\rm esc}$ (see \S\ref{sec:escape}) and $k_{\rm dust}$ in an inhomogeneous, clumpy medium could account for the observed extent of \lya\ signals in the southern nebula \citep[e.g.,][]{Neufeld1991,HansenOh2006}. 

%Since this estimated size is only a third of the observed extent of the southern nebula ($\approx 15$ pkpc, see \S\ref{sec:Lya_NB}), resonant scattering alone is unlikely to be the only emission mechanism for the southern nebula.  Similarly, the observed \lya\ luminosity of the southern nebula of $(3.73\pm0.08)\times 10^{42}\,{\rm erg\,s^{-1}}$ leads to an escape fraction of $\approx 10\%$ if all \lya\ photons are scattered from the star-forming regions in {\it A}2 and {\it A}3, which is in tension with the estimated dust attenuation level of $\approx 99\%$. We therefore argue that the \lya\ signal in the southern nebula is contributed by both resonant scattering and recombination radiation from photo-ionized gas (see \S\ref{sec:escape}).  

For galaxies {\it B}1 and {\it B}2, the uncertainty in $k_{\rm dust}=1 - L_{\rm Ly\alpha}^{\rm obs}/L_{\rm Ly\alpha}^{\rm int}$ is larger, ranging between $\approx 20$--$50\%$. Meanwhile, uncertainties in $A_V$ are also larger, ranging between $A_V\approx 0.05$-0.25 for {\it B}1 and $A_V\approx 0.5$-0.7 for {\it B}2 (see Table \ref{tab:Bagpipes_sedfitting}), corresponding to a wide range of dust attenuation of $\approx 30$--$99\%$ for \lya\ photons, in agreement with the emprical $k_{\rm dust}$ of $\approx 20$--$50\%$. Adopting $k_{\rm dust}=50\%$, Eq.~\ref{eq:Lya_scatter} leads to a distance estimate of $D\approx 3$ pkpc between galaxies {\it B}1/{\it B}2 and the observed \lya\ intensity peak of $7.5\times 10^{-18} \, \mathrm{erg \, s^{-1}\, cm^{-2}\,arcsec^{-2}}$ (the 8-$\sigma$ contour in Figure \ref{fig:img_src_plane_contours_B}). At $\mathrm{SB_{Ly\alpha}}=2.8\times 10^{-18} \, \mathrm{erg \, s^{-1}\, cm^{-2}\,arcsec^{-2}}$ (the 3-$\sigma$ contour in Figure \ref{fig:img_src_plane_contours_B}), we estimate a distance of $\approx 5$ pkpc, which agrees well with the observed extent of the nebula.  

The above exercise shows that resonant scattering of \lya\ photons produced in nearby star-forming regions may be sufficient for explaining the observed \lya\ signals in both systems without invoking additional emission sources. This is in contrast to the recombination radiation scenario discussed in \S\ref{sec:escape}. It shows that even at 100\% escape fraction of ionizing photons, recombination is insufficient for explaining the observed \lya\ flux in System {\it B} and that it would require an escape fraction of $\sim$10\% from {\it A}1 for recombination to contribute significantly to the emission signal.

\subsection{Systematics in interpreting the spatial and spectral profiles of the nebulae}

Due to the clumpy nature of line-emitting gas, the surface brightness profile of extended nebulae is subject to the spatial variation of lensing magnification and its associated uncertainties in the image plane. In principle, gravitational lensing conserves surface brightness of a light-emitting source. However, the conservation of surface brightness does not apply when the lensed image of the source is not resolved in the data.  In our study, the spatial resolution is limited by the size of the seeing disk in ground-based observations.   We see that the source-plane image reconstructed from less magnified regions appear to be fainter (e.g., image {\it a} of System {\it A}) than those reconstructed from more highly magnified images (e.g., images {\it b} and {\it c} of System {\it A}). This surface brightness discrepancy suggests that the individual clumps, even after being magnified by the cluster lens, are still not resolved by the data. Apart from the decrease of surface brightness in image {\it a} of System {\it A} as discussed in \S\ref{sec:Lya_NB}, we also see discrepancy of \lya\ surface brightness near critical curves where the magnification factor is much larger (e.g., in System {\it B}, the \lya\ emitting region that straddles the critical curve between images {\it a} and {\it b} shows the highest apparent surface brightness across the whole lensed arc).  Adopting $\mu\approx 20$ as the fiducial magnification factor near the critical curves, we estimate that the clump size should be $\lesssim 1.5$ kpc in order for the gas clumps to remain unresolved in lensed images recorded under $1\arcsec$ seeing. This upper limit is in agreement with clump sizes of cold gas in the CGM constrained in absorption studies \citep[e.g.][]{Zahedy2019}. Furthermore, small-scale substructures in the lens can also introduce additional perturbations to the lensing effect across an extended source (e.g., the unusually large magnification at the location of image {\it B}1{\it e}, see \S\ref{sec:galaxy_photometry}). In order to accurately quantify the intrinsic surface brightness distribution of extended and clumpy sources in strong lensing fields, a better understanding of the systematic uncertainties of lensing magnification as a function of image position is necessary. 

Systematic uncertainties also remain in the shell model analysis on the \lya\ line profiles and the interpretation of the velocity gradient derived from spatially-varying \lya\ lines in both Systems {\it A} and {\it B}.  For example, our shell model does not include radiative transfer effects inside the galaxies, which would re-shape the input \lya\ line from a single Gaussian into a double-peak profile.  This provides a likely explanation for the large redshift observed at the location of the continuum regions in System {\it A} (see Figure \ref{fig:sysA_vmap}).  However, a clumpy ISM may also be transparent to Lya photons, resulting in a wider Gaussian linewidth instead. 
In addition, because the signal strength is dominated by the much stronger red peak in the \lya\ line (see Figure \ref{fig:lya_profile_and_tlac_models}), the inferred velocity offset could simply represent a shift in the location of the red peak.
The observed blueshifted velocity with increasing projected distance in System {\it A} (see Figure \ref{fig:sysA_vmap}) may also be explained in part due to line-of-sight projection of a uniformly expanding sphere.  While a complete 3D radiative transfer model to consider different possible cloud geometry is beyond the scope of this paper, an initial exercise that explores different cloud geometry and velocity field shows that
%in the velocity map 
%In \S\ref{sec:velocity_shell_fitting}, we have uncovered a velocity gradient in both systems that implies a continuous gas flow from star-forming regions to the halo environment.  
%we note that the observed velocity gradient could also be a result of resonant scattering of \lya\ photons. For an expanding shell/sphere of neutral hydrogen gas with a \lya\ emission source in the center, radiative transfer simulations suggest that 
the emergent spectrum will be increasingly blueshifted (redshifted) from the center to the edge of the cloud with decelerating (accelerating) gas expansions.  We show one example of such exercise in Appendix~\ref{sec:appendix_lya_cloud}, where we extract the emergent \lya\ line profile as a function of projected distance from the center of a spherical cloud that is undergoing expansion with an accelerating or decelerating velocity field.  We therefore argue that System {\it A} is likely decelerating while System {\it B} is accelerating as the gas move outward from the star-forming regions. 

In summary, the observed \lya\ emission morphology in System {\it A} clearly indicates a more complicated gas geometry than what is assumed in current radiative transfer simulations. In addition, significant uncertainties remain in terms of the origin and the spatial distribution of \lya\ emission sources, the effect of local ISM on the \lya\ spectra emergent from the the star-forming regions prior to the scattering of large-scale gas in the CGM, as well as the effect of dust and gas clumpiness. All of these factors can alter the shape of the emerging line profile, the surface brightness profile, and the velocity gradient of \lya\ emission in an extended gas cloud. A more sophisticated radiative transfer model is needed to fully explore the parameter space. % the detailed properties of Systems {\it A} and {\it B} is beyond the scope of this work, and will be explored separately in the future. 

\section{Summary and Conclusions}\label{sec:summary_conclusion}

Combining the strong cluster lensing power with deep wide-field integral field spectroscopic data, we have carried out a detailed analysis of two giant  \lya\ arcs to spatially and spectrally resolve gas flows around two active star-forming regions at $z>3$.  Both \lya\ nebulae are found to be spatially offset from the associated star-forming region and both exhibit a double-peak profile with a significantly enhanced red peak that indicates expansion/outflowing motions.  One of the arcs with \lya\ surface brightness of $3.7\times 10^{-18}\sbunit$, detected at the 3-$\sigma$ level of significance, stretches over $1'$ around the Einstein radius of the cluster, resolving the velocity field of the line-emitting gas on sub-kpc scales around a group of three star-forming galaxies of $0.3$-$1.6\,L_*$ at $z=3.038$.  Based on a lens model constructed from deep {\it HST} images, the de-magnified source-plane \lya\ image exhibits a symmetric double-lobe structure of $\approx 30$ pkpc across, encompassing the galaxy group.  The total integrated \lya\ flux across the nebula is $(6.5\pm0.1)\times 10^{-17}\,{\rm erg\, s^{-1}\, cm^{-2}}$ after correcting lensing magnifications, corresponding to a total \lya\ luminosity of $L_{{\rm Ly}\alpha}=(5.2\pm0.1)\times 10^{42} \, {\rm erg\,s^{-1}}$ at $z\approx 3.038$.  The second arc with \lya\ surface brightness of $2.8\times10^{-18}\, \mathrm{erg \, s^{-1} \, cm^{-2} \, arcsec^{-2}}$ (3-$\sigma$) spans $15\arcsec$ in size, roughly centered around a pair of low-mass dwarf \lya\ emitters of $\approx 0.03\,L_*$ at $z=3.754$.  The total integrated \lya\ flux is $(7.4\pm0.2)\times 10^{-18}\, {\rm erg\, s^{-1}\, cm^{-2}}$, corresponding to a total luminosity of $L_{{\rm Ly}\alpha}=(9.8\pm0.2)\times 10^{41} \, {\rm erg\,s^{-1}}$ at $z\approx 3.754$.  Here we summarize the main findings of our study:

(1) A strong variation in the observed \lya\ surface brightness is clearly seen across both nebulae, suggesting large spatial fluctuations in the underlying gas properties.  While the nebulae at $z=3.038$ is split into northern and southern lobes bracketing the group of luminous star-forming galaxies, the one at $z=3.754$ appears to be more symmetrically distributed around the associated low-mass galaxies. 

(2) Spatial variations in the kinematics profile of the \lya\ emission line are also detected in both nebulae, revealing highly organized velocity fields across the nebulae.  We show that such spatial variations, if unaccounted for in integrated \lya\ profiles, may lead to biased results in constraining the underlying gas kinematics.  By applying a simple expanding shell model to the spatially-varying \lya\ line, we infer a large velocity gradient of $|\Delta\,v/\Delta\,r_\perp| \approx 22-27\,\kms\,{\rm pkpc}^{-1}$ and high neutral hydrogen column density of $\log\,N(\ion{H}{I})/{\rm cm}^{-2}\gtrsim 19.5$ for both nebulae.  The result supports a scenario in which high column density gas is driven outward from the galaxies to beyond 10 pkpc in projected distance into the low-density surroundings.  

(3) Combining known star formation properties of the galaxies and the observed extent and surface brightness of the \lya\ signals, we show that the observed \lya\ photons likely originate from a combination of resonant scattering of \lya\ photons from the nearby star-forming regions and recombination radiation due to escaping ionizing photons, although the relative contribution of these two mechanisms cannot be accurately determined with the current data.

Both nebulae provide clear-cut examples of gas outflows that are thought to be widespread at high redshift and may be responsible for metal enrichment of the \lya\ forest in general.  While the hydrogen \lya\ line, being the strongest emission line in diffuse, photo-ionized gas, enables sensitive studies of spatially extended outflows beyond active star-forming regions, large uncertainties remain due to the resonant nature of this transition.  
Future observations targeting non-resonant transitions, such as $\ion{[O}{II]}\lambda\lambda 3727, \, 3730$, \hbeta $\lambda 4863$, $\ion{[O}{III]}\lambda 5008$, and \halpha $\lambda6565$, within the line-emitting nebulae will provide the necessary discriminating power to resolve the degeneracy between different physical parameters. Based on the observed \lya\ surface brightness in Systems {\it A} and {\it B} and under the assumption that the \lya\ emission arises from recombination radiation of photo-ionized gas, we estimate the expected \halpha\ and \hbeta\ surface brightness to be approximately 3 and $1\times 10^{-19}\,\sbunit$, respectively. The $\ion{[O}{III]}\lambda 5008$ line is expected to be between 3 and 10 times brighter than \hbeta\ in photo-ionized, low-metallicity gas \citep[e.g.][]{Kewley2019}.
%For photoionized gas, the expected \halpha\ flux is approximately an order of magnitude weaker than \lya\ flux, with the ratio $f_{\rm Ly\alpha}/f_{\rm H\alpha}\approx 8$-11, while the \hbeta\ flux is approximately a third of the \halpha\ flux \citep[e.g.,][]{OsterbrockFerland2006,Leibler2018}. A flux ratio $f_{\rm Ly\alpha}/f_{\rm H\alpha}$ much higher than 11, or $f_{\rm Ly\alpha}/f_{\rm H\beta}$ much higher than 30 can be used to infer significant contribution from resonant scattering to the observed \lya\ signal \citep[e.g.,][]{Leibler2018,Li2020}. Apart from hydrogen recombination lines, the $\ion{[O}{III]}\lambda 5008$ oxygen line emission is also a promising feature to detect from photoionized gas, with a line strength comparable with \lya\ in gas clouds with a density $n_{\rm H}\approx 10^{-3}\,{\rm cm^{-3}}$ at a temperature of $T=10^4$ K \citep{Lokhorst2019}. 
While the \halpha\ line is redshifted out of the detection window with existing near-infrared spectrographs on the ground, it is possible to detect \hbeta $\lambda 4863$ and $\ion{[O}{III]}\lambda 5008$ lines in under $\approx 20$ hours, within the reach of current observing facilities.  %exposure on Magellan/FIRE. The rest-frame optical lines of such $z\approx3$-4 sources are also ideal targets for space-based infrared spectrographs, such as NIRSpec on JWST that will be launched and commissioned soon. 
%Despite being limited by the sensitivity of the current instruments, \lya\ emission remains the most promising signal for direct detection of low density gas that contains most of the baryonic matter and traces the large-scale DM distribution. 
We therefore argue that follow-up near-infrared integral field observations, targeting rest-frame optical, non-resonant lines in known \lya\ nebulae, will greatly improve the physical constraints of gas flows around distant star-forming galaxies.

\section*{Acknowledgements}

We thank Erin Boettcher, Fakhri Zahedy, Claude-André Faucher-Giguère and Irina Zhuravleva for helpful discussions.  We also thank an anonymous referee for constructive comments that helped improve this paper. HWC and MCC acknowledge partial support from HST-GO-15163.001A and NSF AST-1715692 grants. MG was supported by NASA through the NASA Hubble Fellowship grant HST-HF2-51409. This research has made use of the services of the ESO Science Archive Facility and the Astrophysics Data Service (ADS)\footnote{\url{https://ui.adsabs.harvard.edu/classic-form}}. The analysis in this work was greatly facilitated by the following \texttt{python} packages:  \texttt{Numpy} \citep{Numpy}, \texttt{Scipy} \citep{Scipy}, \texttt{Astropy} \citep{astropy:2013,astropy:2018}, \texttt{Matplotlib} \citep{Matplotlib}, and \texttt{MPDAF} \citep{MPDAF}.  

\section*{Data Availability}
The data used in this article are available for download through the Mikulski  Archive  for  Space Telescopes  (MAST) and the ESO Science Archive Facility.

%%%%%%%%%%%%%%%%%%%%%%%%%%%%%%%%%%%%%%%%%%%%%%%%%%



\bibliographystyle{mnras}
\bibliography{references} % if your bibtex file is called example.bib

% \begin{thebibliography}{99}
% \bibitem[\protect\citeauthoryear{Author}{2012}]{Author2012}
% Author A.~N., 2013, Journal of Improbable Astronomy, 1, 1
% \bibitem[\protect\citeauthoryear{Others}{2013}]{Others2013}
% Others S., 2012, Journal of Interesting Stuff, 17, 198
% \end{thebibliography}

%%%%%%%%%%%%%%%%%%%%%%%%%%%%%%%%%%%%%%%%%%%%%%%%%%

%%%%%%%%%%%%%%%%% APPENDICES %%%%%%%%%%%%%%%%%%%%%

\appendix

\section{Lens constraints and parameters}
In Table A1, we list the coordinates of all multiple images used as constraints in our lens modeling process, while the best-fit parameters for the fiducial and fine-tuned model are listed in Tables A2 and A3, respectively. 

\begin{table}
    \centering
    \caption{\label{tab:multiple_image_coordinates1}
    Coordinates and redshifts of multiple images included for lens modeling. }
    \begin{tabular}{c c c c}
    \hline
    Image ID & RA & DEC & Redshift \\
    \hline
    {\it A}2{\it a} & 181.562648 & $-$8.796683 & 3.0378 \\
    {\it A}2{\it b} & 181.562497 & $-$8.804908 & 3.0378 \\
    {\it A}2{\it c} & 181.560573 & $-$8.808988 & 3.0378 \\
    {\it A}31{\it a} & 181.562535 & $-$8.796884 & 3.0384 \\
    {\it A}31{\it b} & 181.562501 & $-$8.804524 & 3.0384 \\
    {\it A}31{\it c} & 181.560104 & $-$8.809551 & 3.0384 \\
    {\it A}32{\it a} & 181.562551 & $-$8.796809 & 3.0384 \\
    {\it A}32{\it b} & 181.562492 & $-$8.804617 & 3.0384 \\
    {\it A}32{\it c} & 181.560204 & $-$8.809411 & 3.0384 \\ 
    {\it B}1{\it a} & 181.566558 &  $-$8.804480 & 3.7540 \\ 
    {\it B}1{\it b} & 181.566475 &  $-$8.804733 & 3.7540 \\ 
    {\it B}1{\it c} & 181.566475 &  $-$8.805147 & 3.7540 \\ 
    {\it B}1{\it d} & 181.566275 &  $-$8.806328 & 3.7540 \\
    {\it B}1{\it e} & 181.565591 &  $-$8.807690 & 3.7540 \\
    {\it B}2{\it a} & 181.566605 &  $-$8.804400 & 3.7540 \\
    {\it B}2{\it c} & 181.566494 &  $-$8.805077 & 3.7540 \\ 
    {\it B}2{\it d} & 181.566250 &  $-$8.806446 & 3.7540 \\ 
    {\it B}2{\it e} & 181.565675 &  $-$8.807566 & 3.7540 \\
    1{\it a} & 181.550916 & $-$8.797422 & 1.0121 \\
    1{\it b} & 181.549604 & $-$8.799294 & 1.0121 \\
    1{\it c} & 181.548870 & $-$8.806655 & 1.0121 \\
    3{\it a} & 181.550570 & $-$8.795568 & 1.0433 \\
    3{\it b} & 181.547611 & $-$8.799811 & 1.0433 \\
    3{\it c} & 181.548607 & $-$8.805281 & 1.0433 \\
    4{\it a} & 181.552987 & $-$8.794699 & 1.4248 \\
    4{\it b} & 181.548830 & $-$8.800057 & 1.4248 \\
    4{\it c} & 181.549752 & $-$8.807965 & 1.4248 \\
    5{\it a} & 181.553557 & $-$8.795189 & 1.4254 \\
    5{\it b} & 181.554237 & $-$8.801552 & 1.4254 \\
    5{\it c} & 181.550005 & $-$8.808098 & 1.4254 \\
    6{\it a} & 181.549979 & $-$8.796362 & 1.4255 \\
    6{\it b} & 181.548139 & $-$8.797058 & 1.4255 \\
    6{\it c} & 181.548050 & $-$8.809283 & 1.4255 \\
    8{\it a} & 181.553657 & $-$8.795756 & 1.4864 \\
    8{\it b} & 181.554524 & $-$8.801104 & 1.4864 \\
    8{\it c} & 181.549957 & $-$8.808887 & 1.4864 \\
    9{\it a} & 181.546741 & $-$8.793144 & 1.9600 \\
    9{\it b} & 181.543273 & $-$8.797812 & 1.9600 \\
    9{\it c} & 181.544378 & $-$8.807486 & 1.9600 \\
    10{\it a} & 181.552450 & $-$8.795001 & 2.5393 \\
    10{\it b} & 181.546604 & $-$8.797465 & 2.5393 \\
    10{\it c} & 181.550487 & $-$8.799957 & 2.5393 \\
    10{\it d} & 181.554894 & $-$8.800160 & 2.5393 \\
    10{\it e} & 181.548827 & $-$8.811813 & 2.5393 \\
    12{\it a} & 181.548632 & $-$8.793717 & 3.3890 \\
    12{\it b} & 181.546121 & $-$8.795387 & 3.3890 \\
    12{\it c} & 181.553268 & $-$8.800197 & 3.3890 \\
    \hline
    \multicolumn{4}{p{0.45\textwidth}}{
    We adopt the multiple image identifications from \protect\cite{Caminha2017}, while excluding image systems 2, 7, 13, 21, 24 and 27 (see \S\ref{sec:lens_model} for detailed discussions). We rename their image system 11 to be {\it A}2, and add {\it A}31 and {\it A}32 (the north and south substructures of {\it A}3). Similarly, we rename image system 14 to be {\it B}1, and add {\it B}2 (the fainter structure near {\it B}1 at the same redshift). We also update redshifts for {\it A}2, {\it A}3, {\it B}1 and {\it B}2 to be their best-fit values from fitting the observed emission lines (see \S\ref{sec:analysis_gal}). For the fiducial model, we use all images listed except for {\it A}31, {\it A}32 and {\it B}2. For the fine-tuned model, we only use systems {\it A}2, {\it A}31, {\it A}32, {\it B}1 and {\it B}2, excluding all other lensed systems, in order to optimise the model specifically for {\it A} and {\it B}.
    }
    \end{tabular}
\end{table}
    %Because our goal is not to optimize the lens model for individual cluster members, we exclude those image systems without comprising the accuracy of large-scale cluster lens model. 

\begin{table}
    \centering
    \contcaption{}
    \begin{tabular}{c c c c}
    \hline
    Image ID & RA & DEC & Redshift \\
    \hline
    15{\it a} & 181.555962 & $-$8.791635 & 3.7611 \\
    15{\it b} & 181.557600 & $-$8.803056 & 3.7611 \\
    15{\it c} & 181.551748 & $-$8.810964 & 3.7611 \\
    16{\it a} & 181.554584 & $-$8.791202 & 3.7617 \\
    16{\it b} & 181.546465 & $-$8.799671 & 3.7617 \\
    16{\it c} & 181.556520 & $-$8.802471 & 3.7617 \\
    17{\it a} & 181.556136 & $-$8.795620 & 3.8224 \\
    17{\it b} & 181.556958 & $-$8.799422 & 3.8224 \\
    18{\it a} & 181.555376 & $-$8.796714 & 4.0400 \\
    18{\it b} & 181.555927 & $-$8.798595 & 4.0400 \\
    19{\it a} & 181.562084 & $-$8.794875 & 4.0520 \\
    19{\it b} & 181.561873 & $-$8.805239 & 4.0520 \\
    19{\it c} & 181.559788 & $-$8.809463 & 4.0520 \\
    20{\it a} & 181.547472 & $-$8.800476 & 4.0553 \\
    20{\it b} & 181.556839 & $-$8.803813 & 4.0553 \\
    22{\it a} & 181.544328 & $-$8.791418 & 4.2913 \\
    22{\it b} & 181.540282 & $-$8.796562 & 4.2913 \\
    22{\it c} & 181.540884 & $-$8.806094 & 4.2913 \\
    23{\it a} & 181.563252 & $-$8.796893 & 4.7293 \\
    23{\it b} & 181.563537 & $-$8.803670 & 4.7293 \\
    23{\it c} & 181.559832 & $-$8.811526 & 4.7293 \\
    25{\it a} & 181.559714 & $-$8.796562 & 5.7927 \\
    25{\it b} & 181.560102 & $-$8.800177 & 5.7927 \\
    26{\it a} & 181.550711 & $-$8.803112 & 6.0106 \\
    26{\it b} & 181.551211 & $-$8.803668 & 6.0106 \\
    \hline
    \end{tabular}
    \label{tab:multiple_image_coordinates2}
\end{table}

\begin{table}
    \centering
\caption{Best-fit {\it LENSTOOL} parameters of the fiducial lens model.}
    \begin{tabular}{c c}
    \hline
    First cluster-scale PIEMD halo   &  \\
    \hline
    x ($\arcsec$) & $-1.420_{-0.157}^{+0.314}$\\[0.15cm]
    y ($\arcsec$) & $1.047_{-0.109}^{+0.149}$\\[0.15cm]
    $\epsilon$ & $0.598_{-0.005}^{+0.036}$ \\[0.15cm]
    $\theta$ (deg) & $19.790_{-0.268}^{+1.265}$ \\[0.15cm]
    $r_c$ (kpc) & $35.941_{-1.984}^{+0.814}$\\[0.15cm]
    $\sigma_v$ (km/s) & $986.284_{-8.487}^{+14.338}$\\
    \hline
    Second cluster-scale PIEMD halo & \\
    \hline
    x ($\arcsec$) & $-11.592_{-0.291}^{+0.347}$\\[0.15cm]
    y ($\arcsec$) & $5.729_{-2.367}^{+0.001}$\\[0.15cm]
    $\epsilon$ & $0.429_{-0.0}^{+0.076}$\\[0.15cm]
    $\theta$ (deg)& $100.850_{-1.232}^{+3.853}$\\[0.15cm]
    $r_c$ (kpc) & $212.843_{-28.724}^{+9.732}$\\[0.15cm]
    $\sigma_v$ (km/s) & $1078.762_{-63.885}^{+5.161}$ \\
    \hline
    Third cluster-scale PIEMD halo  & \\
    \hline
    x ($\arcsec$) & $29.401_{-0.437}^{+0.629}$\\[0.15cm]
    y ($\arcsec$) & $-8.171_{-0.239}^{+0.710}$ \\[0.15cm]
    $\epsilon$ & $0.453_{-0.074}^{+0.009}$\\[0.15cm]
    $\theta$ (deg)& $8.895_{-2.322}^{+3.891}$\\[0.15cm]
    $r_c$ (kpc) & $88.386_{-6.934}^{+5.336}$\\[0.15cm]
    $\sigma_v$ (km/s) & $746.233_{-18.050}^{+31.612}$\\
    \hline
    External Shear & \\
    \hline
    $\gamma_{shear}$ & $0.334_{-0.022}^{+0.031}$ \\[0.15cm]
    $\theta_{shear}$ (deg) & $92.177_{-1.357}^{+1.939}$\\
    \hline
    Galaxy members  & \\
    \hline
    $r_{g,t}^0$ (kpc) & $22.940_{-2.840}^{+2.600}$\\[0.15cm]
    $\sigma_{g,v}^0$ (km/s) & $197.907_{-11.880}^{+14.636}$\\
    \hline
    \multicolumn{2}{p{0.45\textwidth}}{Positions x and y are relative to the position of the BCG at ${\rm RA}=181.550648^\circ$ and ${\rm DEC}=-8.800952^\circ$, with positive offsets point to west and north. }
    \end{tabular}
    \label{tab:lenstool_parameters_fiducial}
\end{table}

\begin{table}
    \centering
\caption{Best-fit {\it LENSTOOL} parameters of the fine-tuned lens model. }
\begin{tabular}{c c}
    \hline
    Third cluster-scale PIEMD halo  & \\
    \hline
    x ($\arcsec$) & $13.847_{-9.301}^{+0.497}$\\[0.15cm]
    y ($\arcsec$) & $-5.008_{-1.849}^{+0.063}$ \\[0.15cm]
    $\epsilon$ & $0.573_{-0.098}^{+0.102}$\\[0.15cm]
    $\theta$ (deg)& $9.202_{-1.431}^{+1.418}$\\[0.15cm]
    $r_c$ (kpc) & $97.114_{-11.217}^{+3.834}$\\[0.15cm]
    $\sigma_v$ (km/s) & $799.654_{-43.115}^{+8.018}$\\
    \hline
    Gm1 PIEMD halo  & \\
    \hline
    $r_{g,t}$ (kpc) & $28.104_{-20.0}^{+22.0}$\\[0.15cm]
    $\sigma_{g,v}$ (km/s) & $203.412_{-63.907}^{+68.500}$\\
    \hline
    Gm2 PIEMD halo  & \\
    \hline
    $r_{g,t}$ (kpc) & $24.369_{-20.0}^{+26.0}$\\[0.15cm]
    $\sigma_{g,v}$ (km/s) & $210.705_{-52.603}^{+61.428}$\\
    \hline
    Gm3 PIEMD halo  & \\
    \hline
    $r_{g,t}$ (kpc) & $7.272_{-6.0}^{+12.8}$\\[0.15cm]
    $\sigma_{g,v}$ (km/s) & $83.873_{-23.457}^{+28.112}$\\
    \hline
    \multicolumn{2}{p{0.45\textwidth}}{Positions x and y are relative to the position of the BCG at ${\rm RA}=181.550648^\circ$ and ${\rm DEC}=-8.800952^\circ$, with positive offsets point to west and north. The first and second cluster-scale PIEMD halos, external shear, and galaxy members are fixed to their best-fit values from the fiducial model, as listed in Table \ref{tab:lenstool_parameters_fiducial}. }
    \end{tabular}
    \label{tab:lenstool_parameters_finetune}
\end{table}

\section{Photometry for individual images of Systems {\it A} and {\it B}}
In \S\ref{sec:galaxy_photometry}, we presented the photometric magnitudes of galaxies in Systems {\it A} and {\it B} after correcting the lensing magnification and averaging among multiple images.  Here in Table~\ref{tab:photometry_all}, we list the direct measurements from the data for each individual images without correcting for lensing effect. Note that the Galactic extinction is corrected for each bandpass. 

\begin{table*}
	\centering
	\caption{Photometry from {\it HST} data, directly measured for each individual image without correcting for lensing magnification. The foreground Galactic extinction is corrected (see \S\ref{sec:galaxy_photometry} for details). }
	\label{tab:photometry_all}
	\begin{tabular}{c | c c c c c c c c} % four columns, alignment for each
 		\hline
 		& F330W$^a$ & F390W & F435W & F475W & F606W & F625W & F775W & F814W \\ %& F850LP & F105W & F110W & F125W & F140W & F160W & Spitzer 3.6$\mu$m \\
 		\hline
{\it A}1 & $> 23.94$ & $24.16 \pm 0.08$ & $23.35 \pm 0.03$ & $22.94 \pm 0.02$ & $22.42 \pm 0.07$ & $22.31 \pm 0.01$ & $22.21 \pm 0.01$ & $22.22 \pm 0.01$ \\
{\it A}2{\it a} & $>26.07$ & $25.18 \pm 0.13$ & $24.57 \pm 0.05$ & $24.33 \pm 0.03$ & $24.05 \pm 0.03$ & $23.77 \pm 0.02$ & $23.72 \pm 0.03$ & $23.70 \pm 0.01$  \\
{\it A}2{\it b} & $>26.80$ & $24.87 \pm 0.09$ & $24.22 \pm 0.05$ & $24.07 \pm 0.03$ & $23.53 \pm 0.02$ & $23.31 \pm 0.02$ & $23.19 \pm 0.02$ & $23.17 \pm 0.01$  \\
{\it A}2{\it c} & $>27.26$ & $24.51 \pm 0.32$ & $23.78 \pm 0.04$ & $23.69 \pm 0.03$ & $23.22 \pm 0.02$ & $23.04 \pm 0.02$ & $23.04 \pm 0.02$ & $22.98 \pm 0.01$  \\
{\it A}3{\it a} & $>26.59$ & $25.98 \pm 0.21$ & $25.04 \pm 0.07$ & $24.67 \pm 0.05$ & $24.24 \pm 0.08$ & $24.09 \pm 0.03$ & $24.08 \pm 0.03$ & $24.05 \pm 0.02$  \\
{\it A}3{\it b} & $>27.85$ & $25.45 \pm 0.15$ & $25.03 \pm 0.10$ & $24.52 \pm 0.05$ & $23.83 \pm 0.02$ & $23.70 \pm 0.03$ & $23.51 \pm 0.03$ & $23.43 \pm 0.01$  \\
{\it A}3{\it c} & $>26.49$ & $24.99 \pm 0.11$ & $24.55 \pm 0.08$ & $24.23 \pm 0.04$ & $23.67 \pm 0.02$ & $23.54 \pm 0.03$ & $23.43 \pm 0.03$ & $23.39 \pm 0.02$  \\
		\hline
		& F850LP & F105W & F110W & F125W & F140W & F160W &  &  \\
		\hline
{\it A}1  & $22.20 \pm 0.02$ & $22.22 \pm 0.01$ & $22.20 \pm 0.01$ & $22.21 \pm 0.01$ & $22.03 \pm 0.01$ & $21.86 \pm 0.01$ &  &   \\
{\it A}2{\it a} & $23.67 \pm 0.03$ & $23.82 \pm 0.02$ & $23.86 \pm 0.02$ & $23.90 \pm 0.02$ & $23.73 \pm 0.02$ & $23.68 \pm 0.02$ &  & \\
{\it A}2{\it b} & $23.21 \pm 0.03$ & $23.20 \pm 0.01$ & $23.18 \pm 0.01$ & $23.21 \pm 0.02$ & $23.04 \pm 0.01$ & $22.92 \pm 0.01$ &  & \\
{\it A}2{\it c} & $23.01 \pm 0.03$ & $23.24 \pm 0.02$ & $23.26 \pm 0.01$ & $23.31 \pm 0.02$ & $23.14 \pm 0.01$ & $23.12 \pm 0.01$ &  & \\
{\it A}3{\it a} & $24.08 \pm 0.05$ & $24.16 \pm 0.03$ & $24.14 \pm 0.02$ & $24.22 \pm 0.03$ & $23.96 \pm 0.02$ & $23.86 \pm 0.02$ &  & \\
{\it A}3{\it b} & $23.29 \pm 0.03$ & $23.33 \pm 0.01$ & $23.30 \pm 0.01$ & $23.26 \pm 0.01$ & $23.10 \pm 0.01$ & $22.95 \pm 0.01$ &   & \\
{\it A}3{\it c} & $23.38 \pm 0.03$ & $23.54 \pm 0.02$ & $23.54 \pm 0.01$ & $23.59 \pm 0.02$ & $23.38 \pm 0.02$ & $23.15 \pm 0.01$ &  & \\
		\hline
 		& F450W$^b$ & F475W & F606W & F625W & F775W & F814W & F850LP & F105W \\ %& F850LP & F105W & F110W & F125W & F140W & F160W & Spitzer 3.6$\mu$m \\
 		\hline
{\it B}1{\it a} & $>27.42$ & $ 27.25\pm 0.32 $ & $ 26.48\pm 0.11 $ & $ 26.38\pm 0.16 $ & $ 25.95\pm 0.13 $ & $ 25.98\pm 0.09 $ & $ 25.82\pm 0.15 $ & $ 26.36\pm 0.13 $  \\
{\it B}1{\it c} & $>27.39$ & $ 27.09\pm 0.27 $ & $ 26.34\pm 0.09 $ & $ 26.07\pm 0.11 $ & $ 25.78\pm 0.11 $ & $ 26.00\pm 0.08 $ & $ 26.23\pm 0.21 $ & $ 26.48\pm 0.14 $ \\
{\it B}1{\it d} & $>27.25$ & $ 27.46\pm 0.41 $ & $ 26.43\pm 0.10 $ & $ 26.17\pm 0.13 $ & $ 26.14\pm 0.15 $ & $ 26.26\pm 0.11 $ & $ 26.15\pm 0.21 $ & $ 26.82\pm 0.19 $ \\
{\it B}1{\it e} & $>27.41$ & $ 26.97\pm 0.25 $ & $ 26.17\pm 0.24 $ & $ 25.73\pm 0.09 $ & $ 25.62\pm 0.09 $ & $ 25.77\pm 0.07 $ & $ 25.63\pm 0.13 $ & $ 25.81\pm 0.09 $ \\
{\it B}2{\it a} & $>27.39$ & $ >27.81^c $ & $ 27.04\pm 0.19 $ & $ 26.67\pm 0.21 $ & $ 26.40\pm 0.20 $ & $ 26.23\pm 0.10 $ & $ 26.78\pm 0.37 $ & $ 26.45\pm 0.14 $ \\
{\it B}2{\it c} & $>27.37$ & $ >27.82^c $ & $ 26.74\pm 0.13 $ & $ 26.85\pm 0.24 $ & $ 26.37\pm 0.18 $ & $ 26.36\pm 0.12 $ & $ 25.97\pm 0.17 $ & $ 26.25\pm 0.12 $ \\
{\it B}2{\it d} & $>27.43$ & $ 27.69\pm 0.49 $ & $ 27.24\pm 0.22 $ & $ 27.41\pm 0.41 $ & $ 26.18\pm 0.17 $ & $ 26.48\pm 0.13 $ & $ 26.08\pm 0.20 $ & $ 26.97\pm 0.22 $ \\
{\it B}2{\it e} & $>27.24$ & $ 27.62\pm 0.49 $ & $ 27.22\pm 0.20 $ & $ 26.51\pm 0.19 $ & $ 26.93\pm 0.34 $ & $ 26.80\pm 0.18 $ & $ 26.56\pm 0.30 $ & $ 26.74\pm 0.20 $ \\
        \hline
		& F110W & F125W & F140W & F160W & & & & \\
		\hline
{\it B}1{\it a} & $ 26.41\pm 0.10 $ & $ 26.24\pm 0.13 $ & $ 26.41\pm 0.13 $ & $ 26.69\pm 0.17 $ & & & & \\
{\it B}1{\it c} & $ 26.39\pm 0.09 $ & $ 26.44\pm 0.15 $ & $ 26.44\pm 0.13 $ & $ 26.30\pm 0.12 $ & & & & \\
{\it B}1{\it d} & $ 26.39\pm 0.09 $ & $ 26.67\pm 0.19 $ & $ 26.77\pm 0.17 $ & $ 26.47\pm 0.14 $ & & & & \\
{\it B}1{\it e} & $ 25.74\pm 0.10 $ & $ 25.97\pm 0.10 $ & $ 26.02\pm 0.11 $ & $ 26.26\pm 0.15 $ & & & & \\
{\it B}2{\it a} & $ 26.39\pm 0.10 $ & $ 26.49\pm 0.16 $ & $ 26.29\pm 0.11 $ & $ 26.09\pm 0.10 $ & & & & \\
{\it B}2{\it c} & $ 26.33\pm 0.09 $ & $ 26.22\pm 0.13 $ & $ 26.19\pm 0.10 $ & $ 26.00\pm 0.09 $ & & & & \\
{\it B}2{\it d} & $ 26.84\pm 0.14 $ & $ 26.88\pm 0.23 $ & $ 26.68\pm 0.16 $ & $ 26.66\pm 0.16 $ & & & & \\
{\it B}2{\it e} & $ 27.09\pm 0.10 $ & $ 26.36\pm 0.14 $ & $ 26.44\pm 0.15 $ & $ 26.45\pm 0.15 $ & & & & \\
		\hline
		\multicolumn{9}{l}{$^a$ $2\sigma$ UV flux upper limit, averaged among the F225W, F275W and F336W bandpasses.}\\
		\multicolumn{9}{l}{$^b$ $2\sigma$ UV flux upper limit, averaged among the F225W, F275W, F336W, F390W and F435W bandpasses.}\\
		\multicolumn{9}{l}{$^c$ $2\sigma$ flux upper limit.}
	\end{tabular}
\end{table*}

\clearpage
\section{Fitting a shell model to stacked spectra obtained over a large area with fixed intrinsic \lya\ line width $\sigma_i$}
\begin{figure*}
    \centering
    \includegraphics[width=0.9\linewidth]{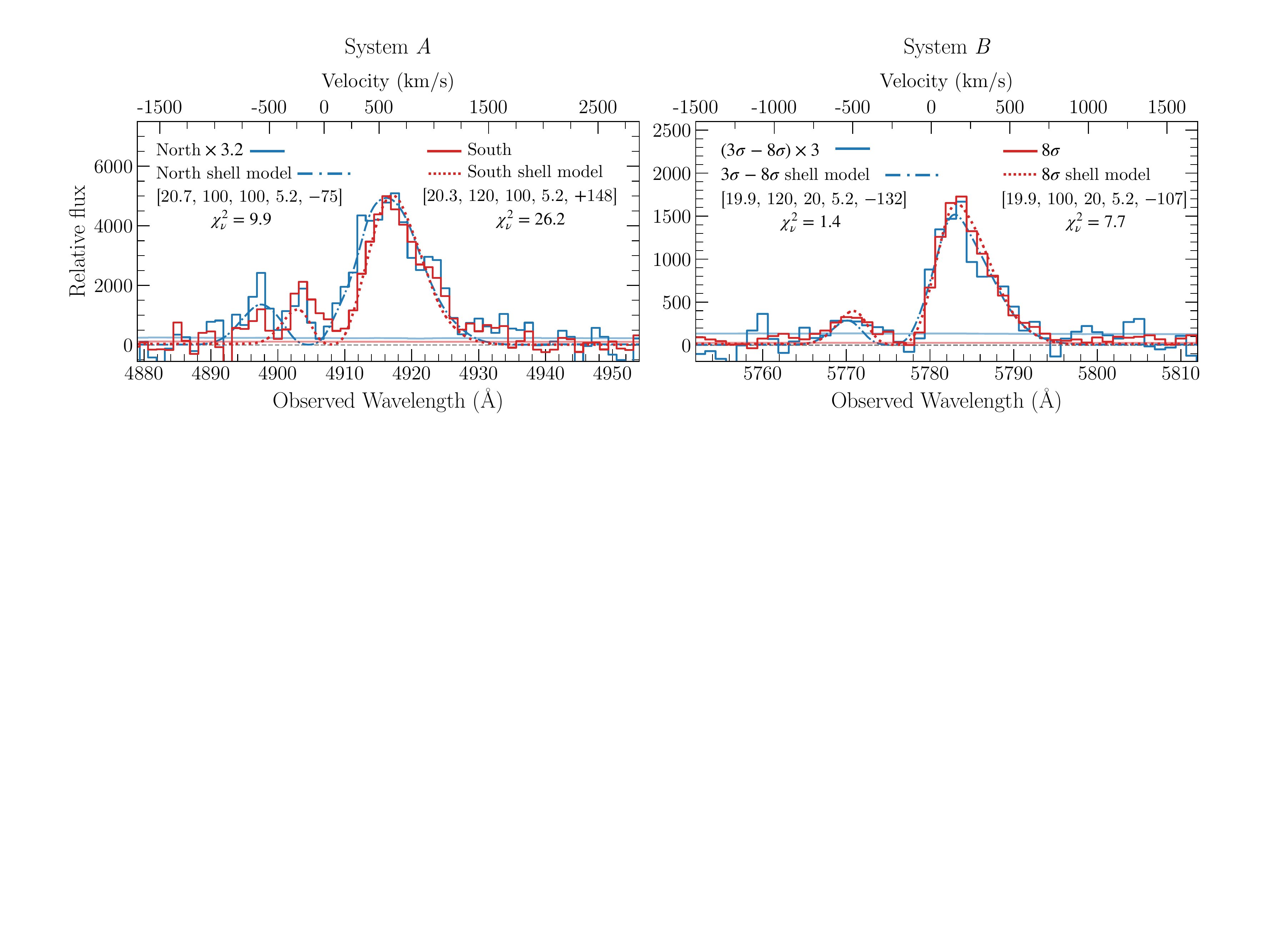}
    \caption{{\it Left:} Stacked spectra from all spaxels within the 3-$\sigma$ contour in System {\it A}, divided into northern and southern nebulae (see Figures~\ref{fig:hst_with_lya_contours_A} and \ref{fig:img_plane_contours_A}). The best-fit models are shown in dash-dotted and dotted curves for the northern and southern nebula, respectively.  To obtain the best-fit models, we fix the intrinsic \lya\ line width $\sigma_i$ to be 95 km/s, corresponding to the observed line width measured from the nebular emission lines (see \S\ref{sec:galaxy_spec} and Table~\ref{tab:line_fitting_HeII_CIII}). Compared with the best-fit models shown in Figure~\ref{fig:lya_profile_and_tlac_models} in the main text where $\sigma_i$ is a free parameter, the models shown here with a fixed $\sigma_i$ provide a worse fit to the data (particularly on the blue peak), which is also reflected with the increased $\chi^2_\nu$. {\it Right: }Stacked spectral from low- and high-surface brightness regions in System {\it B}, extracted from within and outside of the 8-$\sigma$ contours. The best-fit models are shown in dash-dotted and dotted curves for low- and high-surface brightness spectra, respectively. Similar to the models for System {\it A}, we fix $\sigma_i$ to be 20 km/s as measured from the galaxy spectrum. Although these models with fixed $\sigma_i$ also provide a worse fit to the data compared with the models presented in Figure~\ref{fig:lya_profile_and_tlac_models} where $\sigma_i$ is a free parameter, the difference in $\chi^2_\nu$ is not as significant as the difference seen in System {\it A}.  This is consistent with System {\it A} having a steeper velocity gradient across the nebulae, leading to a more significant smearing effect in the stacked spectra from a large area. }
    \label{fig:lya_profile_fixed_sigmai}
\end{figure*}

In \S\ref{sec:velocity_shell_fitting}, we have shown that the best-fit shell models for stacked spectra extracted from a large area in both Systems {\it A} and {\it B} require the intrinsic \lya\ line width $\sigma_i$ to be much larger than the observed nebular emission line width.  We argue that the large $\sigma_i$ is caused by the smearing effect due to the velocity gradient in the nebulae. Here in Figure~\ref{fig:lya_profile_fixed_sigmai}, we show the best-fit models for the same spectra shown in the top row of Figure~\ref{fig:lya_profile_and_tlac_models} in the main text, and demonstrate that by fixing $\sigma_i$ to the observed values from galaxy spectra, the best-fit models provide a worse fit to the data. 

% \clearpage

\section{\lya\ line profile from accelerating and decelerating expanding clouds}\label{sec:appendix_lya_cloud}
We present the \lya\ line profiles emergent at different distances from the center of a spherical cloud undergoing, respectively, an accelerating and decelerating expansion. The physical parameters of the cloud are $\log\,N(\ion{H}{I})/{\rm cm}^{-2}= 20$, $\sigma_i=0$ km/s (i.e., all \lya\ photons are emitted at the same frequency), and $T_{\rm eff} = 10^4$ K. The accelerating cloud has a velocity field changing from 0 km/s at the center to 400 km/s at the outer edge of the cloud with a constant radial acceleration, while the decelerating cloud has a reverse gradient changing from 0 to 400 km/s from the outer edge to the center of the cloud. We extract the emergent \lya\ line profiles in two projected distance bins from the cloud center, with an inner bin corresponding to the distance range [0, $0.5R_{\rm max}$] and an outter bin corresponding to [$0.5R_{\rm max}$, $R_{\rm max}$], where $R_{\rm max}$ is the radius of the cloud. Note that given a fixed $N(\ion{H}{I})$, changing the physical value of $R_{\rm max}$ does not change the shape of the emergent \lya\ profile and therefore $R_{\rm max}$ is not a parameter in the model. 

\begin{figure*}
    \centering
    \includegraphics[width=0.8\linewidth]{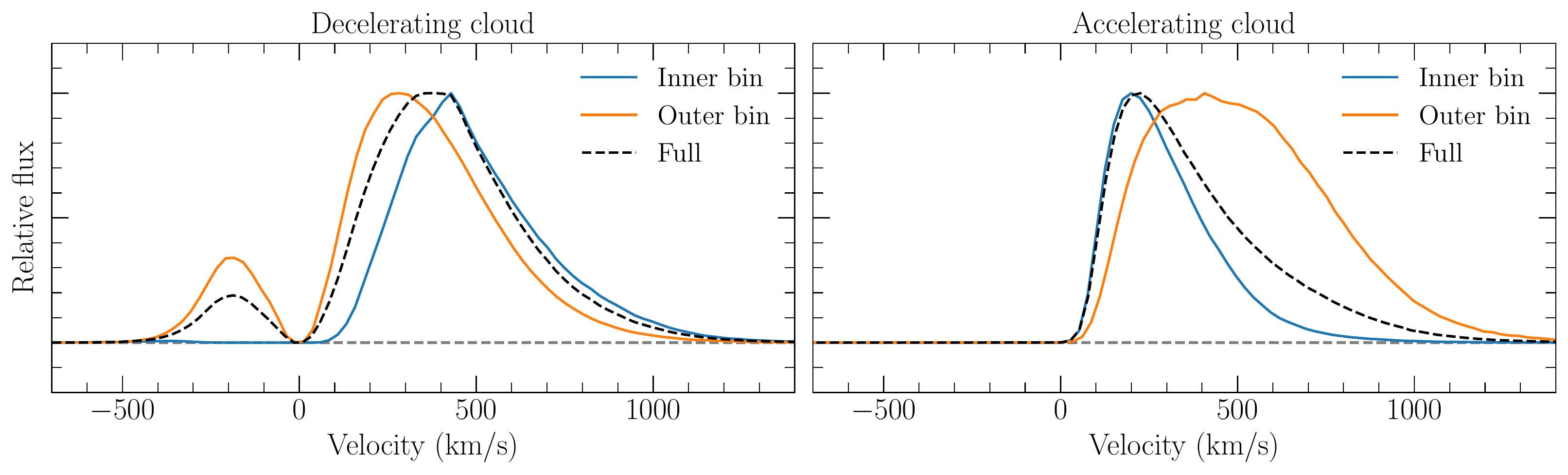}
    \caption{{\it Left:} Emergent \lya\ profiles extracted from the inner and outer bins of a decelerating expanding cloud.  The inner and outer bins correspond to the projected distance range of [0, $0.5R_{\rm max}$] and [$0.5R_{\rm max}$, $R_{\rm max}$] from the center of the cloud, respectively, where $R_{\rm max}$ is the radius of the cloud. The dominant red peak of the \lya\ profiles is more blueshifted in the outer bin. The full spectrum extracted from the entire cloud is shown in the dashed black curve.  {\it Right:} same as the left panel but for an accelerating expanding cloud. The profile is more redshifted in the outer bin, contrary to the trend observed for a decelerating cloud. }
    \label{fig:lya_accel_decel_cloud}
\end{figure*}

The \lya\ profiles from these two bins are shown in Figure~\ref{fig:lya_accel_decel_cloud}. For the decelerating cloud, the dominant red peak is more blueshifted in the outer bin, while the opposite trend is observed for the accelerating cloud. While the amount of shift in velocity and the profile shapes do not match well with the observed \lya\ profiles presented in the this work, this simple exercise demonstrates that differential velocity fields in expanding clouds might be a plausible mechanism to produce velocity gradients seen in spatially-resolved \lya\ profiles. 

%\section{Comparing System {\it B} with the LAE population}
%As discussed in \S\ref{sec:analysis_gal}, the spectrum of {\it B}1 and {\it B}2 combined demonstrate very similar features as other LAEs at redshifts $z\approx 3-6$. Here in Figure~\ref{fig:compare_B_with_Feltre}, we show a comparison between the {\it B}1+{\it B}2 spectrum on the full MUSE wavelength range and a mean spectrum stacked from LAEs with log$_{10} (L_{\rm Ly\alpha}/({\rm erg\,s^{-1}}))\leq 42.05$ identified in the deep MUSE data of the Hubble Ultra Deep Field (HUDF) \citep{Feltre2020}, which we download from the MUSE data release website\footnote{\url{http://muse-vlt.eu/science/data-releases/}}. 
%\begin{figure*}
%    \centering
%    \includegraphics[width=15cm]{fig_compare_B_Feltre+20.pdf}
%    \caption{Comparison between the {\it B}1+{\it B}2 spectrum on the full MUSE wavelength range and a mean spectrum stacked from LAEs with low \lya\ luminosity that are identified in the MUSE data of the Hubble Ultra Deep Field (HUDF). The latter is redshifted to the redshift of System {\it B}, $z=3.7540$. Data spectrum is shown in black and the corresponding error array is shown in blue. The stacked spectrum was presented in \cite{Feltre2020}}.
%    \label{fig:compare_B_with_Feltre}
%\end{figure*}

%%%%%%%%%%%%%%%%%%%%%%%%%%%%%%%%%%%%%%%%%%%%%%%%%%

% Don't change these lines
%\bsp	% typesetting comment
\label{lastpage}
\end{document}